\documentclass[aps,pre,reprint,superscriptaddress,floatfix]{revtex4-2}

\usepackage[utf8]{inputenc}

\usepackage{amssymb,amsmath}

\usepackage{bm}
\usepackage{color}
\usepackage{graphicx}
\usepackage{hyperref}

\usepackage{array}
\newcolumntype{P}[1]{>{\centering\arraybackslash}p{#1}}

\newcommand{\st}{\text{s}}

\newcommand{\ini}{\text{i}}
\newcommand{\alg}{\text{alg}}

\newcommand{\mean}[1]{\left\langle #1 \right\rangle}

\newcommand{\mbf}{m_{\text{bf}}}
\newcommand{\sbf}{\sigma_{\text{bf}}}
\newcommand{\nbf}{n_{\text{bf}}}

\DeclareMathOperator{\Tr}{Tr}

\begin{document}


\title{Strong non-exponential relaxation and memory effects in a
  fluid with non-linear drag}


\author{A. Patrón}%
\email{apatron@us.es}
\affiliation{Física Teórica, Universidad de Sevilla, Apartado de
  Correos 1065, E-41080 Sevilla, Spain}
\author{B. Sánchez-Rey}
\email{bernardo@us.es}
\affiliation{Departamento de Física Aplicada I, E.P.S., Universidad de Sevilla, Virgen de África 7, E-41011 Sevilla, Spain}
\author{A. Prados}
\email{prados@us.es}
\affiliation{Física Teórica, Universidad de Sevilla, Apartado de
  Correos 1065, E-41080 Sevilla, Spain}


\date{\today}

\begin{abstract}
  We analyse the dynamical evolution of a fluid with non-linear drag, for which binary collisions are elastic, {described at the kinetic level by the Enskog-Fokker-Planck equation}. This model system, rooted in the theory of non-linear Brownian motion, displays a really complex behaviour when quenched to low temperatures. Its glassy response is controlled by a long-lived non-equilibrium state, independent of the degree of non-linearity {and also of the Brownian-Brownian collisions rate. The latter property entails that this behaviour persists in the collisionless case, where the fluid is described by the non-linear Fokker-Planck equation.} The observed response, which includes non-exponential, algebraic, relaxation and strong memory effects, presents {scaling} properties: the time evolution of the temperature---for both relaxation and memory effects---falls onto a master curve, regardless of the details of the experiment. To account for the observed behaviour in simulations, it is necessary to develop an extended Sonine approximation for the kinetic equation---which considers not only the fourth cumulant but also the sixth one.
\end{abstract}


\maketitle


\section{Introduction}\label{sec:intro}

Glassy behaviour is typically associated with systems with many strongly interacting units, which give rise to a complex energy landscape with multiple minima separated by barriers~\cite{stillinger_glass_2013,lubchenko_theory_2015,nagel_experimental_2017}. The typical phenomenology of glassy systems includes, among other aspects, strongly non-exponential relaxation~\cite{williams_non-symmetrical_1970,palmer_models_1984,kob_dynamics_1990,brey_stretched_1993,brey_low-temperature_1996,angell_relaxation_2000,brey_slow_2001,richert_physical_2010,paeng_ideal_2015,lahini_nonmonotonic_2017,kringle_structural_2021,nishikawa_relaxation_2021}. The latter facilitates the emergence of memory effects like the Kovacs hump~\cite{kovacs_transition_1963,kovacs_isobaric_1979,buhot_kovacs_2003,bertin_kovacs_2003,arenzon_kovacs_2004,mossa_crossover_2004,aquino_kovacs_2006,bouchbinder_nonequilibrium_2010,prados_kovacs_2010,diezemann_memory_2011,ruiz-garcia_kovacs_2014,lulli_spatial_2020,morgan_glassy_2020,song_activation_2020,peyrard_memory_2020,mandal_memory_2021}. 

In the Kovacs experiment~\cite{kovacs_transition_1963,kovacs_isobaric_1979}, the time evolution of a relevant physical quantity $P(t)$ is monitored. The system is initially equilibrated at the temperature $T_{\ini}$. For $0<t< t_w$, the system is aged at a lower temperature $T_1$. At $t=t_w$, the bath temperature is suddenly changed to $T_w$, such that the instantaneous value of $P$, $P(t_w)$, equals its equilibrium value for $T_w$. The Kovacs effect emerges when $P$, despite having its equilibrium value at $t=t_w$, displays a non-monotonic behaviour for $t>t_w$, i.e. a hump, before returning to equilibrium~\footnote{Figure 1 of Ref.~\cite{prados_kovacs-like_2014} gives a qualitative picture of the Kovacs protocol.}. This is so because the evolution of the system does not only depend on the value of the thermodynamic (or hydrodynamic) variables but also on additional ones, the values of which are determined by the way the system has been previously aged~\cite{bouchaud_weak_1992,cugliandolo_analytical_1993,prados_aging_1997,nicodemi_aging_1999,mossa_crossover_2004,ahmad_velocity_2007,brey_scaling_2007,richert_physical_2010,parravicini_rejuvenation_2012,lahini_nonmonotonic_2017,dillavou_nonmonotonic_2018}.

Aging is also connected with the Mpemba effect~\cite{mpemba_cool_1969}, which has recently been observed in spin glasses~\cite{baity-jesi_mpemba_2019}. In the Mpemba effect, the initially hotter sample cools sooner and the relaxation curves thus cross at a certain time. Only very recently has it been  theoretically investigated, both from a stochastic thermodynamics~\cite{lu_nonequilibrium_2017,klich_mpemba_2019,gal_precooling_2020,kumar_exponentially_2020} and a kinetic theory~\cite{lasanta_when_2017,torrente_large_2019,santos_mpemba_2020, biswas_mpemba_2020,biswas_mpemba_2021,gomez_gonzalez_mpemba-like_2021,takada_mpemba_2021} approach. The former describes the crossing in terms of the Kullback-Leibler distance to equilibrium. The latter describes the crossing in terms of the kinetic temperature, which is closer to the experimental situation. Moreover, it has been succesful in showing that the Mpemba effect comes about in very simple systems like granular gases~\cite{lasanta_when_2017,torrente_large_2019,santos_mpemba_2020,biswas_mpemba_2020,biswas_mpemba_2021,gomez_gonzalez_mpemba-like_2021}. Notwithstanding, the following crucial question remains unanswered: How does the system have to be aged for the Mpemba effect to emerge? This is one key question that we solve in this paper.

We analyse a very general model---rooted in the theory of non-linear Brownian motion~\cite{klimontovich_nonlinear_1994}---for a fluid with non-linear drag force. {From a phenomenological point of view, it can be regarded as the minimal, simplest, model for a fluid with non-linear drag~\cite{klimontovich_statistical_1995,lindner_diffusion_2007,goychuk_nonequilibrium_2021}. From a more fundamental point of view, it arises when an ensemble of Brownian particles, with mass $m$ and particle density $n$, is immersed in an isotropic and uniform background fluid at equilibrium with temperature $T_{\st}$, the particles of which have masses $\mbf$~\cite{ferrari_particles_2007,ferrari_particles_2014}. In the so-called Rayleigh limit, where $\mbf/m\to 0$, the drag force on the Brownian particles is linear in the velocity, $\bm{F}_{\text{drag}}=-m\zeta_0 \bm{v}$, i.e. the drag coefficient $\zeta_0$ is a constant. Still, in a real situation $\mbf/m\ne 0$, and it is thus relevant to consider the corrections to the Rayleigh limit. Specifically, by introducing the first order corrections thereto, i.e. by retaining linear terms in $\mbf/m$ but neglecting $(\mbf/m)^2$ and higher-order terms, the drag force is found to be of the form,
\begin{equation}\label{eq:zeta-v}
    \bm{F}_{\text{drag}}=-m\zeta(v) \bm{v}, \quad 
\zeta(v)=\zeta_0\left(1+\gamma \frac{m v^2}{k_BT_{\st}}\right),
\end{equation}
sometimes called the quasi-Rayleigh limit. The non-linear parameter $\gamma$ is given as a certain integral that includes the Brownian-particle-background-particle differential cross section~\cite{ferrari_particles_2007,ferrari_particles_2014,hohmann_individual_2017}, and typical values are limited to $\gamma\lesssim 0.1-0.2$~\cite{santos_mpemba_2020}.

In this way, the velocity distribution function (VDF) for the Brownian particles obeys the Fokker-Planck (FP) equation
\begin{equation}
\label{eq:FP}
\partial_t f(\bm{v},t)=\frac{\partial}{\partial \bm{v}}\cdot\left[
\zeta(v)\left(\bm{v} +\frac{k_B T_{\st}}{m}\frac{\partial}{\partial
    \bm{v}}\right)f(\bm{v},t)\right].
\end{equation}
The interaction between the Brownian and the background fluid particles gives rise to both the nonlinear drag force
$\bm{F}_{\text{drag}}=-m\zeta(v)\bm{v}$ and the white-noise stochastic force $\bm{F}_{\text{wn}}$. Its correlation $\langle \bm{F}_{\text{wn}}(t) \bm{F}_{\text{wn}}(t')\rangle= 2mk_{B}T_{\st}\zeta(v)\delta(t-t')$, where $k_{B}$ is Boltzmann's constant, follows from the fluctuation-dissipation relation~\cite{klimontovich_nonlinear_1994} and ensures that the only stationary solution of Eq.~\eqref{eq:EFP} is the equilibrium Maxwellian, $f_\st(\bm{v})=n\left(2\pi k_B T_{\st}/m\right)^{-d/2} \exp(-mv^2/2k_B T_{\st})$.} 

{The model described above can be visualised as a mixture of two fluids: a fluid of Brownian particles moving in a background fluid acting as a thermal bath, with the masses of the Brownian and the background fluid particles being comparable. In fact, this is the physical situation for the mixture of ultracold atoms in Ref.~\cite{hohmann_individual_2017}, in which an ensemble of $^{133}$Cs atoms moves in a dilute background cloud of $^{87}$Rb atoms. Despite the very low temperatures involved---in the $\mu$K range---the low density makes it possible to describe the system with the tools of classical statistical mechanics, namely the FP equation \eqref{eq:FP}. 

However, the FP description does not take into account Brownian-Brownian collisions.  Here we consider that the Brownian particles are $d$-dimensional hard spheres and model their dynamics via the Enskog-Fokker-Planck (EFP) equation 
\begin{equation}
\label{eq:EFP}
\partial_t f(\bm{v},t)=\frac{\partial}{\partial \bm{v}}\cdot\left[
\zeta(v)\left(\bm{v} +\frac{k_B T_{\st}}{m}\frac{\partial}{\partial
    \bm{v}}\right)f(\bm{v},t)\right] +J[\bm{v}|f,f].
\end{equation}
The Enskog collision operator $J[\bm{v}|f,f]$ accounts for the collisions among the mutually interacting Brownian particles,
\begin{align}
    \label{collision-operator}
    J[\bm{v}_1|f,f] \equiv \sigma^{d-1} g(\sigma &) \!\! \int \!\! d \bm{v}_2 \! \int \!\!
    d \widehat{\bm{\sigma}} \Theta (\bm{v}_{12}\cdot \widehat{\bm{\sigma}})  \bm{v}_{12}\cdot \widehat{\bm{\sigma}} \nonumber \\ & \;\times\!\!\left[f(\bm{v}_1',t)f(\bm{v}_2',t)\!-\!f(\bm{v}_1,t)f(\bm{v}_2,t)\right].
\end{align}
Above, $g(\sigma) = \lim_{r\rightarrow \sigma^+} g(r)$ is the contact value of the pair correlation function $g(r)$, $\Theta$ is the Heaviside function, $\bm{v}_{12} \equiv \bm{v}_1 - \bm{v}_2$ is the relative velocity, and $\bm{v}_1' = \bm{v}_1 - (\bm{v}_{12}\cdot \widehat{\bm{\sigma}})\widehat{\bm{\sigma}}$, $\bm{v}_2' = \bm{v}_2 + (\bm{v}_{12}\cdot \widehat{\bm{\sigma}})\widehat{\bm{\sigma}}$ are the post-collisional velocities. 
}

{The EFP equation \eqref{eq:EFP} has been previously employed for describing both molecular fluids and heated granular gases~\cite{van_noije_velocity_1998,montanero_computer_2000,poschel_granular_2001,garcia_de_soria_universal_2012,marconi_about_2013,prados_kovacs-like_2014,lasanta_when_2017,santos_mpemba_2020}. It can be considered as a reasonable model that interpolates between two limiting cases---the FP equation and the Enskog (or Boltzmann) equation. In particular, the EFP equation reduces to the FP equation \eqref{eq:FP} in the limit of vanishing (Brownian-Brownian) collision rate.}

The energy landscape of the Brownian particles is very simple, its energy being only kinetic. Still, there appears a strong non-exponential relaxation when the system is quenched to low enough temperatures. Moreover, this non-exponential relaxation is universal in the sense that, after a suitable rescaling of the variables, it does not depend on the initial and final temperatures, nor on the degree of non-linearity, nor on the relevance of the collision term. Interestingly, it is also closely linked to the existence of a long-lived non-equilibrium state (LLNES). Therein, the higher cumulants of the VDF are basically time-independent while the temperature is algebraically decaying. Besides, the LLNES rules the emergence of strong memory effects. Specifically, we investigate the Mpemba and the Kovacs effects, which are also shown to display {scaling features.}

{The glassy behaviour described above---non-exponential relaxation and strong memory effects, linked to the LLNES---will be obtained using the framework of the EFP equation. Though, we will show that these physically appealing results also hold in absence of the collision term, i.e. for the FP equation. In this way, the relevance of the LLNES and its associated glassy behaviour is reinforced.}

{The paper is organised as follows. In Sec.~\ref{sec:evol-eqs-ext-Sonine} we put forward the evolution equations for the temperature and the cumulants in the extended Sonine framework. The quench to low temperatures is analysed in Sec.~\ref{sec:quench-low-temp}. First, in Sec.~\ref{sec:approx-evol-eqs}, we derive the approximate system of evolution equations in this limit. Second, we show how the LLNES and the strongly non-exponential relaxation emerge in Sec.~\ref{sec:univ-nonexp-relax}. Memory effects are the focus of Sec.~\ref{sec:memory-effects}, \ref{sec:Mpemba} for the Mpemba effect and \ref{sec:Kovacs} for the Kovacs effect. Section~\ref{sec:collisionless} is devoted to the study of the relevance of collisions and the Fokker-Planck limit. The main conclusions of our work and a physical discussion of our results are presented in Sec.~\ref{sec:conclusions}. Finally, the Appendices deal with some technical aspects and complementary material, non-essential for the understanding of the results in the main text.
}

\section{Evolution equations for the temperature and the cumulants}\label{sec:evol-eqs-ext-Sonine}

{In this section, we derive the evolution equations for the relevant physical variables.}  The kinetic temperature $T(t)$ is given by $\mean{v^2}=d k_B T/m$. It is useful for our purposes to scale velocities with the thermal velocity $v_T(t)$ by defining 
\begin{equation}
    \bm{c}\equiv \bm{v}/v_T(t), \quad v_T(t) \equiv \sqrt{2k_BT(t)/m},
\end{equation}
which implies $\mean{c^2}=d/2$. In addition, we introduce dimensionless temperature and time, 
\begin{equation}\label{eq:T-t-dimensionless}
    \theta=T/T_{\st}, \quad t^{*}=\zeta_{0}t,
\end{equation}
we drop the asterisk in the following to simplify the notation. For isotropic states, the reduced VDF $\phi(\bm{c},t)\equiv  n^{-1}v_T^d(t) f(\bm{v},t)$ can be expanded in a complete set of orthogonal polynomials as
\begin{equation}
    \label{eq:phiC-Sonine}
    \phi(\bm{c},t) =\pi^{-d/2}e^{-c^2}\left[1 + \sum_{l=2}^{\infty} a_l(t) L_l^{\frac{d-2}{2}}(c^2) \right],
\end{equation}
where $L_l^{(\alpha)}$ are the generalised Laguerre (or Sonine) polynomials~\cite{abramowitz_handbook_1988}. In the simplest---and usual---first Sonine approximation, only the fourth cumulant or excess kurtosis $a_{2}$, 
\begin{equation}
a_{2}\equiv -1+ \frac{4}{d(d+2)}\mean{c^{4}},
\end{equation}
is retained and higher-order cumulants are neglected. Unfortunately, this approximation fails to reproduce  the  behaviour observed in the simulations~\footnote{This approximation was employed in  Ref.~\cite{santos_mpemba_2020} to analytically investigate the Mpemba effect.}, as shown in Appendix~\ref{app:Sonine-expansion}. Then, we must consider an \textit{extended} Sonine approximation, in which not only  $a_{2}$ but also the sixth cumulant
\begin{equation}
a_{3}\equiv 1+3a_{2}-\frac{8}{d(d+2)(d+4)}\mean{c^{6}}
\end{equation} 
are retained. 

{The parameter $\zeta_{0}$ that we have employed to non-dimensionalise time marks one of the two characteristic times in this system: the time $\zeta_0^{-1}$ over which the Brownian particles feel the drag due to the background fluid, $\zeta_{0}^{-1}\propto T_{\st}^{-1/2}$~\cite{ferrari_particles_2007,ferrari_particles_2014,hohmann_individual_2017}. The other characteristic time is set by the collision frequency among the Brownian particles at the steady state $\nu_s \equiv g(\sigma)n\sigma^{d-1}\sqrt{2k_BTs/m}$~\cite{resibois_classical_1977}. The average time between collisions at the steady state is  $\tau_{\st}\equiv \nu_{\st}^{-1}\propto T_{\st}^{-1/2}$~\cite{santos_mpemba_2020}. The dimensionless average time between Brownian-Brownian collisions is thus given by
\begin{equation}\label{eq:xi-def}
    \xi\equiv\zeta_{0}\tau_{\st}.
\end{equation}
Equivalently,  $\xi^{-1}$ is the dimensionless Brownian-Brownian collision rate. This parameter $\xi$ is independent of $T_{\st}$---see also Appendix~\ref{app:Sonine-expansion}.}

Within the extended Sonine approximation, the following evolution equations for $(\theta,a_{2},a_{3})$ hold
\cite{megias_unpublished},
\begin{subequations}\label{eq:evol-eqs-with-a3}
  \begin{align}
\label{evol-eq-T-a3}
    \dot{\theta}&=2(1-\theta)\left[ 1 +\gamma (d+2)\theta \right] - 2\gamma (d+2)\theta^2 a_2,
\\
\label{evol-eq-a2-a3}
    \dot{a}_2 &= 8\gamma (1-\theta)-\! \left[ \frac{4}{\theta} - 8\gamma + 4\gamma (d+8)\theta + \frac{8(d-1)}{d(d+2)}\frac{\sqrt{\theta}}{\xi} \right]\! a_2 \nonumber
    \\
    & \quad+ 2\!\left[2\gamma \theta (d+4) + \frac{(d-1)}{d(d+2)}\frac{\sqrt{\theta}}{\xi} \right]\!a_3,
\\
\label{evol-eq-a3-a3}
  \dot{a}_3 &=   12\!\left[-4\gamma+6\gamma\theta+
                 \frac{(d-1)}{d(d+2)(d+4)}
                 \frac{\sqrt{\theta}}{\xi}\right]\!a_{2}\nonumber\\
                 & \quad+6\!\left[4\gamma-\frac{1}{\theta}-\gamma\theta(d+14)-
                   \frac{(d-1)(4d+19)}{2d(d+2)(d+4)}
                   \frac{\sqrt{\theta}}{\xi}
                   \right]\! a_{3},
\end{align}
\end{subequations}
{Substituting $\xi=\infty$ into Eq.~\eqref{eq:evol-eqs-with-a3} gives the evolution equations for the collisionless EFP equation, i.e. for the FP equation~\eqref{eq:FP}}. In other words, collisions among the Brownian particles are basically negligible when the dimensionless average time between them is very long, i.e. $\xi\gg 1$~\footnote{For the ultracold  gas mixture considered in Ref.~\cite{hohmann_individual_2017},  $\xi\simeq 674$, and the system thus corresponds to this limit.}. The equilibrium solution of this system is $(\theta_{\st}=1,a_{2}^{\st}=a_{3}^{\st}=0)$, the equilibrium VDF is Gaussian, for all values of the parameters $\gamma$ and $\xi$.

For linear drag, $\gamma=0$, the temperature obeys Newton's law of cooling, $\dot{\theta}=2(1-\theta)$. Therefore, it relaxes exponentially to equilibrium, $\theta(t)=1+\left[\theta_{\ini}-1\right]e^{-2t}$, for all $\theta_{\ini}\equiv\theta(0)$. Moreover, the VDF remains Gaussian, $a_{2}(t)=a_{3}(t)=0$. For non-linear drag, one typically has $\gamma\lesssim 0.1$~\footnote{In the three-dimensional case, $\gamma=0.1$ for self-diffusion (equal masses). For the ultracold gas mixture considered in Ref.~\cite{hohmann_individual_2017}, $\gamma\simeq 0.067$.}. {If the initial and final temperatures are of the same order, $\theta_{\ini}=O(1)$, small values of the cumulants and mild deviations from the exponential behaviour are observed, see Appendix~\ref{app:Sonine-expansion}. Therefrom,} one might guess that both the deviations from the exponential relaxation and the Gaussian VDF should always be small: we show in the following that this intuition is utterly wrong. There emerges a strong non-exponential relaxation together with quite large, time- and {$(\gamma,\xi)$-independent}, cumulant values when the system is quenched to a low temperature.

\section{Quench to low temperatures}\label{sec:quench-low-temp}

\subsection{Scaled evolution equations}\label{sec:approx-evol-eqs}

{Glassy behaviour, slow non-exponential relaxation functions, and their associated memory effects such as the Kovacs or Mpemba effects ~\cite{kovacs_isobaric_1979,kovacs_transition_1963,mpemba_cool_1969}, usually arise for low enough temperatures. For a review, see for instance~\cite{ritort_glassy_2003}. For the case of our concern, this translates into considering a quench to low temperatures, i.e. we consider the limit $\theta_{\ini} = T(0)/T_{\st} \gg 1$~\footnote{Let us note that this limit is analogous to the ``cooling'' protocol considered in the literature to investigate the emergence of the Kovacs hump in a uniformly heated granular gas~\cite{prados_kovacs-like_2014}.}.
 } 

{In order to look into the limit $\theta_{\ini}\gg 1$, it is convenient to define the scaled temperature
\begin{equation}
    \label{eq:scaled-temperature}
    Y = \theta / \theta_{\ini}.
\end{equation}
Initially $Y(0)=1$ and  $Y$ remains of the order of unity for not too long times. In fact, this quantity gives the overall relaxation of the temperature. If one defined a normalised relaxation function in the standard way,
\begin{equation}
    \varphi(t)\equiv\frac{T(t)-T_{\st}}{T_{\ini}-T_{\st}}=\frac{\theta(t)-1}{\theta_{\ini}-1}, \qquad \varphi(0)=1, \quad \varphi(\infty)=0,
\end{equation}
we have that $\varphi(t)\simeq Y(t)$ as long as $\theta(t)\gg 1$. They only differ for very long times, when $\theta$ is close to the steady state and takes order of unity values, in fact $\lim_{t\to\infty}Y(t)=\theta_{\ini}^{-1}\ll 1$.
}

{Insertion of this scaling into the evolution equations leads to
\begin{subequations}\label{eq:evol-eqs-dominant}
\begin{align}
    \label{evol-eq-T-dominant}
    \dot{Y} = &-2\gamma \theta_{\ini} (d+2) Y^2(1+a_2) + O(1) + O(\gamma),
    \\
    \dot{a}_2  =& -4\gamma \theta_{\ini} Y \left[ (d+8) (a_2 - a_2^r) - (d+4)(a_3-a_3^r) \right] \nonumber \\ & + O(\gamma) + O(\sqrt{\theta_{\ini}}/\xi),
    \\
  \dot{a}_3  =& -6\gamma \theta_{\ini} Y \left[-12(a_2-a_2^r) + (d+14)(a_3-a_3^r)\right] \nonumber \\
  & + O(\gamma) + O(\sqrt{\theta_{\ini}}/\xi),
\end{align}
\end{subequations}
the dominant terms on the rhs are  the order of $\gamma\theta_{\ini}\gg 1$~\footnote{These dominant terms correspond to the quadratic in $\theta$ ones in Eq.~\eqref{evol-eq-T-a3} and linear in $\theta$ ones in Eqs.~\eqref{evol-eq-a2-a3} and \eqref{evol-eq-a3-a3}.}. The above system of coupled ODEs suggests that the relevant time scale is no longer $t$, but a new \textit{scaled time} $s$ given by
\begin{equation}
    \label{eq:scaled-time}
    s = \gamma \theta_{\ini} t.
\end{equation}
Retaining only the dominant terms in Eqs.~\eqref{eq:evol-eqs-dominant}, one gets the approximate system}
\begin{subequations}\label{eq:evol-eqs-s-scale}
\begin{align}
  \frac{dY}{ds}&=-2(d+2)Y^{2}(1+a_{2}),
  \label{eq:evol-eqs-s-scale-a}\\
  \frac{da_{2}}{ds}&=-4Y \left[(d+8)(a_{2}-a_{2}^{r})-
                     (d+4)(a_{3}-a_{3}^{r}) \right],
  \label{eq:evol-eqs-s-scale-b}\\
  \frac{da_{3}}{ds}&=-6Y
  \left[-12(a_{2}-a_{2}^{r})+ (d+14)(a_{3}-a_{3}^{r}) \right],
  \label{eq:evol-eqs-s-scale-c}                 
\end{align}
\end{subequations}
where
\begin{equation}
  \label{eq:a2r-a3r}
  a_{2}^{r}\equiv-\frac{2(d+14)}{d^{2}+10d+64}, \qquad a_{3}^{r}\equiv
  -\frac{24}{d^{2}+10d+64},
\end{equation}
are the pseudostationary values obtained by imposing $da_2/ds=da_3/ds=0$. Specifically, for $d=2$, $a_{2}^{r}\simeq -0.36$ and
$a_{3}^{r}\simeq -0.27$. 

Note that the right hand side of Eqs.~\eqref{eq:evol-eqs-s-scale} does not depend on $\gamma$; such dependence has been absorbed into the time scale $s$. {In addition, nor does it depend on $\xi$, i.e. these equations are also valid for the collisionless case $\xi=\infty$, where the FP equation~\eqref{eq:FP} applies. 
}

\subsection{Universal non-exponential relaxation and long-lived non-equilibrium state}\label{sec:univ-nonexp-relax}

\begin{figure}
  \centering
  \includegraphics[width=3.25in]{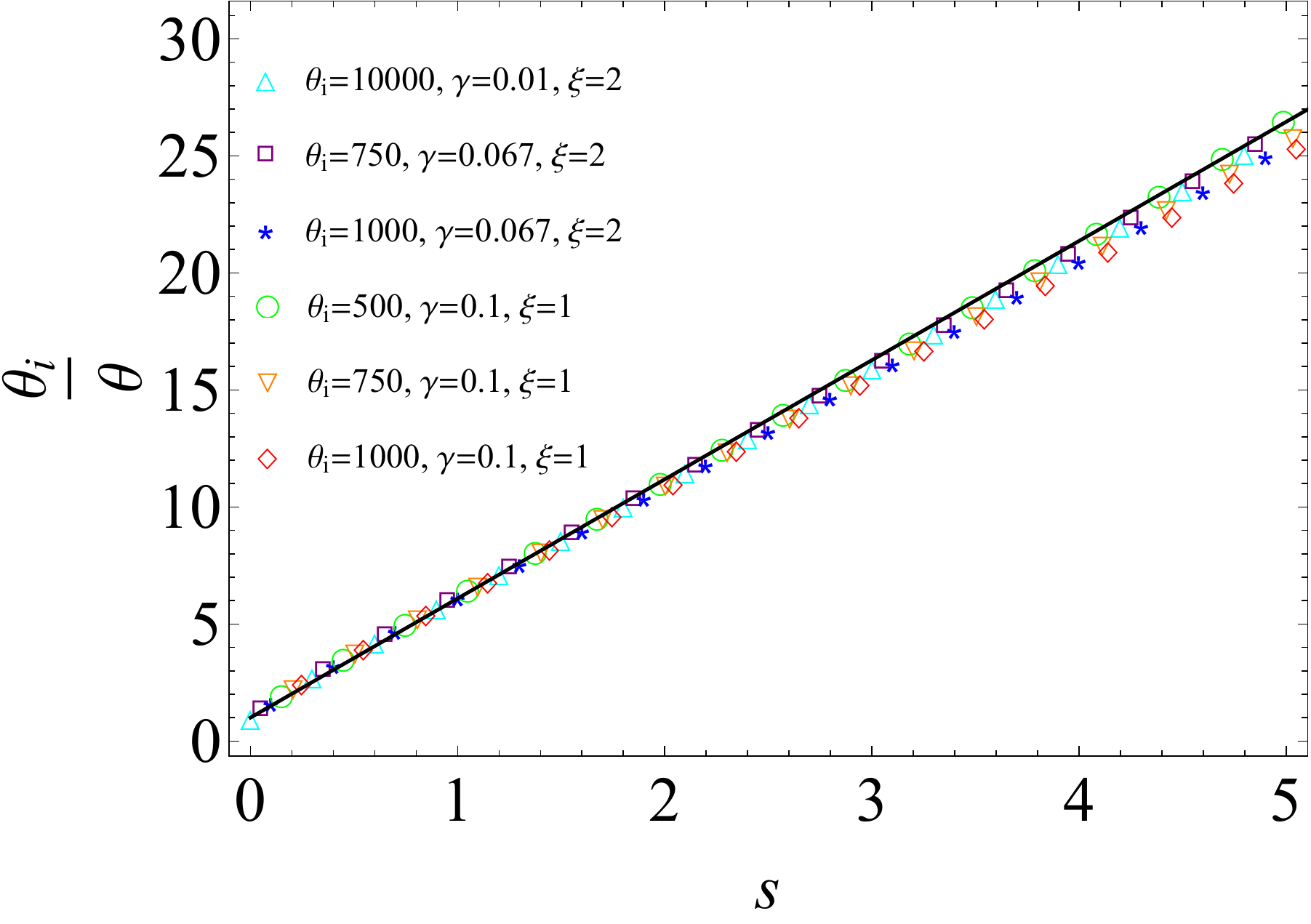}
  \caption{\label{fig:universal-behaviour-temp} Relaxation after a
    quench to a low temperature. Specifically, we plot
    $1/Y=\theta_{\ini}/\theta$ as a function of the scaled time
    $s=\gamma\theta_{\ini}t$. Data from DSMC correspond to parameters $(\theta_{\ini},\gamma,\xi)$, as specified in the legend, and $d=2$. Also plotted is the theoretical prediction in Eq.~\eqref{eq:Y(s)} (solid line). The linear behaviour of $1/Y$ means that the temperature relaxes algebraically, basically as $t^{-1}$.
    }
\end{figure}

The relaxation of the system is universal in the following sense: all the relaxation curves of the temperature should be superimposed when $Y=\theta/\theta_{\ini}$ is plotted against $s=\gamma\theta_{\ini} t$, {independently of the 
values of $\gamma$ and $\xi$}.  This universality is checked in Fig.~\ref{fig:universal-behaviour-temp}, in which several relaxation curves are shown. They have been obtained by numerically solving the kinetic equation with the Direct Simulation Monte Carlo (DSMC) method~\cite{bird_g_a_molecular_1994,montanero_monte_1996}. Specifically, we plot $1/Y$ versus $s$, for values of $\gamma$ and $\theta_{\ini}$ such that  $50\leq\gamma\theta_i\leq 100$, $0.01\leq\gamma\leq 0.1$, and $1\leq\xi\leq 2$. A clear linear behaviour arises, i.e. $Y(s)$ shows a very slow algebraic decay, basically proportional to $s^{-1}$ or, equivalently, $t^{-1}$. A similar behaviour has been recently found for the relaxation dynamics of several glass-forming models to their inherent structures~\cite{nishikawa_relaxation_2021}.

This strongly non-exponential relaxation can be theoretically understood as follows:  the cumulants rapidly tend (over the $s$ scale) to their reference values $a_{2}^{r}$ and $a_{3}^{r}$, {as shown below}. Setting $a_{2}=a_{2}^{r}$ in Eq.~\eqref{eq:evol-eqs-s-scale-a}, we get
\begin{equation}\label{eq:Y(s)}
  Y(s)=Y_{\alg}(s)\equiv\frac{1}{1+2(d+2)(1+a_{2}^{r})s}.
\end{equation}
This theoretical prediction is also plotted in Fig.~\ref{fig:universal-behaviour-temp}, where it is neatly observed that the agreement with the numerical results is excellent. Looking at Eq.~\eqref{evol-eq-T-a3}, one sees that $\dot{\theta}$ is basically proportional to $\theta^2$ for $\theta\gg 1$: this is the reason why the algebraic $t^{-1}$ relaxation emerges.

\begin{figure} 
{\centering \includegraphics[width=1.625in]{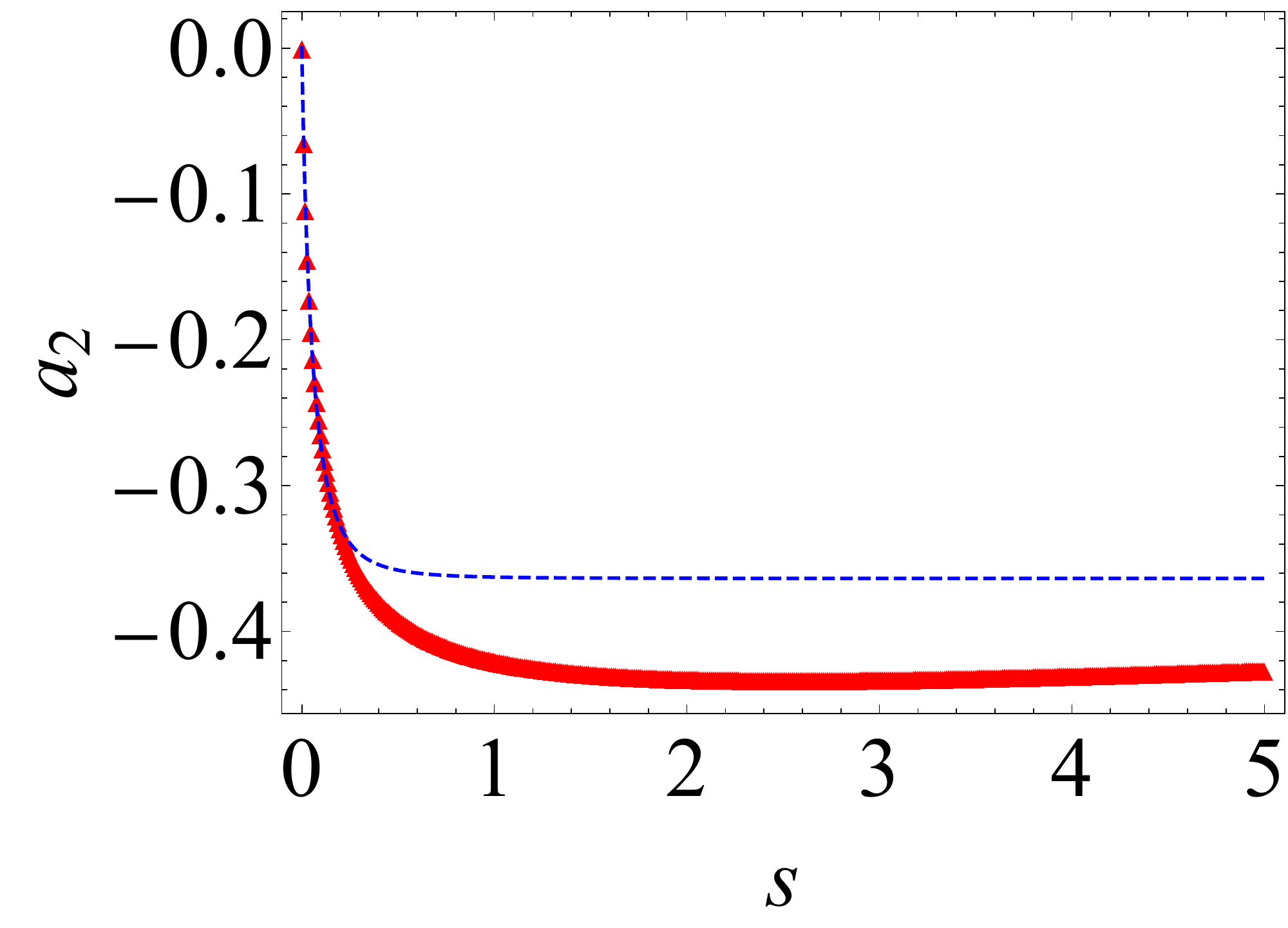} \includegraphics[width=1.625in]{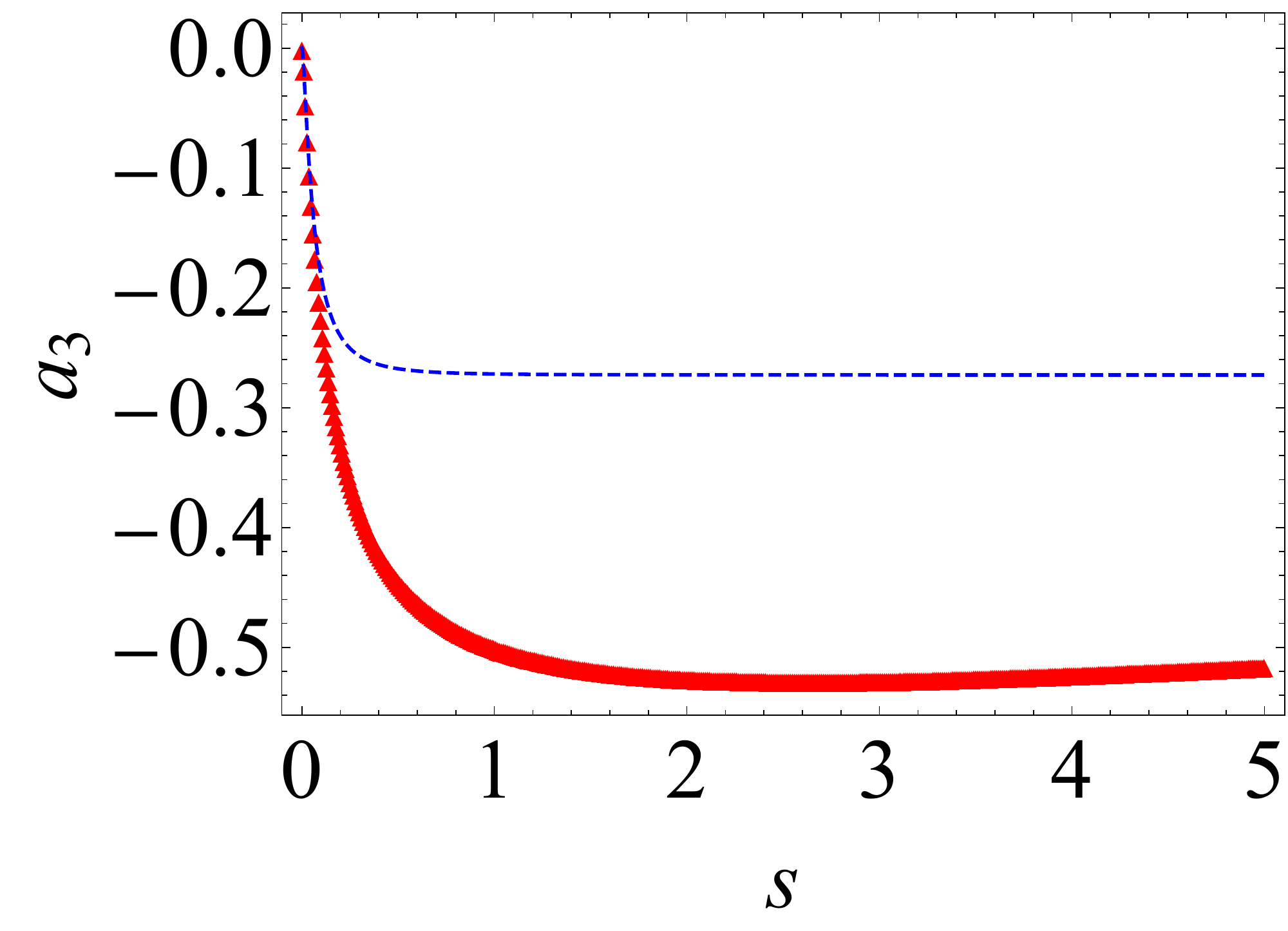} }
\caption{Relaxation of the excess kurtosis (left panel) and the sixth cumulant (right panel). Both $a_2$ and $a_3$ decay towards their respective reference values, which characterise the LLNES. Symbols correspond to DSMC data for $\theta_{\ini} = 1000$, while the dashed lines correspond to the numerical integration of the scaled evolution equations \eqref{eq:evol-eqs-s-scale}. Additional employed parameters are $d=2$, $\gamma = 0.1$ and $\xi = 1$. The actual LLNES obtained through DSMC is characterised by larger (in absolute value) values of the cumulants than those predicted by the extended theory. In particular, the extended Sonine approximation underestimates $a_2^r$ by approximately 15 per cent.}
    \label{fig:LLNES-curves}
\end{figure}
{ Substituting $a_{2}$ with its pseudo-stationary, reference, value $a_{2}^{r}$ is justified by looking into the time evolution, over the $s$ scale, of the cumulants---see also Appendix~\ref{app:fast-relax-LLNES}. This is done in Fig.~\ref{fig:LLNES-curves}, which shows the same time window $0\leq s\leq 5$ of Fig.~\ref{fig:universal-behaviour-temp}.  Both data from DSMC simulations and the numerical integration of the approximate system~\eqref{eq:evol-eqs-s-scale} are plotted. It is neatly seen that both cumulants, $a_2$ and $a_3$, rapidly become negative and quite large, being roughly constant for $s\geq 1$.  Note that, on the other hand, the temperature is reduced by a factor of $25$ from its initial value $\theta_{\ini}$ in Fig.~\ref{fig:LLNES-curves}.
}

{In Fig.~\ref{fig:LLNES-curves}, there appear some discrepancies between the DSMC data and the prediction from the extended Sonine approximation for the cumulants. These discrepancies mainly stem from the truncation done in the latter---i.e. our neglecting of $a_n$ for $n\geq 4$~\footnote{For a more detailed discussion of this issue, see Appendix~\ref{app:Sonine-expansion}.}. Still, we must keep in mind that the---rather slight---discrepancy in the reference value of $a_2$ has very little impact on the predicted behaviour of the kinetic temperature.
}

{The above analysis}  means that the system remains in a LLNES for most of the relaxation in the low-temperature quench. Over the LLNES, the cumulants $a_{2}$ and $a_{3}$ equal their reference values \eqref{eq:a2r-a3r}, whereas the temperature decays algebraically following Eq.~\eqref{eq:Y(s)}. This state only breaks for very long times, for which $1/Y$ does not diverge but saturates to its equilibrium value~\footnote{For the values of the parameters in  Fig.~\ref{fig:universal-behaviour-temp}, this takes place for very small values of $\theta/\theta_{\ini}$, namely   $\theta/\theta_{\ini}\lesssim 0.04$ ($1/Y\gtrsim 25$).}.

\section{Memory effects}\label{sec:memory-effects}

The just described non-exponential relaxation opens the door to the emergence of strong memory effects. We have shown that there exists a regime, $\theta_{\ini}\gg 1$ or, in other words, a quench to low enough temperatures, for which the system moves over the far-from-equilibrium LLNES state. The strength of possible memory effects roughly depends on the values of the cumulants, which measure the deviations from equilibrium. If their value is small (large), the VDF is close to (far from) the Gaussian shape and weak (strong) memory effects appear. Therefore, if we age the system to the LLNES, strong memory effects are expected.  In the following, we analyse the Mpemba and Kovacs effects separately.

\subsection{Mpemba effect}\label{sec:Mpemba}

{

We start the analysis with the Mpemba effect. In the Mpemba effect, the initially hotter fluid sample (A, initial temperature $\theta_{\ini A}$) cools sooner than the one initially cooler (B, initial temperature $\theta_{\ini B}$). Therefore, the ``cooling rate'' of the hotter system should be larger: since the cooling rate  increases with the excess kurtosis $a_{2}$, as follows from Eq.~\eqref{eq:evol-eqs-with-a3}~\footnote{A similar tendency of the cooling rate with the excess kurtosis has been found in other systems described at a kinetic level, both with inelastic and elastic collisions~\cite{lasanta_when_2017,torrente_large_2019,gomez_gonzalez_mpemba-like_2021,santos_mpemba_2020,takada_mpemba_2021}.}, the Mpemba effect is maximised when the hotter (cooler) sample has the largest (smallest) possible value of $a_2$. In such a way, the hotter (colder) samples cools as fast (slow) as possible.

Here, not only do we show that for large enough difference $\Delta a_{2\ini}\equiv a_{2\ini,A}-a_{2\ini,B}$ the Mpemba effect emerges, but (i) how to maximise the effect and (ii) how the system has to be previously \textit{aged} to get such an initial preparation of the samples. As stated above, $a_{2\ini,A}$ ($a_{2\ini,B}$) must take its largest (smallest) possible value to optimise the Mpemba effect. A rigorous mathematical derivation of the extrema (maximum and minimum) values of $a_2$ compatible with the fluid dynamics  makes it necessary to employ the tools of optimal control theory~\cite{pontryagin_mathematical_1987,liberzon_calculus_2012}. The quite lengthy calculation is outside the scope of this paper and thus will be published elsewhere~\cite{patron_unpublished}.  However, the result is physically appealing and compatible with the more intuitive  analysis performed in Appendix~\ref{app:optimisation-kurtosis}.

On the one hand, the minimum value $a_2^{\min}=a_2^r$ of the excess kurtosis is obtained for a quench to a very low temperature, i.e. when $\theta_{\ini}\gg 1$ and the system is cooled to the $(\gamma,\xi)$-independent LLNES described in the previous section. On the other hand, the maximum value of $a_2$ is obtained for the somehow ``opposite process'', i.e. for $\theta_{\ini}\ll 1$ that corresponds to  a heating to a much higher temperature. In Appendix~\ref{app:optimisation-kurtosis}, we show that $a_2^{\max}$ is proportional to $\gamma$ and much smaller than $|a_2^{\min}|$. For example, in the case $(\gamma=0.1,d=2,\xi=1)$ we have that $a_2^{\max}\simeq 0.04$ whereas $a_2^r=-0.36$. An even larger absolute value of $a_2^r$ is found in DSMC simulations, as illustrated in Fig.~\ref{fig:LLNES-curves}.


For maximising the Mpemba effect, then one should age the samples in the following way. The hot sample A must be aged by heating  from a much lower temperature, so that $a_2$ takes its maximum value and the sample has the largest possible cooling rate. The cold sample B must be aged by cooling  from a much higher temperature, so that $a_2$ takes its minimum, reference, value over the LLNES and the sample has the smallest possible cooling rate. Still, since $a_2^{\max}$ is quite small, a practical and very close to optimal procedure is to take the hot sample A at equilibrium, for which $a_2=0$. In this way, the difference $\Delta a_2 \equiv a_{2\ini,A} - a_{2\ini,B}$ is around 90 per cent of the optimal value $a_2^{\text{max}} - a_2^{\text{min}}$. This is the initial preparation that we employ throughout this work.

These samples $A$ and $B$ are put in contact with a common thermal reservoir at a much lower temperature, so Eqs.~\eqref{eq:evol-eqs-s-scale} govern the evolution of our system for a long time and, in particular, are capable of describing the \textit{universal} Mpemba effect observed. The initially hotter sample cools with $a_{2}$ decreasing from zero towards $a_{2}^{r}$, i.e.
\begin{equation}\label{eq:YA}
  Y_{A}(s_{A})=\frac{\theta_{A}(s_{A})}{\theta_{\ini,A}}=f(s_{A}),
  \quad s_{A}=\gamma\theta_{\ini,A}t,
\end{equation}
where $f$ is a certain function, independent of  $\theta_{\ini,A}$,
the exact form of which is irrelevant for our discussion. The
initially colder sample cools following Eq.~\eqref{eq:Y(s)}, i.e.
\begin{equation}\label{eq:YB}
  Y_{B}(s_{B})=\frac{\theta_{B}(s_{B})}{\theta_{\ini,B}}=Y_{\alg}(s_{B}),
  \quad s_{B}=\gamma\theta_{\ini,B}t.
\end{equation}
The Mpemba effect takes place when $\theta_{A}=\theta_{B}$ for some
crossing time $t_{\times}$.

\begin{figure}
  \centering
  \includegraphics[width=3.25in]{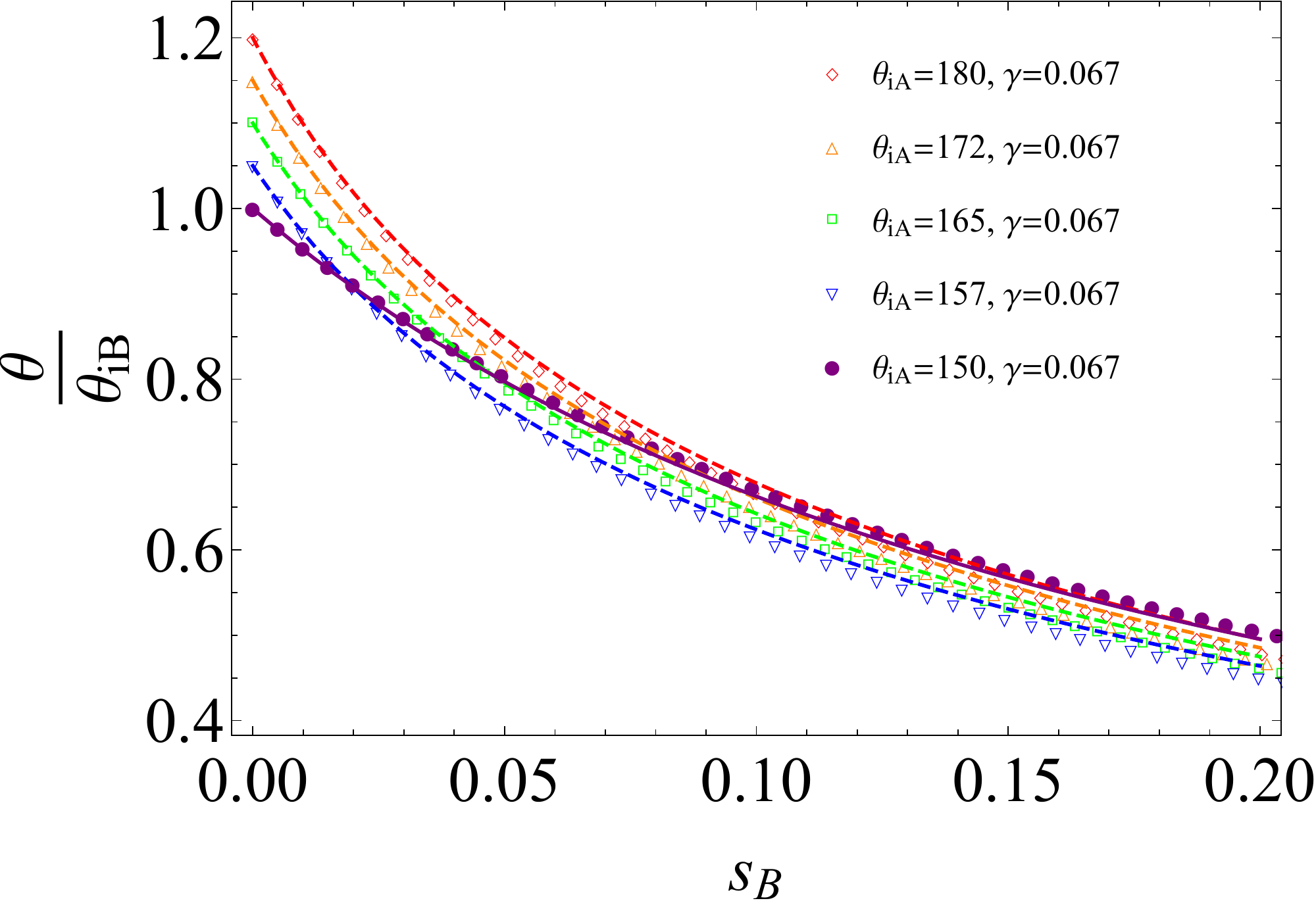}
  \caption{Mpemba effect for different initial temperature ratios
    $R_{AB}$. Specifically, we consider four values of $R_{AB}$,
    $R_{AB}=1.05$, $1.1$, $1.15$ and $1.2$. Additional parameters employed are $d=2$ and $\xi = 1$. We plot
  $\theta/\theta_{\ini,B}$ as a function of $s_{B}$, from the DSMC
  simulation and the theoretical prediction stemming from
  Eq.~\eqref{eq:evol-eqs-s-scale}. The relaxation curve of the cold
  sample B (circles DSMC, solid line theory), starting from
  $\theta_{\ini,B}=100$ with $a_{2\ini,B}=a_{2}^{r}$, is crossed by the curves
  for the hot samples A (empty symbols DSMC, dashed lines theory), which
  start from $\theta_{\ini,A}=R_{AB}\theta_{\ini,B}$ with
  $a_{2\ini,A}=0$ (i.e. at equilibrium).}
  \label{fig:large-Mpemba}
\end{figure}
\begin{figure}
\includegraphics[width=3.25in]{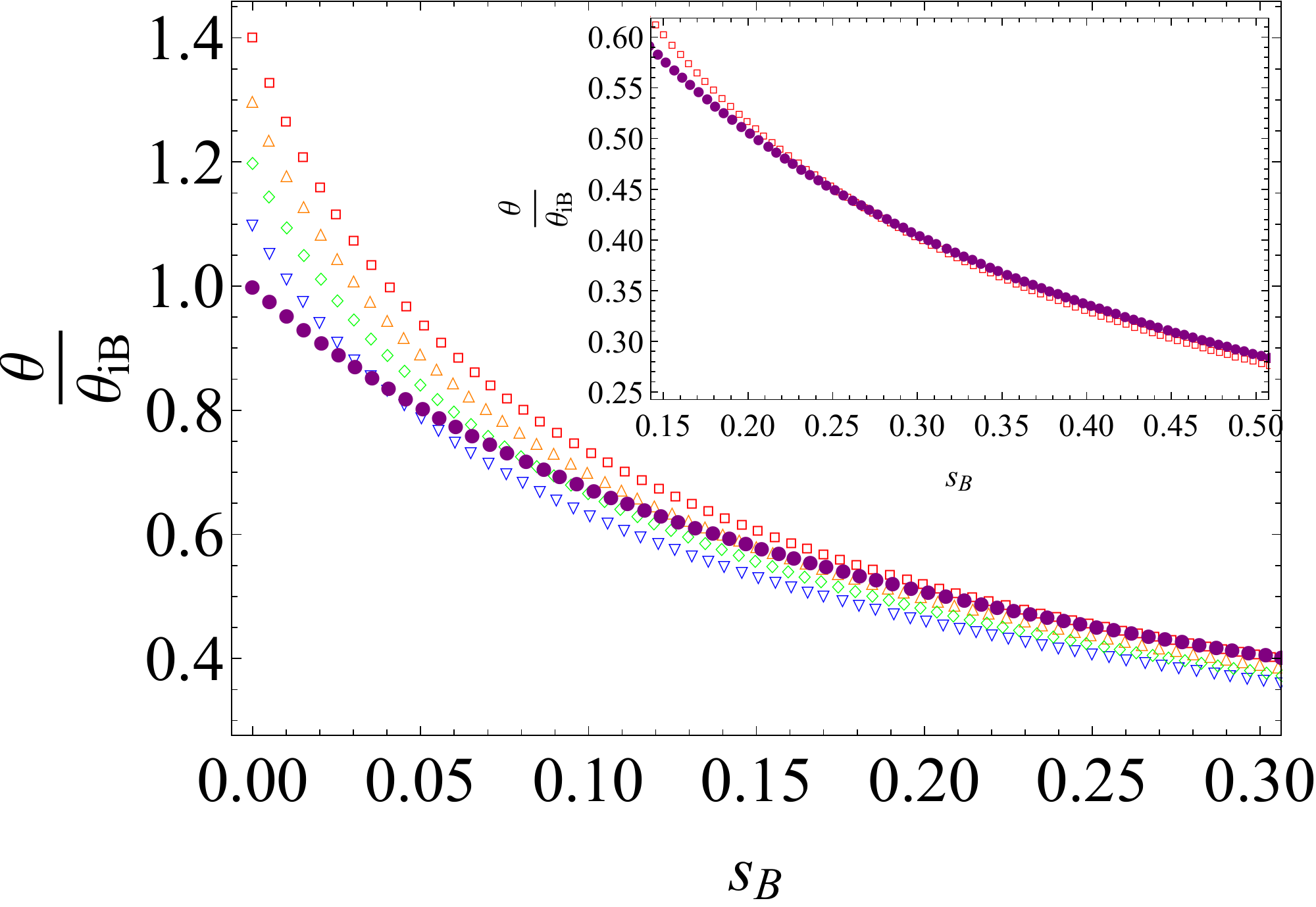}
\caption{Same as in Fig.~\ref{fig:large-Mpemba}, but for larger temperature ratios. The curves correspond to $R_{AB}=1.1$ (down-triangles), $1.2$ (circles), $1.3$ (up-triangles), and $1.4$ (squares). Within the figure, an inset has been plot in order to appreciate the Mpemba effect for $R_{AB} = 1.4$ (40\% initial temperature difference).}
\label{fig:supp-mat-mpemba}
\end{figure}
Figure~\ref{fig:large-Mpemba} shows the large Mpemba effect we observe. Since both the $Y$ and $s$ variables depend on the initial conditions, we plot $Y_{B}=\theta/\theta_{\ini,B}$ vs. $s_{B}=\gamma\theta_{\ini,B}t$. After defining the initial temperature ratio $R_{AB}\equiv\theta_{\ini,A}/\theta_{\ini,B}>1$, $Y_{A}=Y_{B}/R_{AB}$ and $s_{A}=R_{AB}s_{B}$. Specifically, we consider one $B$ sample, with $\theta_{i,B}=100$, and four different $A$ samples, with $R_{AB}=1.05,1.1,1.15,1.2$. Symbols correspond to DSMC simulations of the system and lines to the theoretical prediction stemming from Eqs.~\eqref{eq:evol-eqs-s-scale}. The temperature curves cross at a certain time $s_{B,\times}$, which corresponds to $t_{\times}$ in the original time scale, $s_{B,\times}=\gamma\theta_{\ini,B}t_{\times}$. For $s_{B}>s_{B,\times}$, the curve for the initially hotter sample lies below that of the initially colder. The Mpemba effect is even neatly observed for $R_{AB}=1.2$  (i.e. 20 per cent initial temperature difference). In fact, it is still present up to 40 per cent initial temperature difference, i.e. $R_{AB}=1.4$, as illustrated by Fig.~\ref{fig:supp-mat-mpemba}.

\begin{figure}
  \centering
  \includegraphics[width=3.25in]{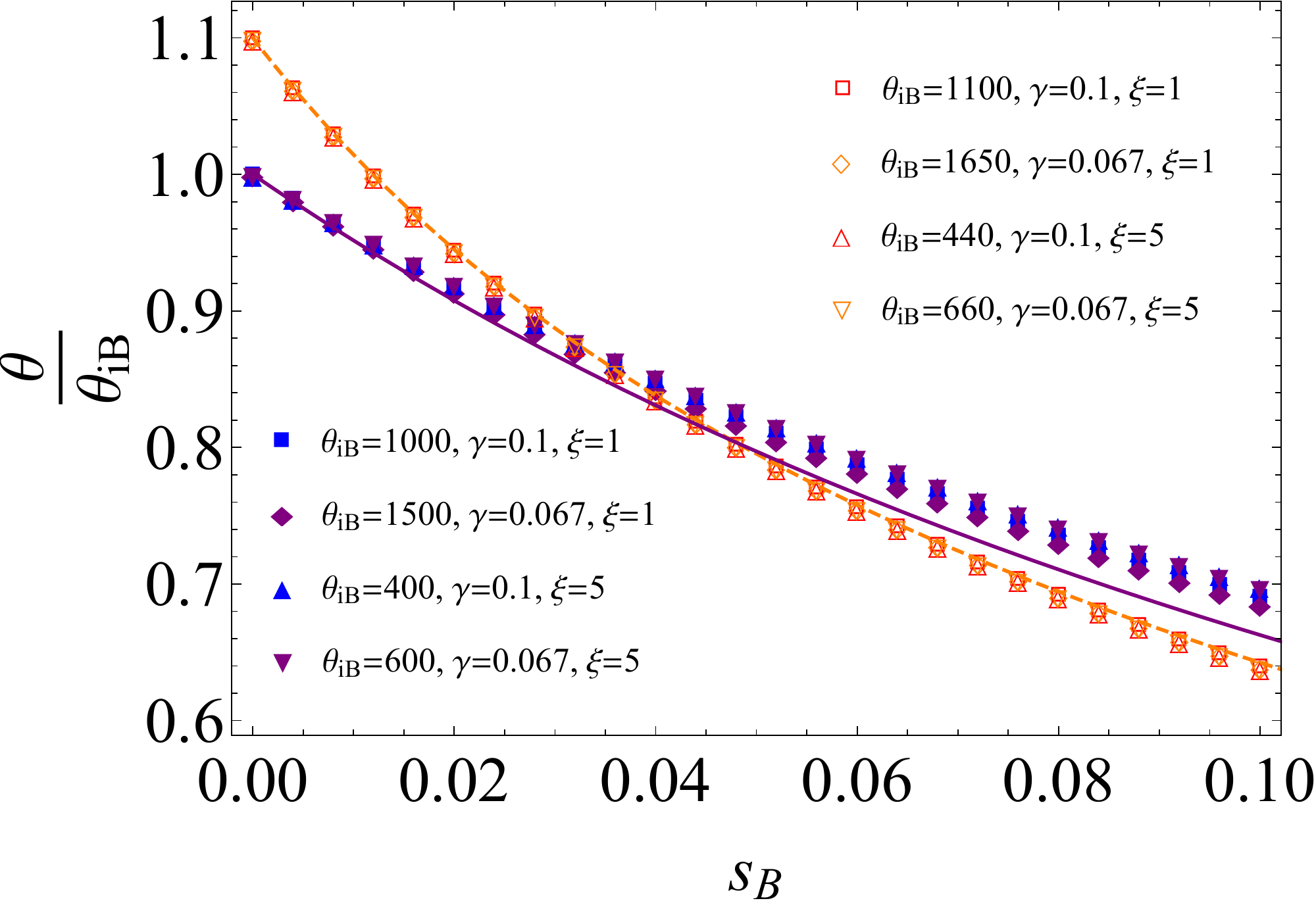}
  \caption{Universal Mpemba effect for different initial preparations and parameters $(\gamma,\xi)$. In particular, we plot $\theta/\theta_{\ini,B}$, i.e. the temperature in units of the initial temperature of the colder sample, as a function of the scaled time $s_{B}$ for the colder sample, defined in Eq.~\eqref{eq:YB}. For a fixed value of the initial temperature ratio $R_{AB}$, all the curves corresponding to different sets of $(\theta_{\ini,B},\gamma,\xi)$ superimpose, both for the hotter (A) (open symbols) and colder (B) samples (filled symbols). There are eight simulation curves: four corresponding to hot samples with $R_{AB}=1.1$ and the corresponding four curves for the cold samples. Dashed and full curves are the solutions of Eq.~\eqref{eq:evol-eqs-s-scale} for ($a_{2\ini}$,$a_{3\ini}$)= ($0,0$) and ($a_2^r$,$a_3^r$), respectively.
  }
  \label{fig:universal-Mpemba}
\end{figure}
The Mpemba effect is moreover universal in the following sense. Let us consider a fixed value of the ratio $R_{AB}$, but different values of the the initial temperatures $\theta_{\ini,A}$ and $\theta_{\ini,B}$, the non-linearity parameter $\gamma$,  and the average time between collisions $\xi$. If we plot $\theta/\theta_{\ini,B}$ vs. $s_{B}$, all the curves corresponding to the colder temperatures superimpose, as Fig.~\ref{fig:universal-Mpemba} shows. Besides, also the curves corresponding to the hotter temperatures superimpose, because $s_{A}=R_{AB}\,s_{B}$ and Eq.~\eqref{eq:YA} entails $\theta_{A}(s_{B})=\theta_{\ini,A}f(R_{AB}\,s_{B})$, i.e. $\theta_{A}(s_{B})/\theta_{\ini,B}=R_{AB}\,f(R_{AB}\,s_{B})$. This is neatly shown in Fig.~\ref{fig:universal-Mpemba}, where we have plotted relaxation curves for $R_{AB}=1.1$ and different values of $(\theta_{\ini,B},\gamma,\xi)$, as detailed in the legend. The analytical prediction from Eq.~\eqref{eq:evol-eqs-s-scale} for the colder sample is slightly under the DSMC data, because of our underestimating the excess kurtosis over the LLNES. 

In order to quantify the strength of the Mpemba effect, we introduce the parameter $\text{Mp}$ defined in Ref.~\cite{torrente_large_2019}, which corresponds to the maximum difference between the relaxation curves once they have crossed each other.  We have computed the numerical values of $\text{Mp}$ from the DSMC simulation. Since the strength of the Mpemba effect is proportional to $\theta_{\ini,B}$, we have specifically computed $\text{Mp}/\theta_{\ini,B}$. For the curves shown in Fig.~\ref{fig:supp-mat-mpemba}, the values are $\text{Mp}/\theta_{\ini,B}=0.059$, $0.045$, $0.034$ and $0.026$ for initial temperature ratios $R_{AB}=\theta_{\ini,A}/\theta_{\ini,B}=1.1$, $1.2$, $1.3$ and $1.4$, respectively.  As expected, $\text{Mp}$ decreases with the initial temperature difference $\theta_{\ini,A}-\theta_{\ini,B}$---or, equivalently, with $R_{AB}$. Since $\theta_{\ini,B}\gg 1$, the actual values of $\text{Mp}$ for the this system are typically larger than unity. In the figure, $\theta_{\ini,B}=100$, so $\text{Mp}$ ranges from $2.6$ to $5.9$, values that are indeed higher than those for the large Mpemba-like effect  reported in Ref.~\cite{torrente_large_2019} for a  rough granular gas.

\begin{figure}
\includegraphics[width=3.25in]{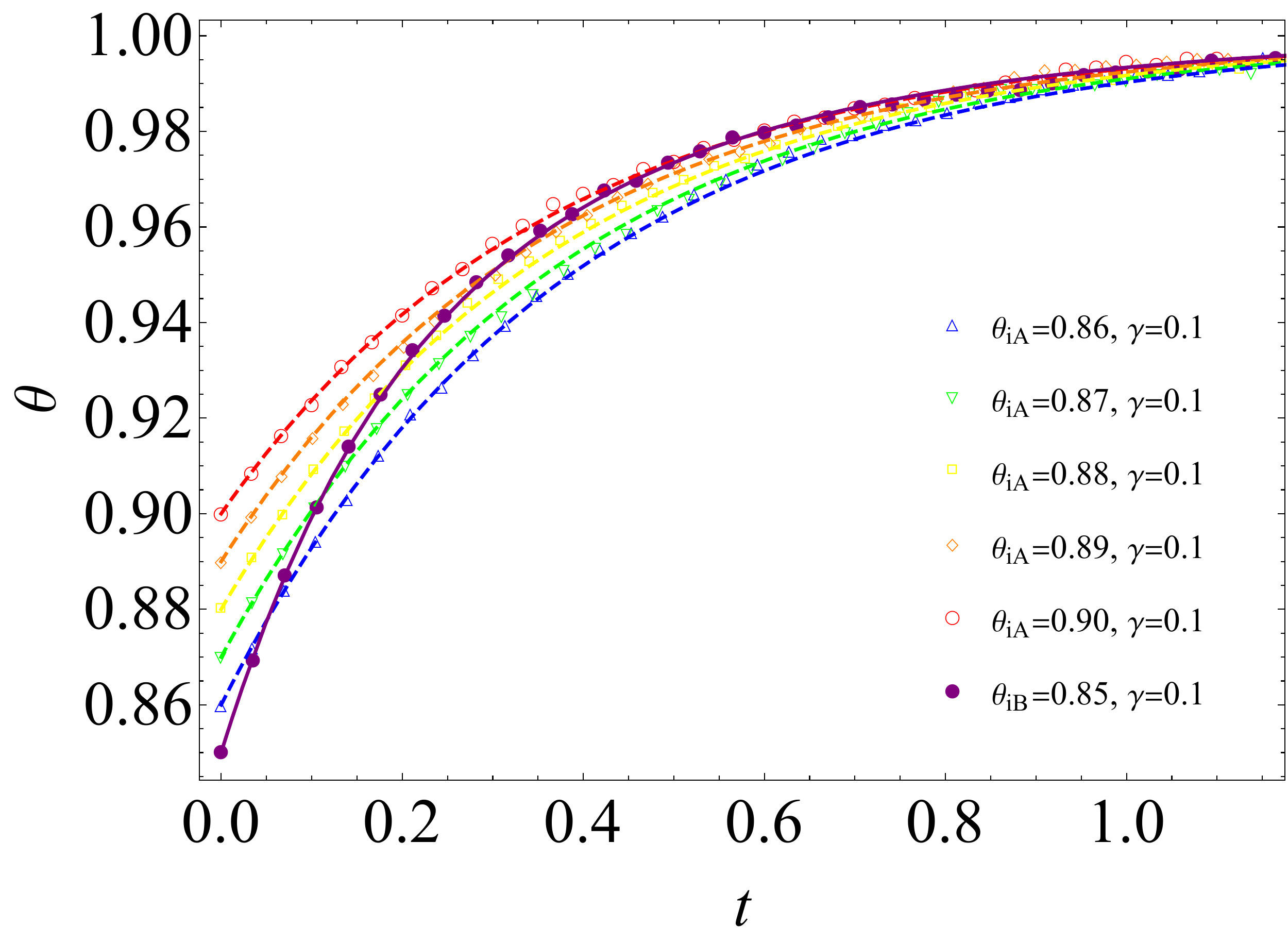}
\caption{Inverse Mpemba effect for different initial temperatures for the hotter sample. Specifically, we consider hotter samples with temperatures $\theta_{\ini,A} = 0.86$, $0.87$, $0.88$, $0.89$ and $0.90$, while the colder sample departs from a temperature of $\theta_{\ini,B}= 0.85$. Additional employed parameters are $\xi = 1$ and $d = 2$. Empty (filled) symbols correspond to DSMC data for the hotter (colder) samples, while the dashed (full) lines correspond to the numerical integration of the evolution equations Eqs.\eqref{eq:evol-eqs-with-a3} for the hotter (colder) samples.}
    \label{fig:inv-mpemba}
\end{figure}
It is also interesting to study the inverse Mpemba effect, in which the initially colder sample heats sooner than the  initially hotter one, which has also been observed in a wide variety of systems~\cite{lu_nonequilibrium_2017,klich_mpemba_2019,gal_precooling_2020,lasanta_when_2017,torrente_large_2019,santos_mpemba_2020, biswas_mpemba_2020,biswas_mpemba_2021,gomez_gonzalez_mpemba-like_2021,takada_mpemba_2021}. Now, samples $A$ (initially hotter) and $B$ (initially colder) are put in contact with a thermal reservoir at a larger temperature. If sample A heats more slowly  than sample B, the inverse Mpemba effect emerges. But heating more slowly is basically equivalent to cooling faster: in both cases, we want to have $\dot{\theta}$ as large as possible. Therefore, we would like again to have the initially hotter sample with the maximum possible value of $a_2$ and the initially colder one with the minimum possible value, exactly the same preparation as for the normal case. 

Following the  reasoning in the previous paragraph, we study the inverse Mpemba effect when the initially cooler sample departs from the LLNES while the hotter one departs from equilibrium~\footnote{Again, the optimal procedure would be to follow a heating protocol such that $a_{2\ini,A}=a_2^{\max}$ but, since $a_{2}^{\max}$ is very small, our initial preparation is nearly optimal and more practical.} In Fig.~\ref{fig:inv-mpemba} 
we may observe that the initial temperature differences are smaller than those for the normal Mpemba effect. Here, the maximum value of the parameter $R_{AB}$ is $1.06$, i.e. a 6\% maximum initial temperature difference, whereas in the normal case it was 40\%. Consistently, the strength of the inverse Mpemba effect is smaller than that of the normal one: the values of the $\text{Mp}$ parameter range between $0.001$ and $0.013$ in this case.

}

\subsection{Kovacs effect}\label{sec:Kovacs}

Next, we look into the Kovacs effect. In our system, the relevant physical quantity is the kinetic temperature. The Kovacs hump will come about if the cumulants are non-zero at the waiting time $t_w$. Therefore, to maximise the effect the (absolute) value of $a_2$ and $a_3$ have to be in turn maximised. This entails that the optimal aging protocol is a quench to a much lower temperature, i.e. $T_1\ll T_{\ini}$, over which the system reaches the LLNES. Equations~\eqref{eq:evol-eqs-with-a3} govern the time evolution of the system for $t>t_{w}$, with $\theta=T/T_{w}$ and initial conditions $\theta(t_{w})=1$, $a_{2}(t_{w})=a_{2}^{r}$, $a_{3}(t_{w})=a_{3}^{r}$.  
\begin{figure}
  \centering
  \includegraphics[width=3.25in]{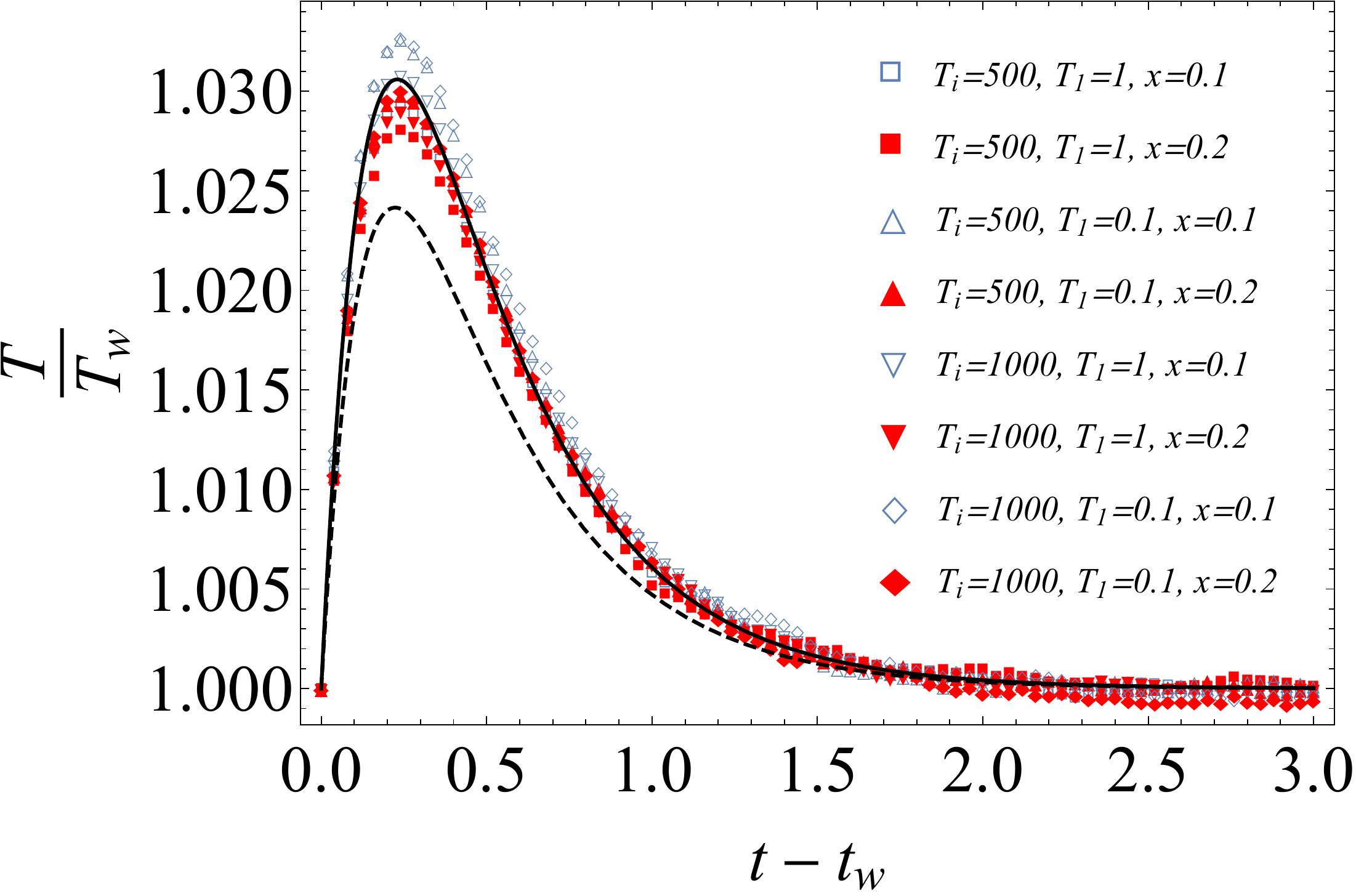}
  \
  \caption{Evolution of the temperature in the Kovacs protocol. Parameter values are $\gamma = 0.1$, $d=2$,  and $\xi = 1$. Eight simulation curves are shown for different combinations of the initial ($T_{\ini}$), aging ($T_1$), and final temperature ($T_w$). 
Writing the aging temperature as $T_{w}=T_{1}+x(T_{\ini}-T_{1})$, the data shown correspond to $x=0.2$ (filled symbols) and $0.1$ (open symbols).  Curves for smaller values of $x$ are basically superimposed with those for $x=0.1$. The dashed (solid) line corresponds to the numerical integration of  Eqs.~\eqref{eq:evol-eqs-with-a3} with the theoretical (simulation) values for the cumulants over the LLNES.}
  \label{fig:Kovacs}
\end{figure}

The resulting Kovacs response also has {scaling} properties, although somehow weaker than those of the temperature relaxation and the Mpemba effect. The initial conditions and, therefore, the subsequent Kovacs hump do not depend on $(T_{\ini},T_{w},T_{1})$. Yet, it does depend on $\gamma$ and $\xi$.
Figure~\ref{fig:Kovacs} illustrates {the scaled} Kovacs hump, we plot $\theta=T/T_{w}$ as a function of $t-t_{w}$, for $t>t_{w}$. Indeed, the triplet $(T_{\ini},T_{w},T_{1})$ does not affect the Kovacs hump measured in DSMC simulations. Here, for the sake of simplicity, we have taken one of the aging temperatures as unity~\footnote{In the relaxation experiment and the Mpemba memory effect, the unit of temperature was formally the steady temperature $T_s$.}. Moreover, our theory quantitatively describes the numerical results: the agreement is very good, especially when the simulation value of $a_2^r$ is employed~\footnote{Our theory underestimates $a_2^r$ by roughly 15 per cent, as shown by Figure~\ref{fig:LLNES-curves}.}.

{To further study the Kovacs effect, a perturbative analysis can be carried out---see Appendix~\ref{app:kovacs} for details. It gives that
\begin{widetext}
\begin{align}
\label{eq:kovacs-hump-first-order}
    K(t) \equiv \theta(t)-1 = -\gamma a_2^r \frac{2(d+2)}{\lambda_+-\lambda_-}\Bigg[ & \frac{M_{11}+M_{12}+|\lambda_-|}{|\lambda_+|-\alpha}\left(e^{-\alpha (t-t_w)}-e^{-|\lambda_+|(t-t_w)}\right) \nonumber \\ & - \frac{M_{11}+M_{12}+|\lambda_+|}{|\lambda_-|-\alpha}\left(e^{-\alpha (t-t_w)}-e^{-|\lambda_-|(t-t_w)}\right)\Bigg]+O((\gamma a_2^r)^2).
\end{align}
where $\alpha=2[1+\gamma(d+2)]$, $M_{ij}$ are the elements of a $2\times 2$ matrix $\bm{M}$,
\begin{subequations}\label{eq:Mij}
\begin{align}
        M_{11} = &-4 \left[1+\gamma(d+6) + \frac{2(d-1)}{d(d+2)\xi} \right], & M_{12} =& 2 \frac{a_{3}^r}{a_{2}^r} \left[2\gamma (d+4) + \frac{d-1}{d(d+2)\xi} \right],
        \\
        M_{21} = &12\frac{a_{2}^r}{a_{3}^r}\left[2\gamma + \frac{d-1}{d(d+2)(d+4)\xi} \right], & M_{22} =& -6\left[\gamma (d+10) +1 + \frac{(d-1)(4d+19)}{2d(d+2)(d+4)\xi} \right].
\end{align}
\end{subequations}
\end{widetext}
and $\lambda_{\pm}$ are the eigenvalues of the matrix $\bm{M}$, 
\begin{equation}\label{eq:lambda-pm}
    \lambda_{\pm} = \frac{\Tr{\bm{M}} \pm \sqrt{(\Tr{\bm{M}})^2-4\det{\bm{M}}}}{2} < 0.
\end{equation}
The Kovacs effect is always normal, as it must be in a molecular system~\cite{prados_kovacs_2010}, since $a_2^r<0$. Note that $M_{12}$, as defined by Eq.~\eqref{eq:Mij}, depends on the cumulants, in particular on the ratio $a_{3}^r/a_{2}^r$. Had we aged the system in a different manner, $a_2^r$ and $a_3^r$ would have been substituted with $a_2(t_w)$ and $a_3(t_w)$~\footnote{$a_2(t)<0$ when the system is cooled, as observed in Fig.~\ref{fig:phase-1-gamma-approach}, so that $a_2(t_w)<0$ and the effect remains to be normal.}. 

The accuracy of our perturbative expansion is checked by comparing Eq.~\eqref{eq:kovacs-hump-first-order}  for the Kovacs hump to DSMC data. This is done in Fig.~\ref{fig:kovacs-diff-gamma}, where we plot the function $K(t)$ for three different values of $\gamma$, namely $\gamma=0.1$, $\gamma=0.05$, and $\gamma=0.025$. Again, we write $T_{w}=T_{1}+x(T_{\ini}-T_{1})$ and the data shown correspond to $x=0.1$. We compare the DSMC data with  Eq.~\eqref{eq:kovacs-hump-first-order}, both employing the theoretical predictions for $a_2^r$  and $a_3^r$ (dashed line) and their simulation values  (solid line). The mild discrepancies basically stem from the difference between the theoretical and DSMC value of the excess kurtosis, as illustrated by the very good agreement observed for the solid lines.
\begin{figure}
\includegraphics[width=3.25in]{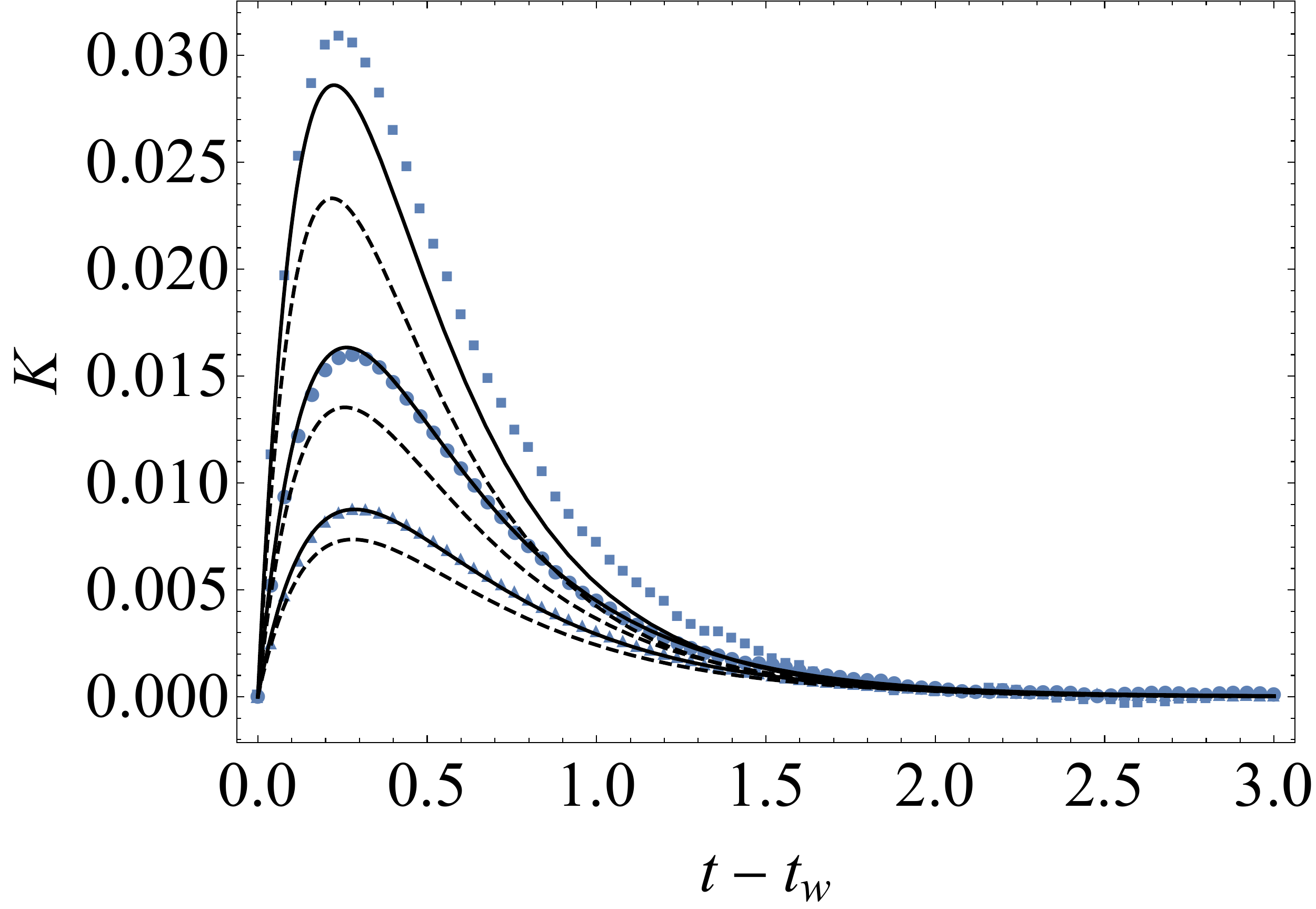}
\caption{Dependence of the Kovacs hump on the non-linearity parameter $\gamma$. Three sets of data are plotted: both correspond to the triplet $(T_i = 1000, T_1 = 0.1, x =0.1)$ for three different values of $\gamma$, specifically $\gamma = 0.1$ (squares), $0.05$ (circles) and $0.025$ (up triangles). Additional parameter values are $d=2$  and $\xi = 1$.  The dashed (solid) lines correspond to the first order perturbative expression~\eqref{eq:kovacs-hump-first-order} with the theoretical (DSMC) values of $a_2^r$ and $a_3^r$.}
    \label{fig:kovacs-diff-gamma}
\end{figure}

Let us analyse the position and the height of the maximum, which we denote by $t_M$ and $K_M\equiv K(t_M)$, respectively. The values of $t_M$ and $K_M$ corresponding to the curves in Fig.~\ref{fig:kovacs-diff-gamma} are given in Table~\ref{table}. Specifically, we give their values stemming from the theoretical expression \eqref{eq:kovacs-hump-first-order}, again both employing the theoretical predictions for $a_2^r$ and $a_3^r$ and their simulation values. The agreement between the theory and the simulation is very good, especially when the DSMC values of the cumulants are inserted into the theoretical expression. The maximum position $t_M$ depends very weakly on $\gamma$, whereas its height $K_M$ is roughly proportional to it.
\begin{table*}
\centering
 \begin{tabular}{|| P{1.5 cm} | P{4.8 cm} | P{4.8 cm} | P{2.6 cm} |} 
 \hline
  & Eq.~\eqref{eq:kovacs-hump-first-order} \ ($a_2^r, a_3^r$ from Sonine) & Eq.~\eqref{eq:kovacs-hump-first-order} ($a_2^r, a_3^r$ from DSMC) & DSMC data \\ [0.5ex] 
 \hline
  $\gamma = 0.025$ & (0.281,0.007) & (0.286,0.009) & (0.280,0.009) \\ [1ex]
 \hline
 $\gamma = 0.05$ & (0.257,0.014) & (0.262,0.016) & (0.280,0.016) \\ [1ex]
 \hline
 $\gamma = 0.1$ & (0.219,0.023)  & (0.225,0.029) & (0.240,0.031) \\ [1ex]
 \hline
 \end{tabular}
\caption{Values of the maximum coordinates $(t_M,K_M)$ for the Kovacs hump. Specifically, the reported values correspond to the curves  plotted in Fig.~\ref{fig:kovacs-diff-gamma}.}
\label{table}
\end{table*}
}

{
\section{Relevance of collisions and the Fokker-Planck limit}
\label{sec:collisionless}

The relevance of the Enskog collision term in the EFP equation is modulated by the dimensionless average time between Brownian-Brownian collisions $\xi$. In previous sections, we have typically considered order of unity values of $\xi$, for which the drag force and collisions act over the same time scale. As already stated below the evolution equations~\eqref{eq:evol-eqs-with-a3}, the limit $\xi=\infty$ corresponds to the collisionless case, in which the EFP equation simplifies to the FP equation. Now, motivated by recent work in binary mixtures of ultracold atoms~\cite{hohmann_individual_2017}, we investigate how the existence of the LLNES and the associated slow algebraic relaxation is affected in the limit as $\xi\gg 1$.

In Ref.~\cite{hohmann_individual_2017}, the behaviour of a binary mixture of Cs and Rb atoms is investigated. Quantum effects are negligible---despite temperatures being in the $\mu$K
range, due to the low densities of both the Brownian (Cs atoms) and background  (Rb atoms) fluids. Therefore, the motion of the Cs atoms is described by means of a Langevin equation---or the equivalent non-linear FP equation~\eqref{eq:FP}---with non-linear drag force, because the masses of the Cs ($m_{\text{Cs}}$) and Rb ($m_{\text{Rb}}$) atoms are comparable. For the mixture of Cs and Rb atoms, the parameters for our EFP equation framework are $\gamma = m_{\text{Rb}}/(10 m_{\text{Cs}}) \approx 0.067$ and the dimensionless characteristic time $\xi = 674.17$---see Appendix~\ref{app:Sonine-expansion}.

We show below that the high value of the dimensionless average time between collisions $\xi$ in Ref.~\cite{hohmann_individual_2017} entails that the predictions for the EFP equation and the FP equation are basically equivalent. In other words, collisions are so infrequent that the Enskog collision term can be completely disregarded in that case. Also, we show that the existence of the LLNES and thus of a wide time window over which the temperature relaxes algebraically---for a quench to low temperatures---is independent of the value of $\xi$; systems with $\xi=1$, $\xi=674$ and $\xi=\infty$ display exactly the same behaviour in the time scale $s$. 

We present the results for the relaxation of the temperature in Fig.~\ref{fig:gas-frio}. Symbols correspond to (i) the numerical simulations for the EFP equation for two different values of the characteristic time $\xi$, $\xi=1$ and $\xi=674$, and (ii) the FP equation ($\xi=\infty$). The line corresponds to the algebraic relaxation~\eqref{eq:Y(s)}, with the theoretical value $a_2^r=-0.33$ for $d=3$. It is neatly observed that all the curves are basically superimposed. Specifically, there is no difference between the simulation results for the EFP equation with $\xi=674$ and the FP equation. Also, the agreement between these two simulation curves and the theoretical prediction~\eqref{eq:Y(s)} is better than that of the case $\xi=1$, which is already very good. In fact, the terms involving $\xi$ in Eq.~\eqref{eq:evol-eqs-dominant}---which have been neglected when writing \eqref{eq:evol-eqs-s-scale}---vanish for $\xi=\infty$, so Eq.~\eqref{eq:evol-eqs-s-scale} was expected to give a better description for the collisionless case.
\begin{figure}
\begin{minipage}{3.25in}
\includegraphics[width=3.25in]{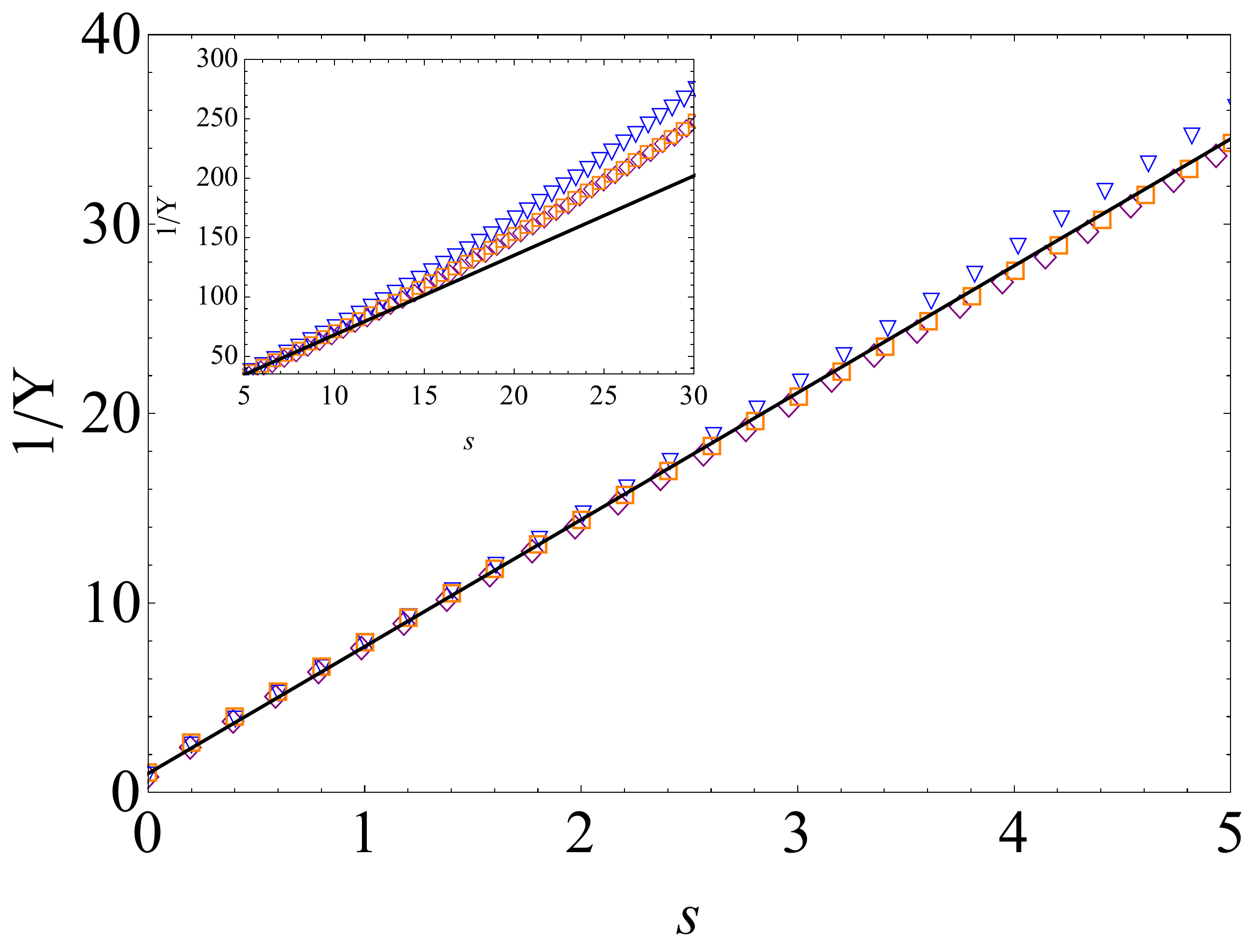}
\end{minipage}
\caption{Dependence of the LLNES on the collision rate. DSMC data for the relaxation of after a quench to a low temperature, specifically with $\theta_{\ini}=1000$, are plotted: $\xi=1$ (down triangles), $\xi=674$ (diamonds), and $\xi=\infty$ (squares)---the first two correspond to the EFP equation, whereas the latter correspond to the FP equation. The simulation data are compared with our theoretical prediction for the LLNES, Eq.~\eqref{eq:Y(s)} (solid line). The agreement theory-simulation is very good for all curves but especially for the cases $\xi=674$ and $\xi=\infty$, which are basically superimposed. The inset shows the relaxation curves for longer times, $5\leq s\leq 30$, whereas in the main panel $0\leq s\leq 5$. Therein, it is observed how the system starts to depart from the LLNES: the smaller the collison rate $\xi^{-1}$, the smaller the separation from the LLNES.} 
\label{fig:gas-frio}
\end{figure}

The inset in Fig.~\ref{fig:gas-frio} shows the relaxation of the temperature for longer times. Therein, we clearly observe that the LLNES persists for a longer time when collisions are infrequent ($\xi=674$) or inexistent ($\xi=\infty$). This is reasonable from a physical point of view. The collision term does not directly affect the time evolution of the temperature, because collisions are elastic and kinetic energy is conserved. However, collisions indeed affect the time evolution of of the VDF through higher-order cumulants like $a_2$ and $a_3$: they favour the ``mixing'' of velocities and thus make the relaxation to equilibrium faster. Accordingly, the relaxation curve of $1/Y$ for $\xi=1$ is always above than those for $\xi=674$ and $\xi=\infty$.
}

\section{Conclusions}\label{sec:conclusions}

The molecular fluid with non-linear drag shows a very complex relaxation behaviour. The leading role is played by the LLNES reached by the system when quenched to a low temperature.  Over it, the temperature displays a very slow, algebraic, decay and the VDF neatly separates from the Maxwellian shape. The strong non-Gaussianities are characterised by  large (absolute) values of the fourth and sixth cumulants, which we have termed their ``reference'' values $a_{2}^{r}$ and $a_{3}^{r}$.

{Both the own existence of the LLNES and the physical properties over it---algebraic decay of the temperature and reference values of the cumulants---do not depend on the degree of non-linearity, as measured by $\gamma$, nor on the  Brownian-Brownian collision rate, as measured by $\xi$. It must be remarked that, in particular, the LLNES survives in the limit $\xi=\infty$, when the Enskog collision term is not present and the velocity VDF for the Brownian particles obey the FP equation~\eqref{eq:FP} with non-linear drag.}

This LLNES also rules the emergence of large memory effects, both Mpemba-like and Kovacs-like.  On the one hand, not only have we shown that a large Mpemba effect---present for temperature differences up to 40 per cent---comes about but also how the hot and cold samples have to be prepared. The identification of the aging procedure is important for the experimental reproducibility of the Mpemba effect: here, the hot sample starts from equilibrium whereas the cold sample starts from the LLNES. The strongly non-exponential relaxation associated with the Mpemba effect is quite unique, since the relaxation is basically exponential in the majority of systems in which the Mpemba effect has been studied. On the other hand, it is the relaxation following the quench to a low temperature that has to be interrupted to maximise the Kovacs effect, once the system has reached the LLNES. The reported Kovacs hump, of the order of $3$ per cent in Fig.~\ref{fig:Kovacs}, is quite large as compared to typical values. For example, it is one of order of magnitude larger the original observation by Kovacs~\cite{kovacs_transition_1963,kovacs_isobaric_1979}, $2-3$ times larger than its value in a Lennard-Jones fluid~\cite{mossa_crossover_2004}, and of the same order of magnitude of the recently reported results in a disordered protein construct~\cite{morgan_glassy_2020}.

Both the non-exponential relaxation and the memory effects present {scaling} features. When properly scaled, all relaxation curves corresponding to the quench to a low temperature superimpose. Not only does the relaxation in scaled variables not depend on the initial temperature $\theta_{\ini}$ {but also is independent of the degree of non-linearity $\gamma$ and the average time $\xi$ between Brownian-Brownian collisions. This is why we employ the term universal to refer to the observed relaxation of the temperature.}

For the Mpemba effect, a similar scaling entails that all curves corresponding to a given initial temperature ratio also superimpose, independently of the value of other parameters: initial temperatures of the hot and cold samples and also $(\gamma,\xi)$. {In this sense, we also speak about a universal Mpemba effect. The Kovacs effect also displays scaling properties, although  weaker: the hump depends on $(\gamma,\xi)$ but not on the initial, final, and aging temperatures.}

{The LLNES naturally emerges  when the system is quenched from a very high temperature $\theta_{\ini}\gg 1$, and thus the temperature $\theta\gg 1$ over a---quite wide---time window. Would the LLNES still be relevant for other, more general, protocols, in which the temperature of the bath followed a certain program $T_s(t)$? Looking back at the evolution equations \eqref{eq:evol-eqs-with-a3} in the second Sonine approximation, Eqs.~\eqref{evol-eq-a2-a3} and \eqref{evol-eq-a3-a3} would remain unchanged whereas \eqref{evol-eq-T-a3} would have an additional term $-\theta d\ln T_s(t)/dt$ on its rhs. This implies that, as long as $\theta(t)\equiv T(t)/T_s(t)\gg 1$, Eqs.~\eqref{evol-eq-a2-a3} and \eqref{evol-eq-a3-a3} for the time evolution of the cumulants are still valid and the cumulants would tend to their reference values, characteristic of the LLNES, in this more general situation. As for the temperature, Eq.~\eqref{evol-eq-T-a3} would have an additional term $-Y d\ln T_s/ds$ making, quite logically, the time evolution of $\theta$ depend on the considered program. The analysis of the behaviour of the fluid with non-linear drag under such a time-dependent program for the bath temperature is an interesting perspective for future work.

Another relevant question is the robustness of the LLNES for other, more general forms, of the non-linear drag. The results derived in this paper are specific for the quadratic non-linearity in  Eq.~\eqref{eq:zeta-v} but, what about higher-order non-linearities? For instance, let us think of the next correction in the systematic expansion in powers of the mass ratio $\mbf/m$ introduced in Refs.~\cite{ferrari_particles_2007,ferrari_particles_2014}, which incorporates a  quartic, proportional to $v^4$, term. Incorporating it  would result in the coupling of the time evolution of the temperature not only with $\mean{v^4}$, which gives rise to the term proportional to $\theta^2 a_2$, but also with $\mean{v^6}$, which would give rise to a new term proportional to $\theta^3 a_3$---dominant for a quench to low temperatures, where $\theta\gg 1$. This entails that the third-order Sonine approximation would be necessary to describe the evolution of the temperature, since so is quantitatively predicting $a_3$. Still, a LLNES would appear in which $a_2$, $a_3$, and $a_4$ would tend to \textit{pseudostationary} reference values $a_2^r$, $a_3^r$, and $a_4^r$. The temperature would also have an algebraic decay but with a different exponent, since we would have $\dot{\theta}\propto -\theta^3$ (instead of $-\theta^2$) for $\theta\gg 1$ and therefore $\theta\propto t^{-1/2}$ (instead of $t^{-1}$).
}

{
In this work, we have employed the extended---or second---Sonine approximation, retaining not only the excess kurtosis $a_2$ but also the sixth cumulant $a_3$. This stems from the evolution equation of the temperature $\theta$ being directly coupled with $a_2$, whereas $a_3$ only appears in the evolution equation of $a_2$. In our study, the $n$-th-order Sonine approximation---i.e. retaining $(\theta,a_2,\ldots,a_{n+1})$---allows for quantitatively describing the behaviour up to the second to last kept cumulant $a_{n}$ as the initial temperature is increased. The discrepancies between the theory and the DSMC simulations slightly increase with the order of the cumulant---i.e. when one goes from $\theta$ to $a_n$. However, it ``only'' gives a qualitative account of the behaviour of the last kept cumulant $a_{n+1}$. This makes it necessary to consider the extended, second-order, Sonine approximation when considering a quench to low temperatures, because an accurate prediction for the time evolution of the excess kurtosis is needed.
}

{The most rigorous approach to analyse the mixture of Brownian and background fluids would be writing down the Boltzmann (or Enskog) equation for the two species. Comparing the results of this framework with those from the EFP equation---for order of unity Brownian-Brownian collision rate $\xi$---is an interesting perspective for future work. It is worth recalling that both frameworks give rise to the FP equation in the limit $\xi\to\infty$, in which we have shown that the glassy behaviour found for the EFP equation persists.}

Our work opens the door to investigating aging phenomena and glassy behaviour in ultracold atoms. A key  result of this work is the role played by the quench to a much lower temperature that leads the system to the LLNES, which controls the emergence of non-exponential relaxation and the associated memory effects (both Mpemba- and Kovacs-like). Since the model employed here describes mixtures of ultracold atoms, like that in Ref.~\cite{hohmann_individual_2017}, the central role of the LLNES may be checked  in real experiments. 

\begin{acknowledgments}
We acknowledge financial support from project PGC2018-093998-B-I00, funded by: FEDER/Ministerio de Ciencia e Innovación--Agencia Estatal de Investigación (Spain). A.~Patr\'on acknowledges support from the FPU programme trhough Grant FPU2019-4110. Also, we would like to thank D. Gu\'ery-Odelin, A. Meg\'{\i}as and A. Santos for useful discussions.
\end{acknowledgments}

\appendix

\section{Sonine expansion}\label{app:Sonine-expansion}

Here we summarise the main features of the so-called Sonine expansion of the VDF, which makes it possible to---by introducing suitable approximations---truncate the infinite hierarchy of equations for the cumulants. Also, we compare the theoretical predictions of the first Sonine approximation and the extended Sonine approximation with DSMC simulations of the EFP equation~\eqref{eq:EFP}.

For the scaled VDF introduced in Eq.~\eqref{eq:phiC-Sonine}, the EFP equation becomes~\cite{santos_mpemba_2020}
\begin{align}
    \label{apA-EFP-dimensionless}
    \partial_t  \phi(\bm{c} & ,t) = \frac{1}{\xi}\sqrt{\theta} I[\bm{c}|\phi,\phi] \nonumber
    \\
    & +\frac{\partial}{\partial \bm{c}}\! \cdot \!\left[ \frac{\dot{\theta}}{2\theta}\bm{c} + \left( 1 + 2\gamma \theta c^2\right)\!\cdot\!\left(\bm{c}+\frac{1}{2\theta}\frac{\partial}{\partial \bm{c}}\right) \right]\phi(\bm{c},t),
\end{align}
where $\theta$ and $t$ are the dimensionless temperature and time defined in Eq.~\eqref{eq:T-t-dimensionless}---recall that we have dropped the asterisk to simplify the notation, 
\begin{align}
    \label{eq:collision-operator-dimensionless}
    I[\bm{c}_1|\phi,\phi] = &\int d \bm{c}_2\int d \widehat{\bm{\sigma}} \ \Theta (\bm{c}_{12}\cdot \widehat{\bm{\sigma}}) \ \bm{c}_{12}\cdot \widehat{\bm{\sigma}}  \nonumber
    \\
     & \times \left[\phi (\bm{c}_1')\phi (\bm{c}_2')-\phi (\bm{c}_1)\phi (\bm{c}_2)\right]
\end{align}
is the dimensionless Enskog collision operator, and $\xi$ is the parameter defined in Eq.~\eqref{eq:xi-def}.

The parameter $\xi$ measures the relative relevance of the nonlinear drag force---i.e. collisions between the background fluid particles and the Brownian ones---and the Brownian-Brownian collisions. The regime $\xi \gg 1$ implies that collisions act over a much longer time scale than the drag force. When the background fluid is also composed of hard-spheres---therefore, $d=3$---of density $\nbf$ and diameter $\sbf$, it has been shown that~\cite{hohmann_individual_2017,santos_mpemba_2020}
\begin{equation}
  \label{apA-eq:zeta0-*}
  \xi=\frac{2\nbf}{3n}\left(1+\frac{\sbf}{\sigma}\right)^{2
  }\frac{\sqrt{5\gamma}}{1+10\gamma}.
\end{equation}
In the case of self-diffusion, $\mbf=m$, $\nbf=n$, and $\sbf=\sigma$, we have that $\xi=0.9428$ for $\gamma=0.1$, i.e. very close to unity. This is the reason why  we have often chosen $\xi=1$ in this work.

For isotropic states, the reduced VDF $\phi(\bm{c},t)$ is expanded in Sonine polynomials, as given by Eq.~\eqref{eq:phiC-Sonine}.  The coefficients with $l=2$ and $l=3$ correspond to the cumulants $a_2$ and $a_3$, respectively. The $n$-th order Sonine approximation consists in retaining up to the $(n+1)$-th cumulant in the above expansion and neglecting higher order ones, for these are assumed to be small. Moreover, nonlinear combinations of the cumulants are also usually dropped, because of their smallness. In this Appendix, we consider two possibilities: the first and the  second---or extended---Sonine approximations.

Under the first Sonine approximation, a closed set of differential equations for the variables $\theta$ and $a_2$ is obtained, since $a_3$ and higher-order cumulants are neglected (also nonlinear terms in $a_2$). From Eq.~\eqref{apA-EFP-dimensionless}, the following evolution equations are derived~\cite{santos_mpemba_2020},
\begin{subequations}\label{eq:evol-eqs-first}
\begin{align}
    \dot{\theta}=& 2(1-\theta)[1+\gamma(d+2)\theta]-2\gamma(d+2)\theta^2a_2,\label{eq:T-evol}\\
    \dot{a}_2= & 8\gamma(1-\theta) \nonumber
    \\
    &-\left[\frac{4}{\theta}-8\gamma+4\gamma(d+8)\theta + \frac{8(d-1)}{d(d+2)}\frac{\sqrt{\theta}}{\xi}\right]a_2, \label{eq:a2-evol}
\end{align}
\end{subequations}
which are linear in $a_2$ but nonlinear in $\theta$. Under the second (or extended) Sonine approximation, the sixth cumulant $a_3$ is incorporated to the picture. Therein, we obtain a closed set of differential equations for the variables $\theta$, $a_2$ and $a_3$, where higher order cumulants---i.e. from $a_4$ on---and non-linear combinations of $a_2$ and $a_3$ are neglected. The result is the system in Eq.~\eqref{eq:evol-eqs-with-a3} of the  main text.
\begin{figure}
\begin{minipage}{1.67in}
\includegraphics[width=\linewidth]{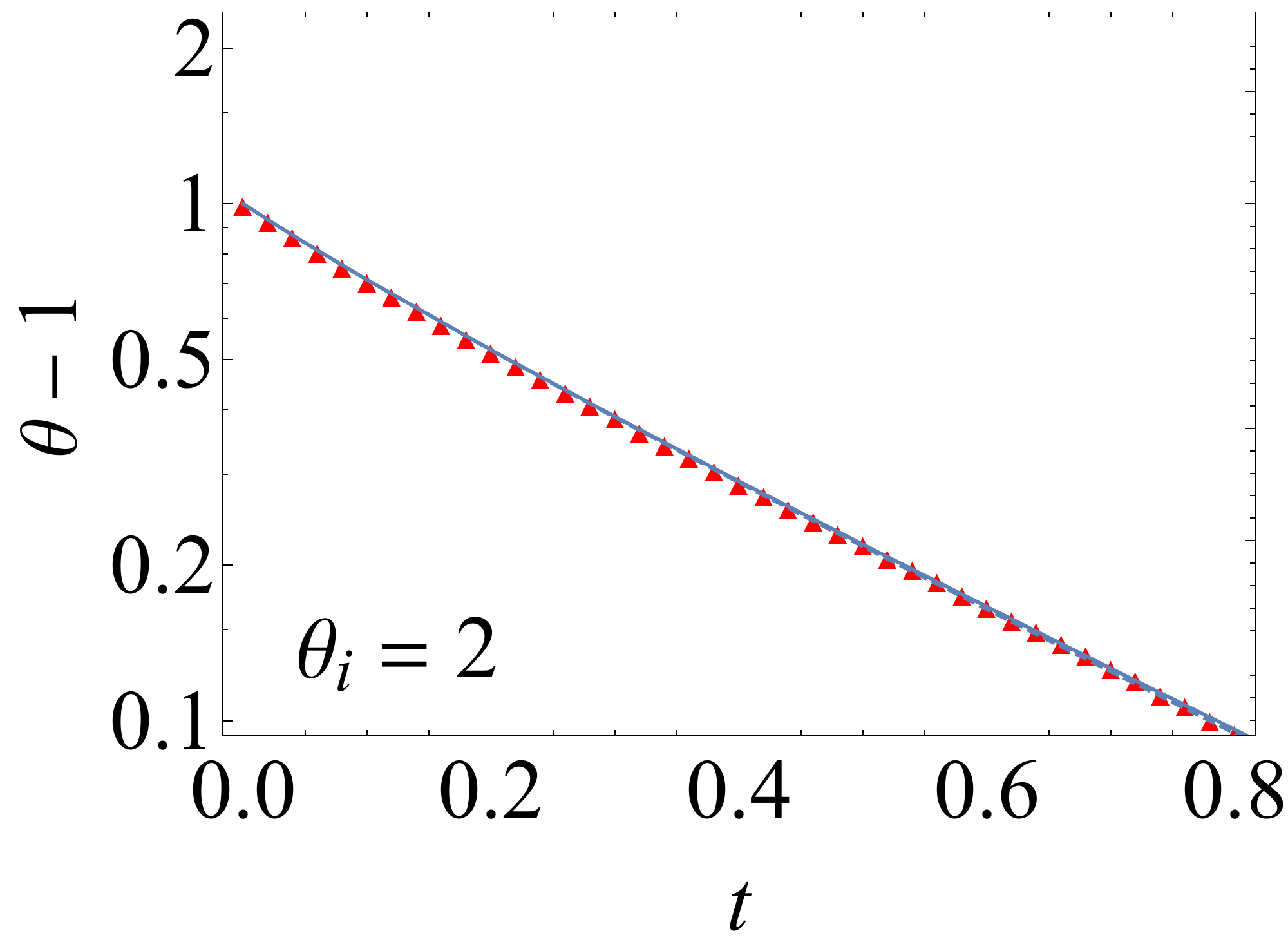}
\label{fig:c1-AA}
\end{minipage}
\begin{minipage}{1.67in}
\includegraphics[width=\linewidth]{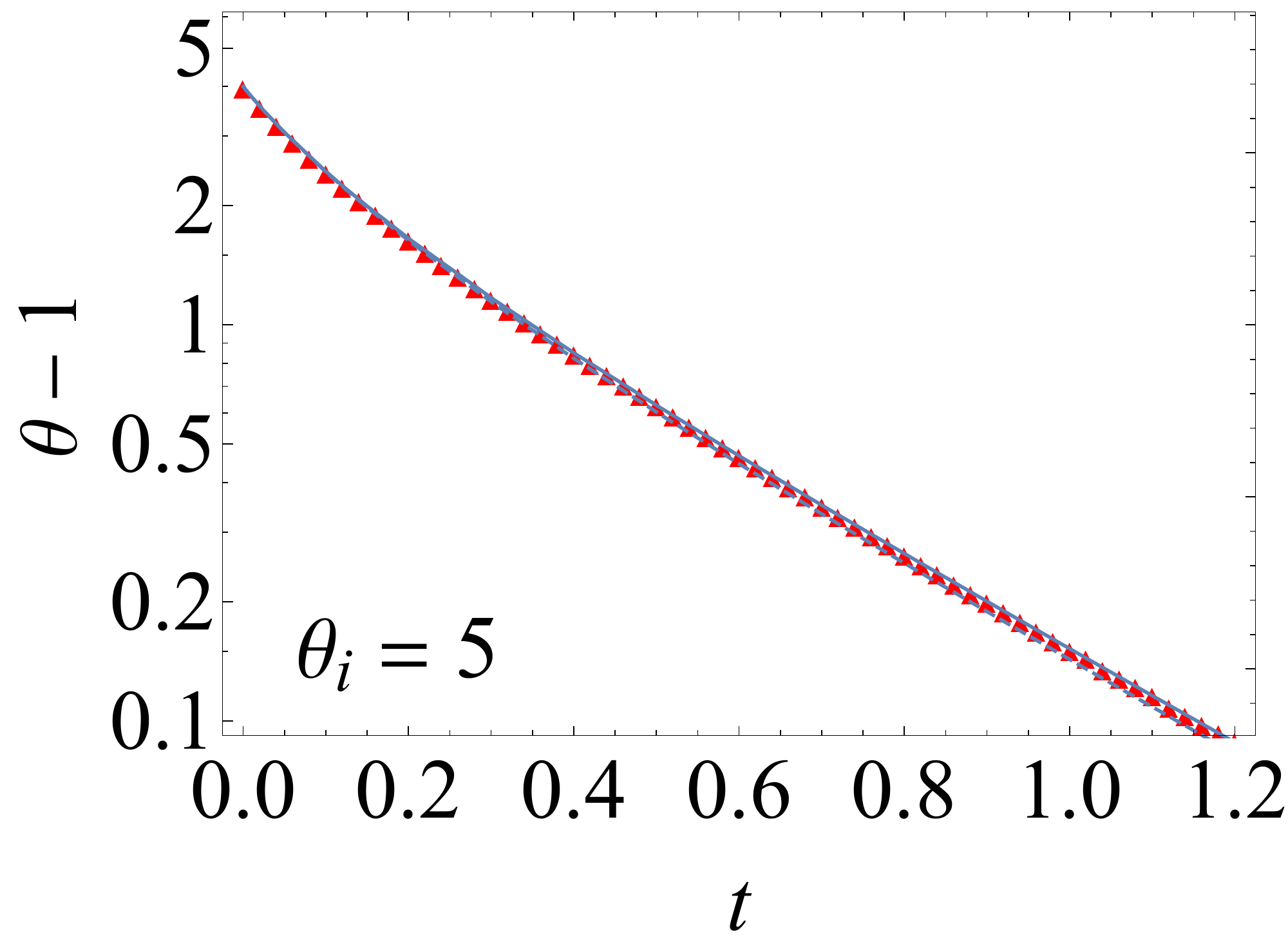}
\label{fig:d1-AA}
\end{minipage}
\medskip
\begin{minipage}{1.625in}
\includegraphics[width=\linewidth]{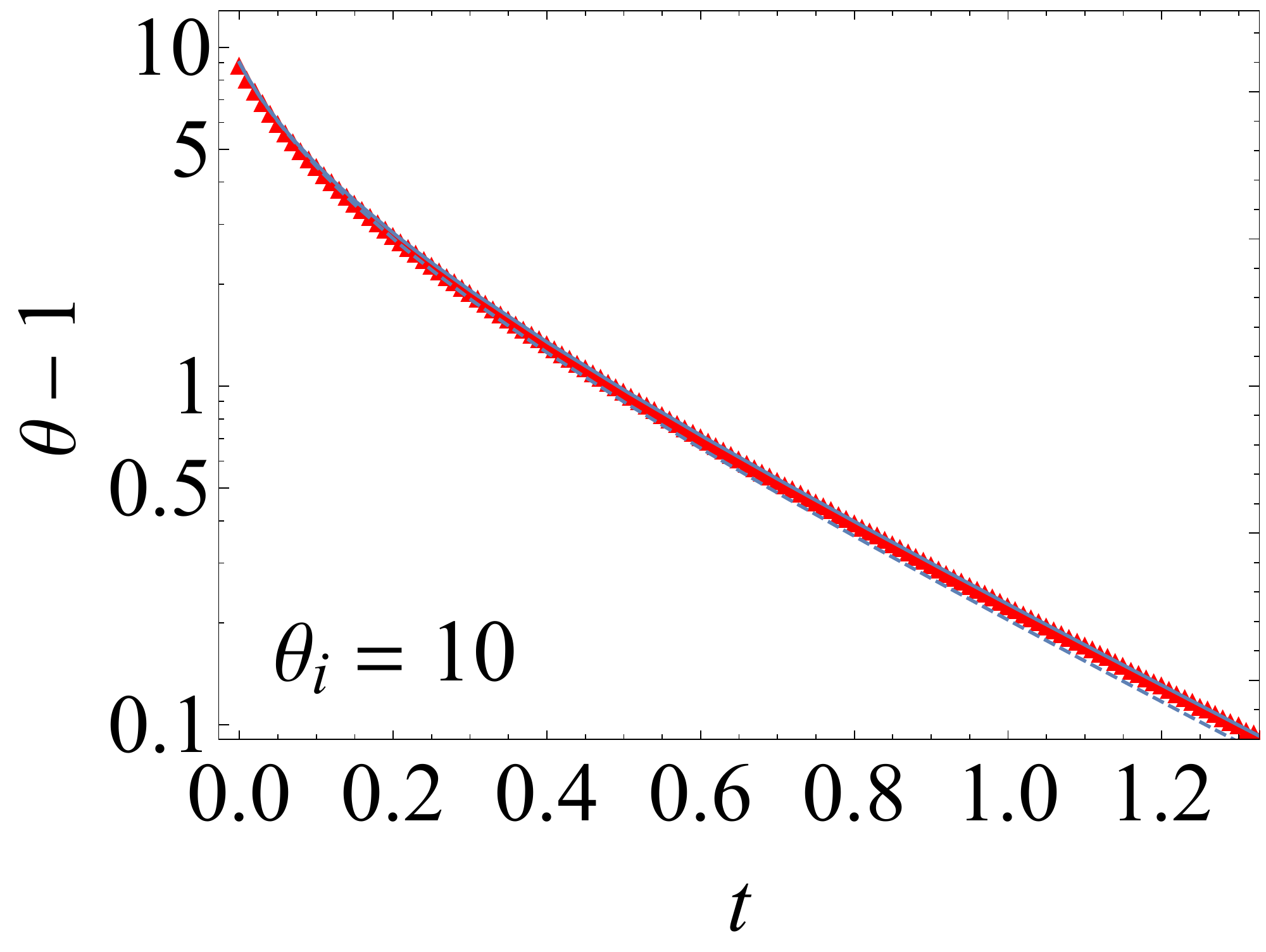}
\label{fig:e1-AA}
\end{minipage}\hspace*{\fill}
\begin{minipage}{1.625in}
\includegraphics[width=\linewidth]{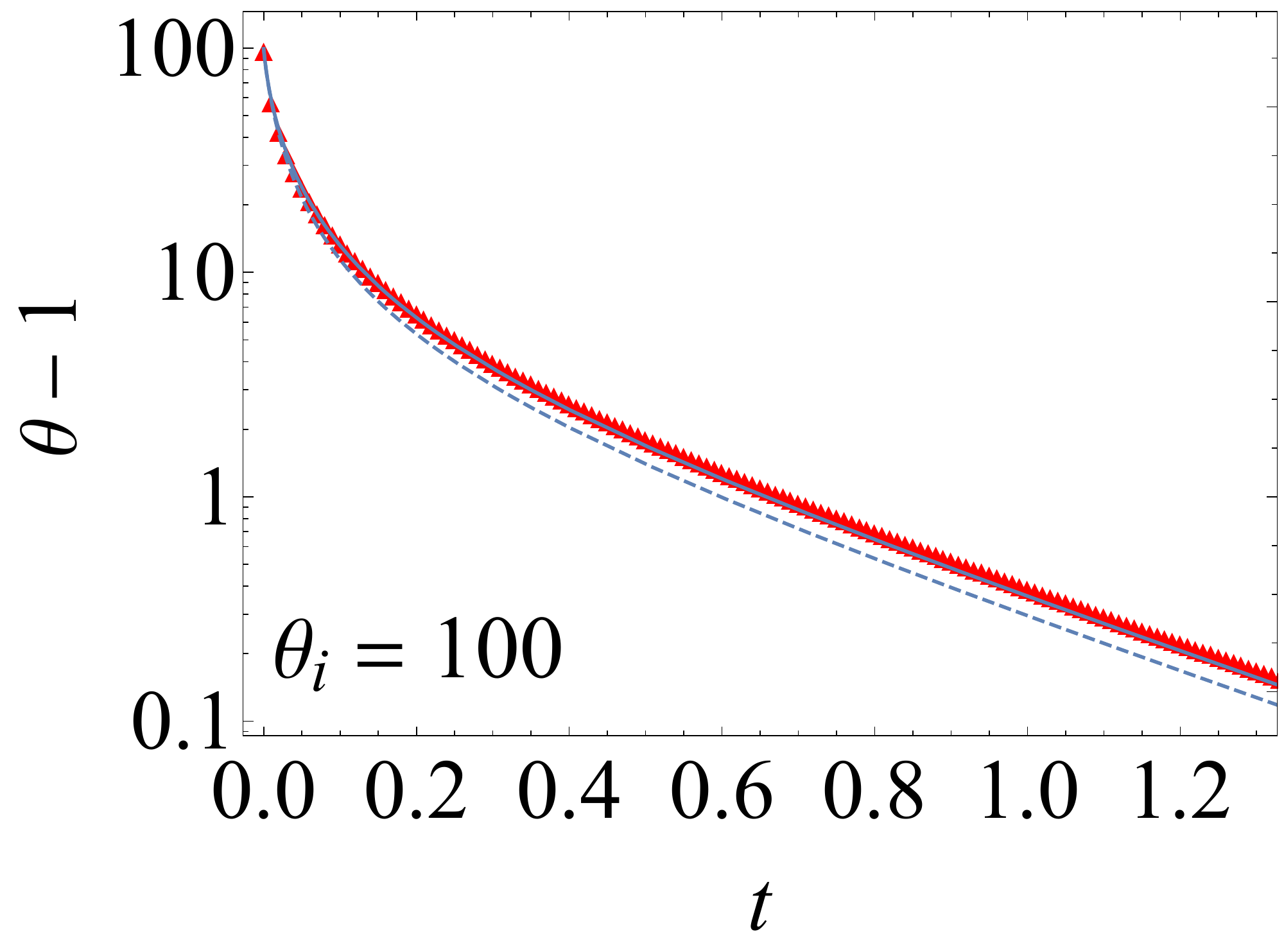}
\label{fig:f1-AA}
\end{minipage}
\caption{Time evolution of the dimensionless temperature for different values of the initial dimensionless temperature. Specifically, we present a logarithmic plot of $\theta-1$, so that a straight line corresponds to an exponential decay to the steady state value $\theta_{\st}=1$. Additional employed parameters are $d=2$, $\xi = 1$ and $\gamma = 0.1$. Symbols correspond to DSMC data. Dashed lines represent the numerical integration of Eqs.\eqref{eq:evol-eqs-first}, for the first Sonine approximation, whereas the full lines correspond to the numerical integration of Eqs.\eqref{eq:evol-eqs-with-a3}, for the extended Sonine approximation.} \label{fig:temp-direct-relax}
\end{figure}

In what follows, we test the validity of the evolution equations provided by the first and the extended Sonine approximations, Eq.~\eqref{eq:evol-eqs-first} and Eq.~\eqref{eq:evol-eqs-with-a3}, respectively. We compare the numerical integration thereof  with DSMC simulations of the EFP equation, which numerically solve it. Specifically, we have considered a two-dimensional system (i.e. hard-discs) with $\gamma=0.1$ and $\xi=1$, which is initially prepared at the equilibrium state corresponding to different values of $\theta_{\ini}$, ranging from $2$ to $100$.

Figure \ref{fig:temp-direct-relax} presents the time evolution of the kinetic temperature. As we may observe, discrepancies between DSMC data and the first Sonine approximation emerge for high enough temperatures. Specifically, they become noticeable over the scale of the figure for $\theta_{\ini}=100$, for which the extended Sonine approximation is clearly superior. As we show in the following, this is due to the cumulants value increasing with $\theta_{\ini}$. Also, it is neatly observed that the relaxation of the temperature changes from being basically exponential for $\theta_{\ini}=2$ and $5$ to strongly non-exponential behaviour for $\theta_{\ini}=100$. We investigate this point in more depth in Appendix~\ref{app:fast-relax-LLNES}.

\begin{figure}
\begin{minipage}{1.645in}
\includegraphics[width=\linewidth]{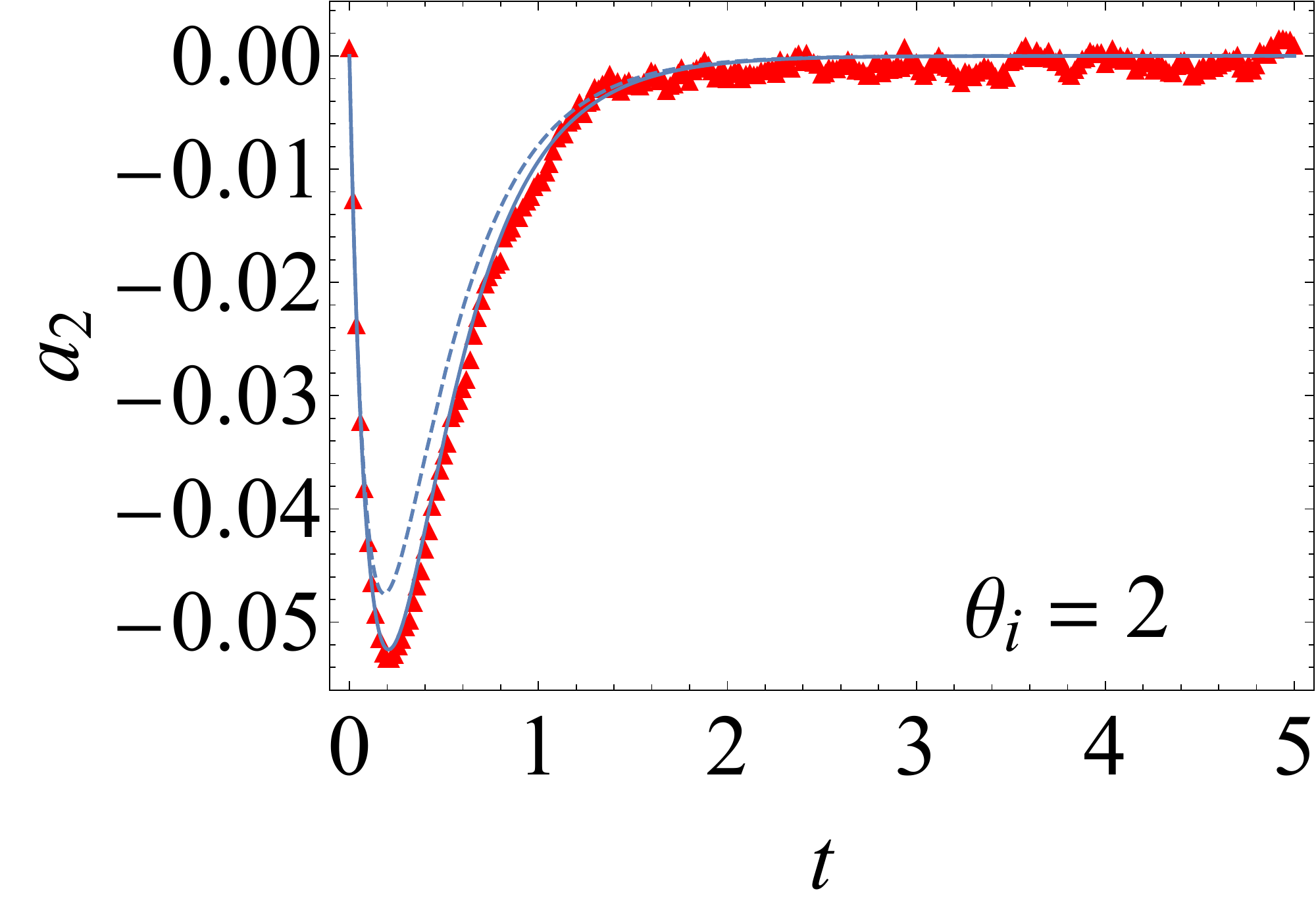}
\label{fig:c1-B}
\end{minipage}\hspace*{\fill}
\begin{minipage}{1.655in}
\includegraphics[width=\linewidth]{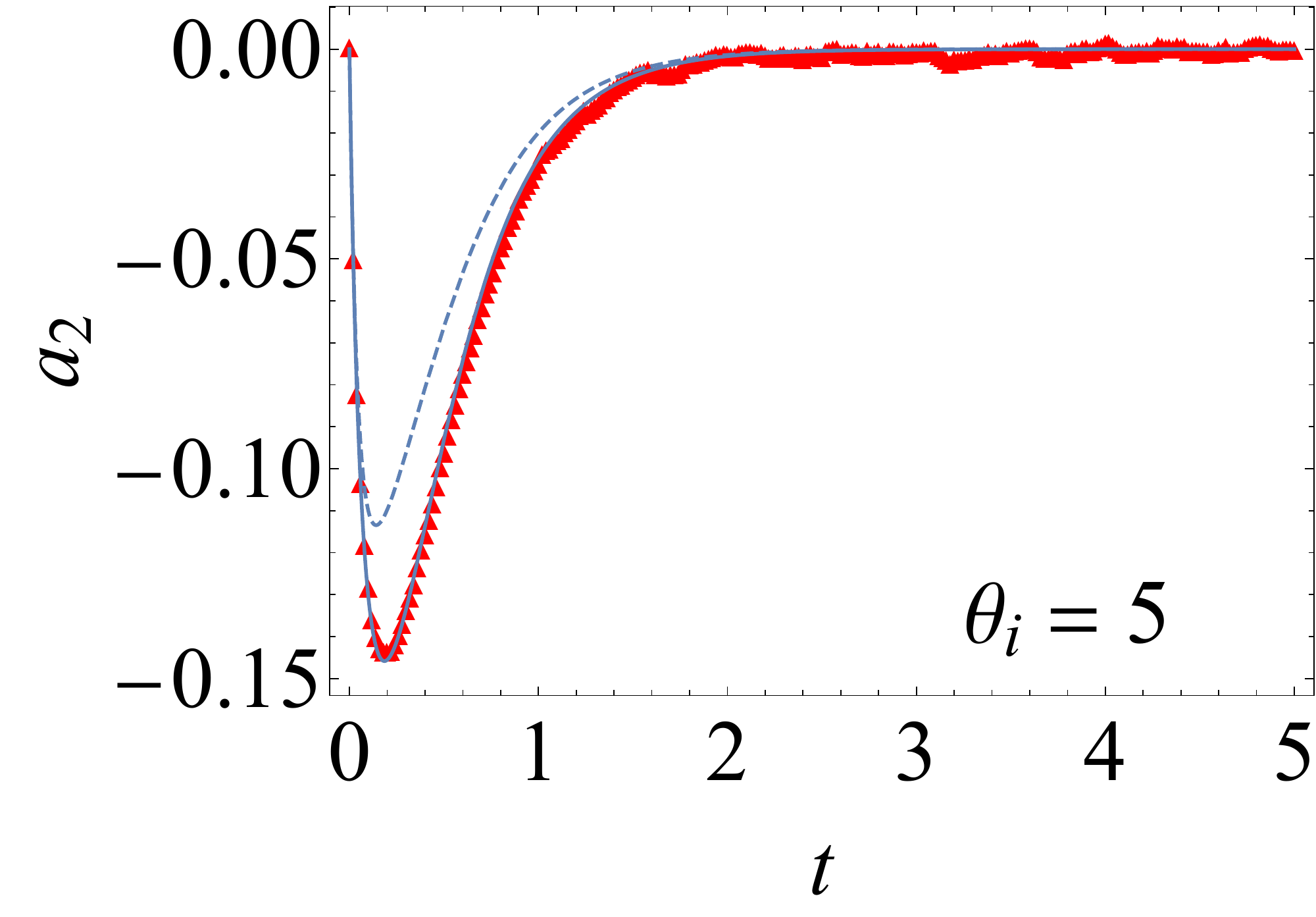}
\label{fig:d1-B}
\end{minipage}
\medskip
\begin{minipage}{1.655in}
\includegraphics[width=\linewidth]{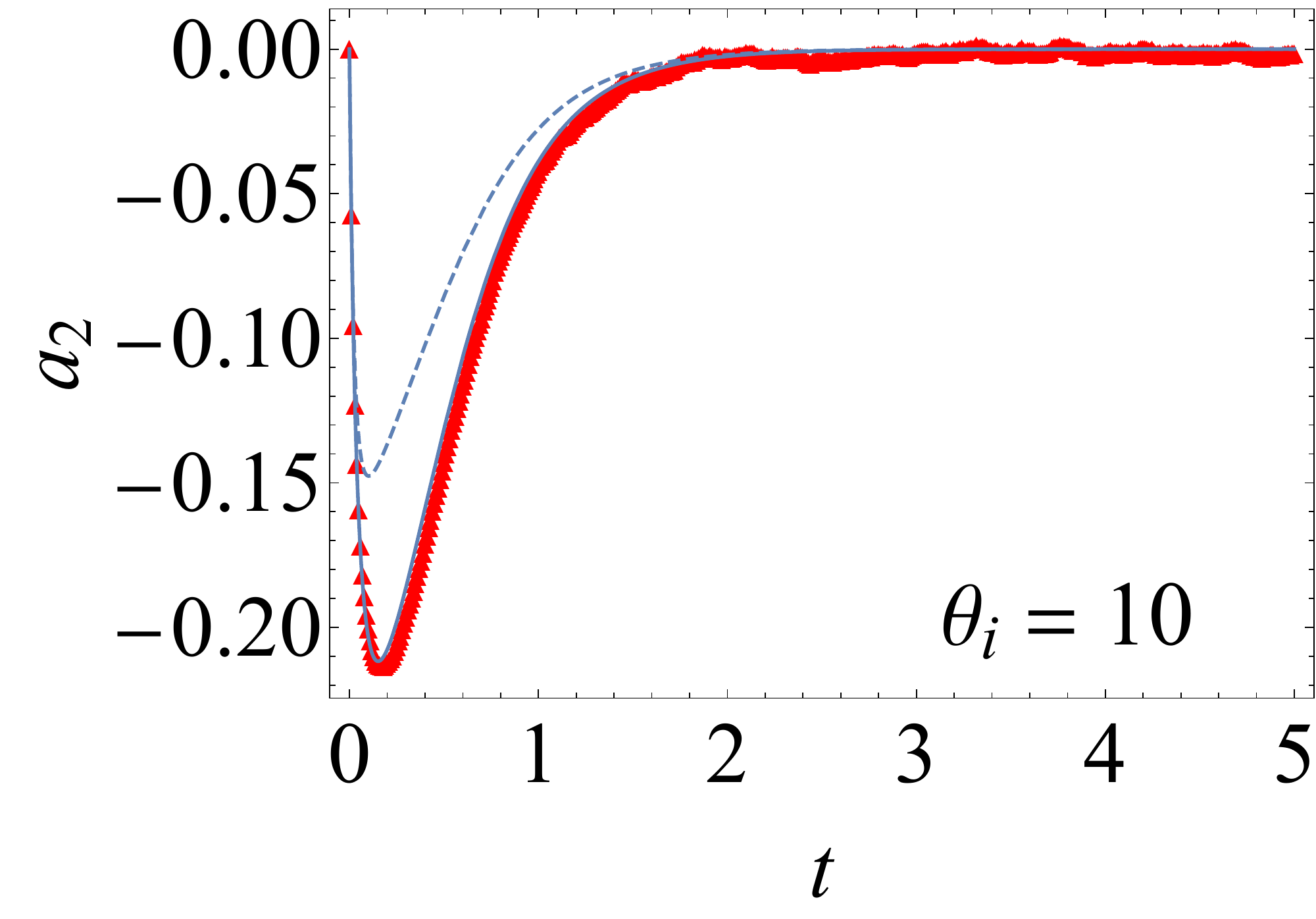}
\label{fig:e1-B}
\end{minipage}\hspace*{\fill}
\begin{minipage}{1.625in}
\includegraphics[width=\linewidth]{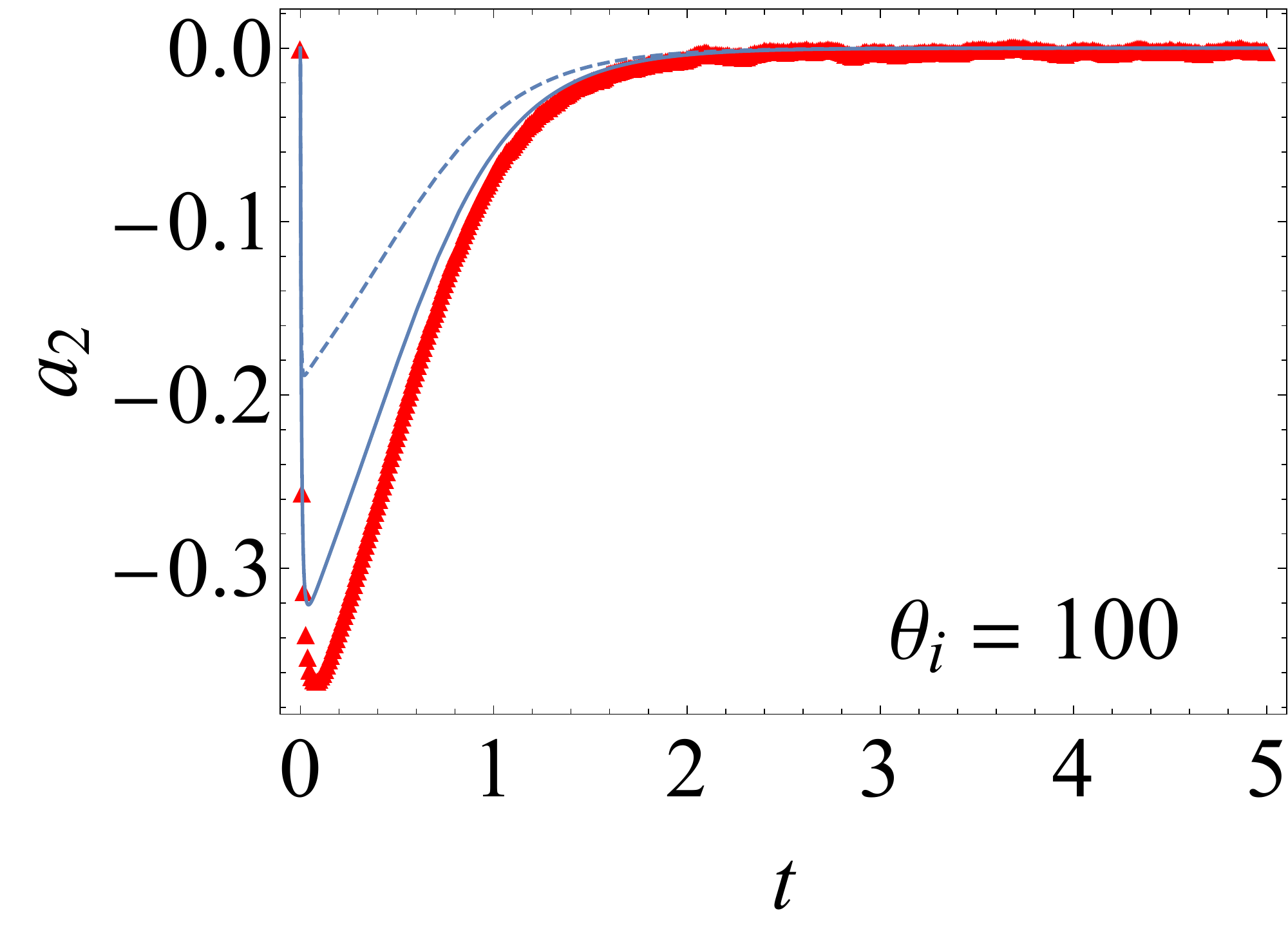}
\label{fig:f1-B}
\end{minipage}
\caption{Time evolution of the excess kurtosis for different values of the initial dimensionless temperature. The panels correspond to the same cases shown in Fig.~\ref{fig:temp-direct-relax} for the temperature, with the same codes for the lines and symbols. It is clearly observed that the extended Sonine approximation (solid line) gives a better description of simulation data (symbols) than the first Sonine approximation (dashed line).
}
\label{fig:phase-1-gamma-approach}
\end{figure}

The differences between the first and second Sonine approximations are even clearer in  Fig.~\ref{fig:phase-1-gamma-approach}, in which we show the time evolution of $a_2$. For the lowest initial temperature, $\theta_{\ini}=2$, both Sonine approximations give quite close results, although it is already observed that the extended Sonine approximation describes the behaviour of the excess kurtosis in a more accurate, quantitative, way. As the initial temperature is increased, the difference between both approaches becomes larger, with the extended Sonine approximation giving always the better description of the actual behaviour of the system. 

For the highest temperature considered in panel (d), $\theta_{\ini}=100$, the minimum value for the excess kurtosis in the first Sonine approximation is roughly one-half of that in DSMC, whereas  the deviation of the theoretical prediction from the DSMC value decreases to $\simeq 10\%$ in the extended Sonine approximation. Looking back at panel (d) of Fig.~\ref{fig:temp-direct-relax}, we note that this slight underestimation of $a_2$ does not impinge on the theoretical prediction for the time evolution of the kinetic temperature, which is the focus of our work.

\begin{figure}
\begin{minipage}{1.665in}
\includegraphics[width=\linewidth]{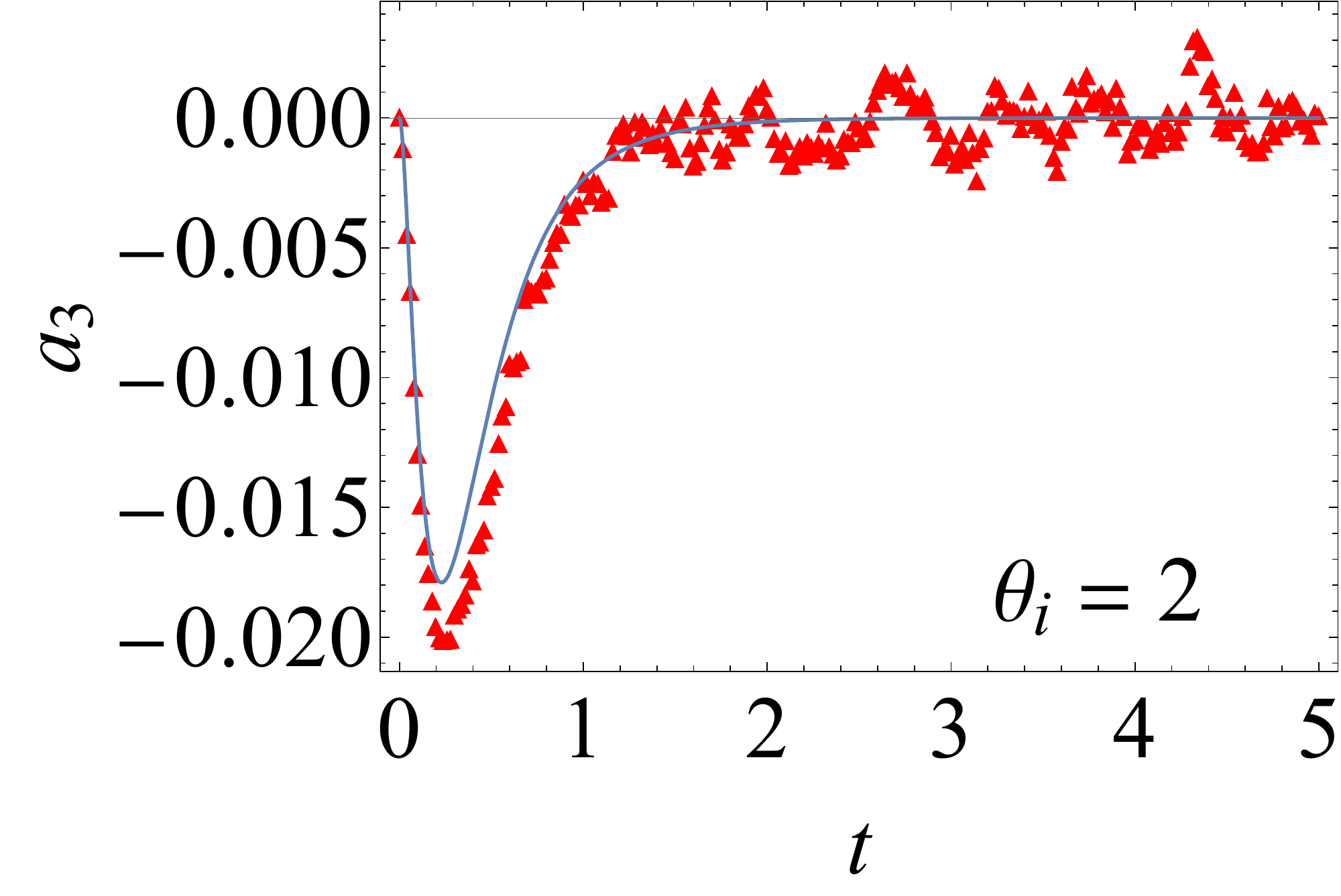}
\label{fig:c1-A}
\end{minipage}\hspace*{\fill}
\begin{minipage}{1.655in}
\includegraphics[width=\linewidth]{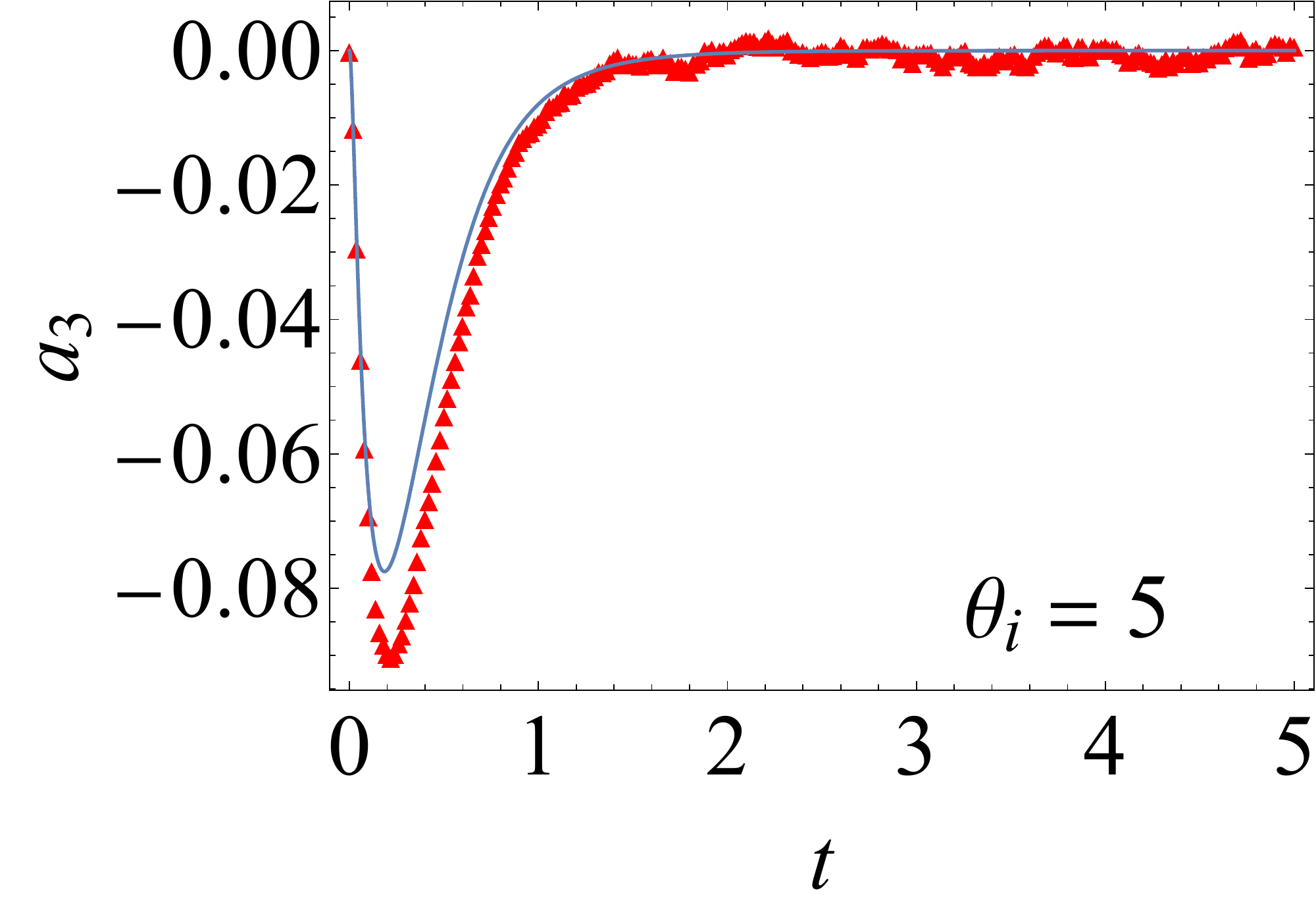}
\label{fig:d1-A}
\end{minipage}
\medskip
\begin{minipage}{1.625in}
\includegraphics[width=\linewidth]{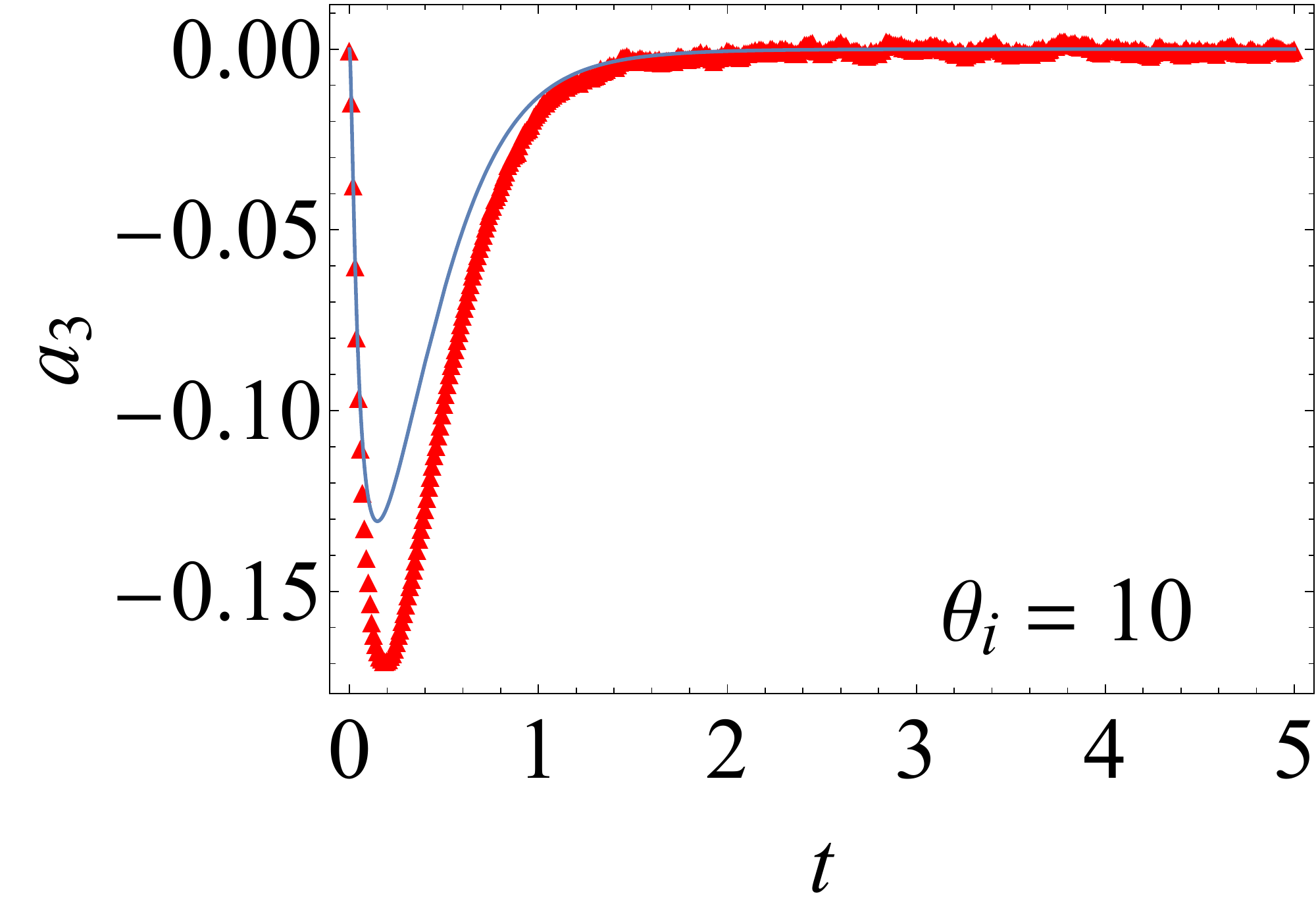}
\label{fig:e1}
\end{minipage}\hspace*{\fill}
\begin{minipage}{1.625in}
\includegraphics[width=\linewidth]{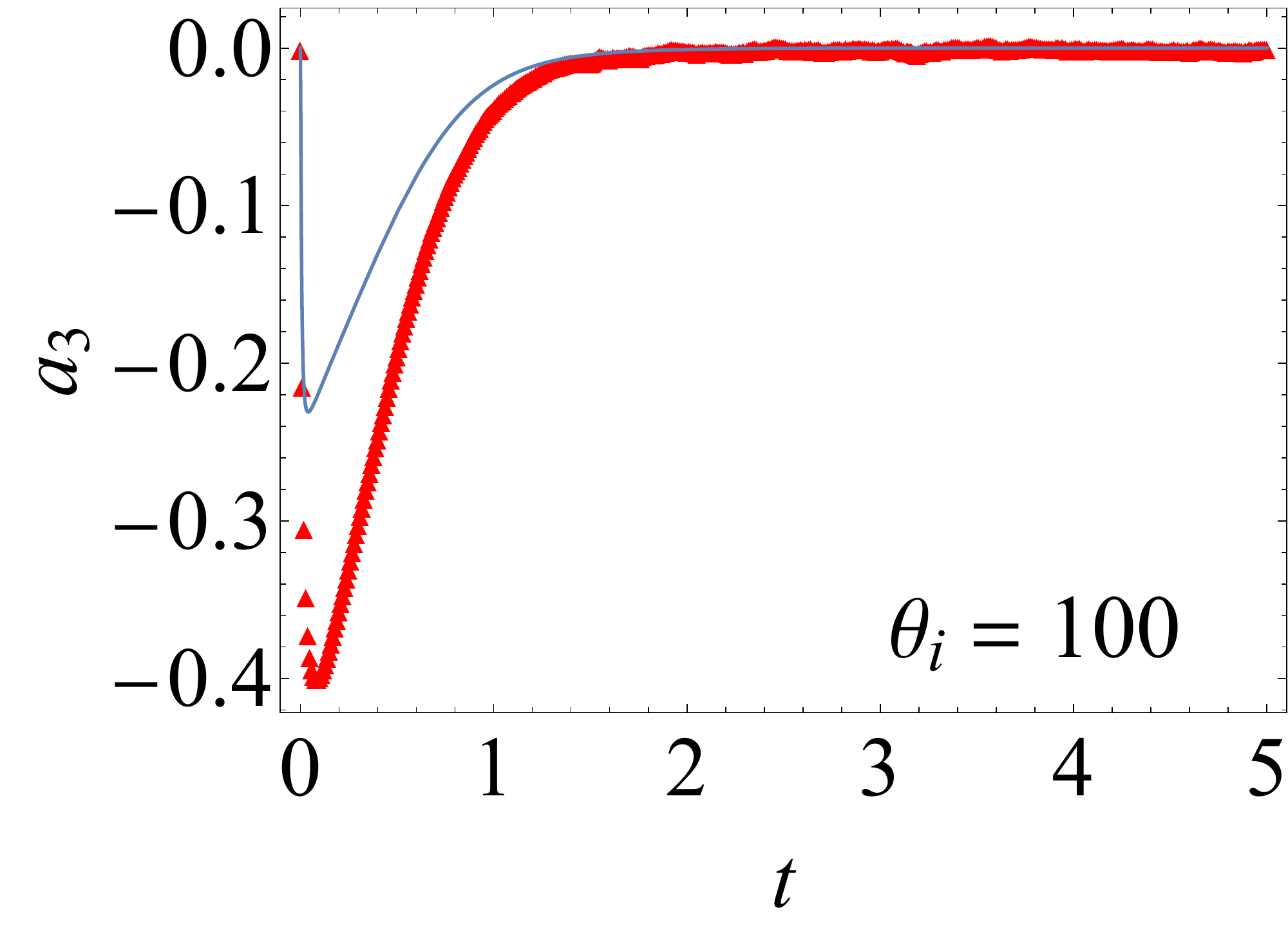}
\label{fig:f1}
\end{minipage}
\caption{Time evolution of the sixth cumulant for different values of the initial dimensionless temperature. The panels correspond to the same cases shown in Fig.~\ref{fig:temp-direct-relax} for the temperature, with the same codes for the lines and symbols.} \label{fig:a3-discrepancies}
\end{figure}

Finally, we show the prediction for $a_3$ in Fig.~\ref{fig:a3-discrepancies}, which makes only sense in the extended Sonine approximation. The discrepancies between the DSMC data and the numerical integration of the evolution equations become more important than for $a_2$, especially as the temperature is increased and the absolute value of $a_3$ also increases. Notwithstanding, the extended Sonine approximation, Eq.~\eqref{eq:evol-eqs-with-a3}, provides the correct qualitative picture. 

Note that, since the temperature is directly coupled to $a_2$ but not to $a_3$, the discrepancies in the sixth cumulant observed in Fig.~\ref{fig:a3-discrepancies} are not relevant for the investigation of the dynamical evolution of the temperature.  To diminish the discrepancies in $a_3$ observed in the second-order Sonine approximation, one could consider a third-order Sonine approximation by introducing the eighth cumulant $a_4$. Within this third-order Sonine approximation, one would expect a qualitative description of $a_4$ and a quantitative account of $a_2$, $a_3$, and the temperature. More specifically, the discrepancies in $a_3$ observed in the second Sonine approximation would be ``transferred'' to $a_4$ in the third Sonine approximation (and those in $a_2$ to $a_3$, those in $\theta$ to $a_2$).

\section{Fast relaxation to the LLNES}\label{app:fast-relax-LLNES}

Here, we show that the cumulants decay to their reference values over a time scale that is shorter than that of the relaxation of the temperature, after a quench to a low temperature. Therefore, the system quickly reaches the LLNES described in the main text, over which the cumulants are basically constant and equal to their reference values and the temperature relaxes algebraically.

According to the approximate evolution equations in the $s$ scale \eqref{eq:evol-eqs-s-scale}, both $Y(s)$ and the cumulants $a_2(s)$ and $a_3(s)$ tend to stationary values for long enough times. On the one hand, $Y\to 0$, which seems counterintuitive at first glance, but we
must not forget that the approximate system of ODEs only remains valid for high enough temperatures, i.e. $Y = O(1)$. On the other hand, the cumulants tend to their respective reference values $a_2^r$ and $a_3^r$.  For longer times, i.e when $Y\ll 1$,  Eq.~\eqref{eq:evol-eqs-s-scale} ceases to be valid, and the whole extended Sonine framework, as described by Eqs.~\eqref{eq:evol-eqs-with-a3} has to be used. It is only over this very long time scale that the temperature actually relaxes towards its stationary value $\theta_{\st}=1$, and all the cumulants tend to zero---for the equilibrium VDF is Gaussian. 

Here we show that the main part of the relaxation of the temperature takes place over the $s$ scale. Moreover, we show that the cumulants quickly relax to their reference values, as given by Eq.~\eqref{eq:a2r-a3r}, while the temperature relaxes in a much slowlier way. To do so, it is useful to start by considering the evolution equations in the $s$ scale in the first Sonine approximation, i.e. when $a_3$ is neglected. Therein, we have the system
\begin{subequations}\label{eq:evol-eqs-dominant-first}
\begin{align}
    \label{evol-eq-T-dominant-first}
    \frac{dY}{ds} & \approx -2 (d+2) Y^2(1+a_2),
    \\
    \frac{da_2}{ds} & \approx -4 Y (d+8) (a_2 - a_2^{\prime r}),
\end{align}
\end{subequations}
in which $a_2^{\prime r}=-2/(d+8)$ is the reference value for the excess kurtosis in the first Sonine approximation. This system of equations can be solved in parametric form, since
\begin{equation}
\label{eq:parametric-ode}
    \frac{da_2}{dY}=\frac{2(d+8)}{Y(d+2)}\frac{a_2-a_2^{\prime r}}{1+a_2},
\end{equation}
is a separable first order ODE with solution
\begin{equation}
\label{eq:parametric-Y}
    Y(a_2) = \left[\frac{d+8}{2}(a_2 - a_2^{\prime r})\right]^{\alpha}\exp \left[\frac{(d+2)a_2}{2(d+8)}\right], 
\end{equation}
where $\alpha = (d+2)(d+6)/[2(d+8)^2]$. Equation~\eqref{eq:parametric-Y} implies that $a_2$ reaches its reference value when the temperature is still relaxing. Let us prove this statement by considering a small perturbation in $a_2$ around its reference value, $a_2 = a_2^{\prime r} + \delta, \delta \ll 1$, and inserting it into \eqref{eq:parametric-Y},
\begin{align}
    Y &\approx \left[ \frac{d+8}{2} \right]^{\alpha} \exp \left[-\frac{d+2}{(d+8)^2}\right]\delta^{\alpha} \nonumber 
    \\
    & \Longrightarrow
    \delta = \frac{2}{d+8}Y^{1/\alpha}\exp \left(\frac{2}{d+6} \right).
\end{align}
Thus, for $Y = 1/2$, we get $\delta \approx 3.37 \times 10^{-3}$ (for $d=2$), which gives a relative error for the excess kurtosis $\delta / |a_2^{\prime r}| \approx 0.02$. 

\begin{figure}
\includegraphics[width=3.25in]{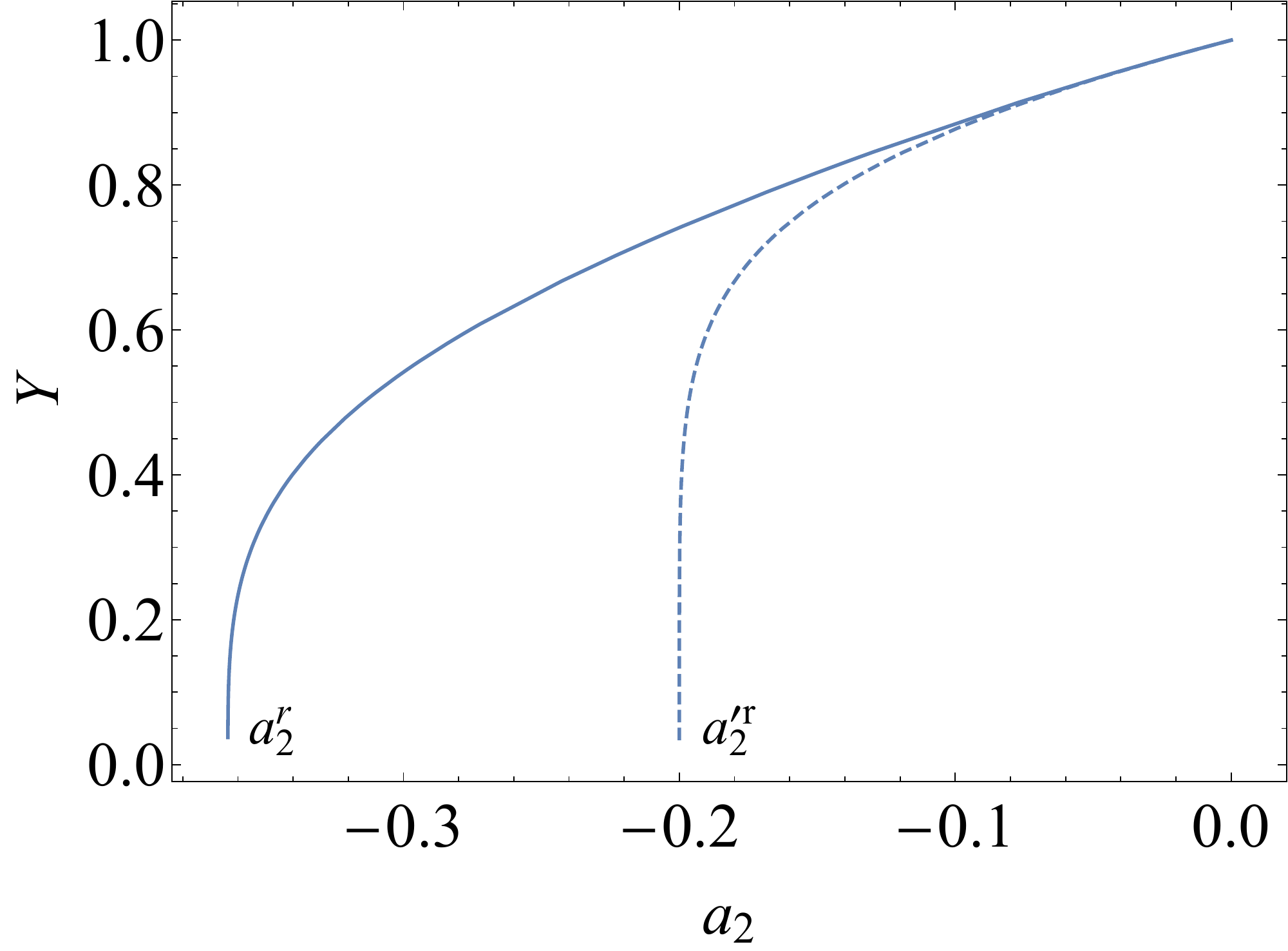}
\caption{Plot of the parametric curves $Y = Y(a_2)$. The curves for both the first (dashed line) and the extended (solid line) Sonine approximations are plotted, for $d=2$. The former is given by Eq.~\eqref{eq:parametric-Y}, while the latter follows from the numerical integration of Eqs.~\eqref{eq:evol-eqs-s-scale}. Also marked are the reference values for the excess kurtosis in both frameworks, $a_2^{\prime r}$ and $a_2^r$.}
    \label{fig:high-theta-parametric-A}
\end{figure}

The general picture outlined above is illustrated in Fig.~\ref{fig:high-theta-parametric-A}. Therein, we plot the parametric solution \eqref{eq:parametric-Y} (dashed line). It is clearly observed that $a_2$ is very close to its reference value  $a_2^{\prime r}$  for $Y\leq 1/2$. Also plotted is the corresponding parametric curve for the extended Sonine approximation (solid line), which has been obtained from the numerical integration of Eq.~\eqref{eq:evol-eqs-s-scale}. The same qualitative picture applies, although the values of $Y$ for which $a_2$ is very close to its reference value $a_2^r$ become smaller, $Y\leq 0.2$. 

\section{Extrema for the cumulants}\label{app:optimisation-kurtosis}

\begin{figure}
\includegraphics[width=3.25in]{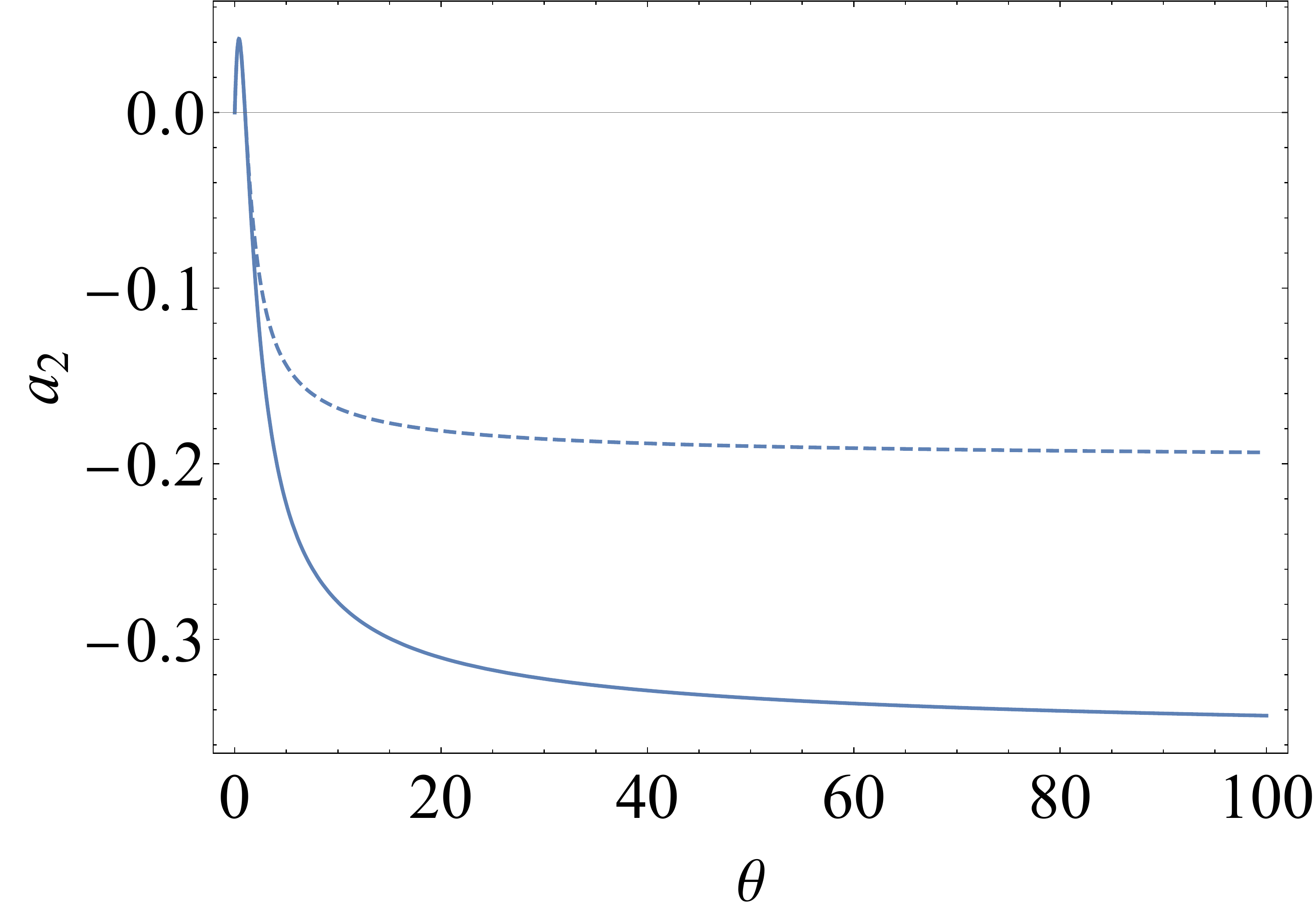}
\includegraphics[width=3.25in]{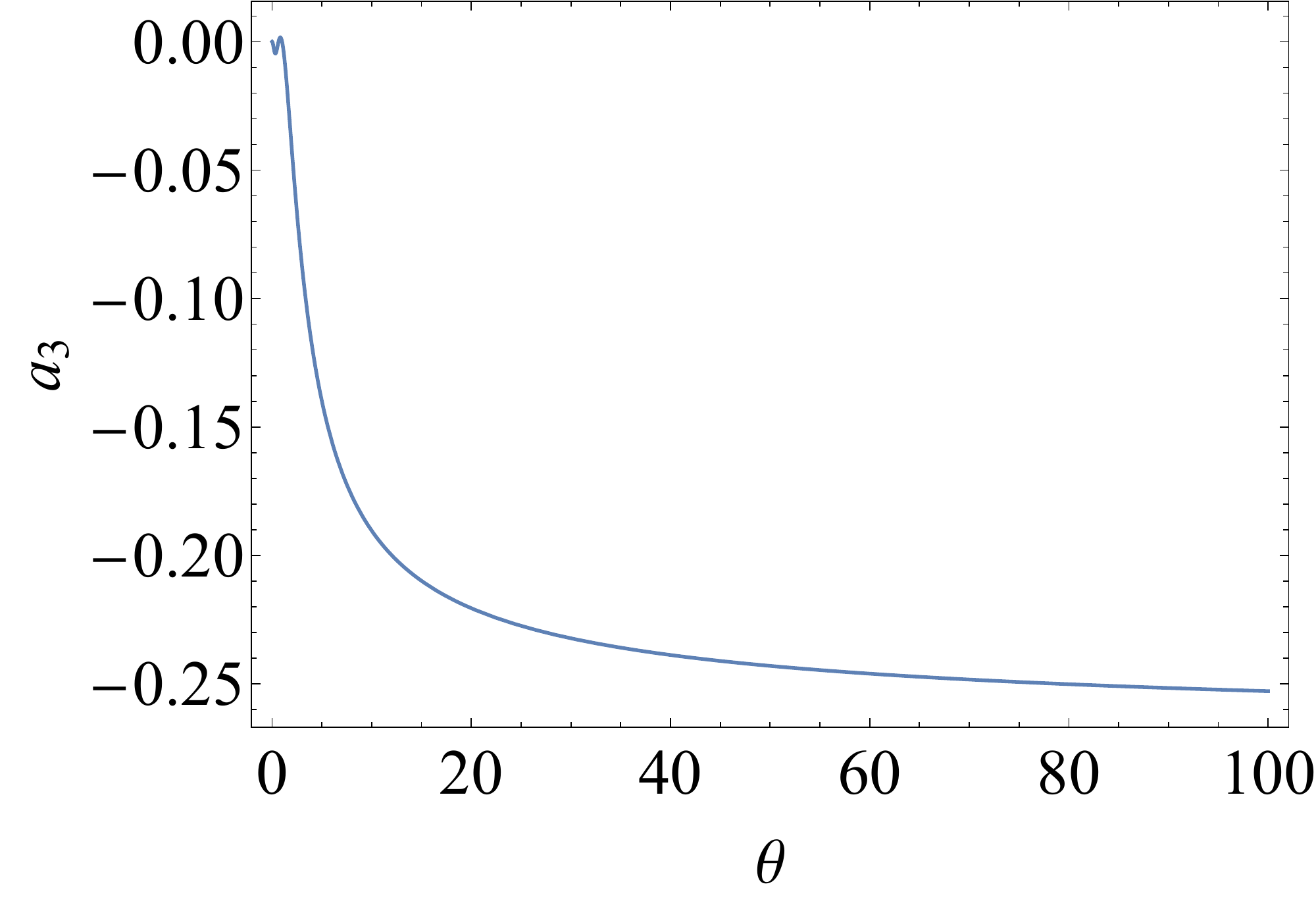}
\caption{Parametric solutions of the extrema for the cumulants as functions of the dimensionless temperature. Specifically, we plot the extremum for the excess kurtosis $a_{2}$ (top panel) and the sixth cumulant $a_{3}$ (bottom panel). In the extended Sonine approximation, they are obtained by imposing $\dot{a}_2 = \dot{a}_3 = 0$ in Eqs.~\eqref{eq:evol-eqs-s-scale} (solid lines). In the first Sonine approximation, only the curve for the excess kurtosis can be plotted (dashed line on panel (a)), which is given by Eq.~\eqref{eq:12ext-1st-Sonine}. Additional employed parameters are $d=2$, $\xi = 1$, and $\gamma = 0.1$. }
    \label{fig:extremes-cumulants}
\end{figure}

In this Appendix, we look into the extreme values---minimum and maximum---of the cumulants $a_2$ and $a_3$. Again, it is instructive to start by considering the first Sonine approximation. Let us focus on \eqref{eq:a2-evol}: at the time such that $a_2$ reaches one of its extrema, we have that $\dot{a}_2=0$, i.e. the corresponding value of the excess kurtosis must verify
\begin{equation}\label{eq:12ext-1st-Sonine}
    a_2^{\text{ext}}=\frac{8\gamma(1-\theta)}{\frac{4}{\theta}-8\gamma+4\gamma(d+8)\theta + \frac{8(d-1)}{d(d+2)}\frac{\sqrt{\theta}}{\xi}}.
\end{equation}
For a given value of $\xi$, $a_2^{\text{ext}}$ is a function of $\theta$. In fact, the asymptotic behaviour of $a_2^{\text{ext}}$ is independent  of the average inter-collision time $\xi$ both in the limits $\theta\to 0^+$ and $\theta\to +\infty$,
\begin{align}
\label{eq:extremes-excess-kurtosis}
    a_2^{\text{ext}}&\sim 2\gamma\theta, \quad \theta\to 0^+, \nonumber
    \\
    a_2^{\text{ext}}&\to a_2^{\min}=-\frac{2}{d+8}, \quad \theta\to +\infty.
\end{align}
Moreover, $a_2^{\text{ext}}=0$ for $\theta=1$, $\forall\xi$. This means that the general qualitative picture of $a_2^{\text{ext}}$ is the following, $\forall\xi$: it vanishes at $\theta=0$, has a maximum in the interval $\theta\in(0,1)$, and decreases to its minimum value $a_2^{\min}$ for $\theta>1$. The specific case $\xi=1$ is presented in panel (a) of  Fig.~\ref{fig:extremes-cumulants} (dashed line). Note that $a_2^{\min}$ is also independent of $\gamma$, in fact it equals the reference value $a_2^{\prime r}$ in the first Sonine approximation. On the other hand, $a_2^{\max}$ is roughly proportional to $\gamma$ and thus quite small: in the case $\xi=\infty$ (FP limit), to the lowest order in $\gamma$ one has $a_2^{\text{ext}}\approx 2\gamma\theta(1-\theta)$ and $a_2^{\max}\approx\gamma/2$.

In the extended Sonine approximation, we impose $\dot{a}_2 = \dot{a}_3 = 0$ in Eqs.~\eqref{eq:evol-eqs-s-scale} to get the extrema of $a_2$ and $a_3$, $a_2^{\text{ext}}$ and $a_3^{\text{ext}}$. The explicit expressions of $a_2^{\text{ext}}$ and $a_3^{\text{ext}}$ as a function of $\theta$ and $\xi$ are quite complicated and not particularly illuminating, so we do not write them here. Yet, the qualitative behaviour of $a_2^{\text{ext}}$ and $a_3^{\text{ext}}$ is similar to the one found for the excess kurtosis in the first Sonine approximation.
Both $a_2^{\text{ext}}$ and $a_3^{\text{ext}}$  vanish at $\theta=0$ and $\theta=1$, 
and tend to their minimum (negative) values $a_2^{\min}$ and $a_3^{\min}$ for $\theta\to +\infty$, independently of the value of $\xi$. Also, both minima $a_2^{\min}$ and $a_3^{\min}$ do not depend on $\gamma$ and coincide with their pseudo-stationary, reference, values $a_2^r$ and $a_3^r$, respectively. A particular case, again for $\xi=1 $, is presented in Fig.~\ref{fig:extremes-cumulants} (solid lines), $a_2$ ($a_3$) in its left (right) panel. The maximum values of both cumulants are again basically proportional to $\gamma$ and thus much lower (in absolute value) than their respective minima. 

\section{Perturbative approach to the Kovacs effect}\label{app:kovacs}

Now we consider the Kovacs effect described in the main text. In the aging time window $0\leq t\leq t_w$, the system relaxes towards the LLNES and therefore the cumulants take their reference values at the end of this stage, $a_2(t_w)=a_2^r$, $a_3(t_w)=a_3^r$. Here, we derive an analytical expression for the non-monotonic behaviour of the temperature, i.e. the Kovacs hump, that arises when the system is put in contact with a thermal bath at temperature $T=T(t_w)$ for $t>t_w$.

The evolution equations \eqref{eq:evol-eqs-with-a3} cannot be exactly solved, but we may resort to a perturbative expansion to get an approximate expression for the time evolution of the temperature. The initial conditions  for the Kovacs experiment are
\begin{equation}\label{eq:initial-conds-Kovacs}
    T(t_w)=T_{\st}, \quad a_2(t_w)=a_2^r, \quad a_3(t_w)=a_3^r.
\end{equation}
A perturbation theory in the cumulants is not expected to give good results, since $a_2^r$ and $a_3^r$ are quite large, as we have already discussed. However, bringing to bear that $\gamma\leq 0.1$, we can develop a perturbative theory in the product $\gamma a_{0}$, where $a_0$ is of the same order as the reference values for the cumulants, i.e. $a_2^r/a_0$ and $a_3^r/a_0$ are both of the order of unity. Then we write
\begin{subequations}
\label{eq:Kovacs-perturbative-a3}
\begin{equation}
    \theta(t) = \theta^{(0)}(t) + \gamma a_{0} \theta^{(1)}(t) + O((\gamma a_{0})^2),
\end{equation}
\begin{equation}
    A_2(t) = A_2^{(0)}(t) + \gamma a_{0} A_2^{(1)}(t) + O((\gamma a_{0})^2),
\end{equation}
\begin{equation}
    A_3(t) = A_3^{(0)}(t) + \gamma a_{0} A_3^{(1)}(t) + O((\gamma a_{0})^2),
\end{equation}
\end{subequations}
in which we have defined
\begin{equation}
    A_2(t) \equiv \frac{a_2(t)}{a_2^r}, \qquad A_3(t) \equiv \frac{a_3(t)}{a_{3}^r},
\end{equation} 
which also are of the order of unity. The above expansions lead to the following hierarchy: to the lowest, $O(1)$, order we have
      \begin{widetext}
    \begin{subequations}
    \begin{equation}
    \label{eq:theta^0-a3}
        \dot{\theta}^{(0)}= 2(1-\theta^{(0)})\left[1+\gamma (d+2)\theta^{(0)}\right],
      \end{equation}
    \begin{equation}
    \label{eq:A2^0-a3}
    \begin{split}
        \dot{A}_2^{(0)}= & \frac{8\gamma}{a_{2}^r}(1-\theta^{(0)}) -\left[\frac{4}{\theta^{(0)}}-8\gamma+4\gamma(d+8)\theta^{(0)}+ \frac{8(d-1)}{d(d+2)}\frac{\sqrt{\theta^{(0)}}}{\zeta_{0}^*}\right]A_2^{(0)}
        \\
        & + 2 \left[ 2\gamma \theta^{(0)} (d+4) + \frac{(d-1)}{d(d+2)}\frac{\sqrt{\theta^{(0)}}}{\xi} \right]\frac{a_{3}^r}{a_{2}^r}A_3^{(0)},
    \end{split}
    \end{equation}
    \begin{equation}
    \label{eq:A3^0-a3}
    \begin{split}
        \dot{A}_3^{(0)}= & 12\left[-4\gamma + 6\gamma \theta^{(0)} + \frac{(d-1)\sqrt{\theta^{(0)}}}{d(d+2)(d+4)\xi}\right]\frac{a_2^r}{a_3^r}A_2^{(0)}
        \\
        & + 6 \left[4\gamma - \frac{1}{\theta^{(0)}} - \gamma \theta^{(0)}(d+14) - \frac{(d-1)(4d+19)\sqrt{\theta^{(0)}}}{2d(d+2)(d+4)\xi}\right]A_3^{(0)},
    \end{split}
    \end{equation}
  \end{subequations}
      \end{widetext}
    and to the first,  $O(\gamma a_{0})$, order
    \begin{align}   
      \dot{\theta}^{(1)}=&-2\theta^{(1)}\left[1+\gamma (d+2)\theta^{(0)}\right]+2\gamma (d+2)\theta^{(1)}[1-\theta^{(0)}] \nonumber \\ & -2(d+2) \frac{a_2^r}{a_0}\left(\theta^{(0)}\right)^2 A_2^{(0)}.
      \label{eq:theta^1-a3}
    \end{align}
    We do not write the equations for $A_2^{(1)}$ and $A_3^{(1)}$ because they are not necessary for the calculation of the temperature to the first order, which is our goal here.

In the scaled variables, the initial conditions are $\theta(t_w)=A_2(t_w) = A_3(t_w) = 1$. This means that, in the perturbative series, $\theta^{(0)}(t_w)=A_2^{(0)}(t_w) = A_3^{(0)}(t_w) = 1$ whereas $\theta^{(j)}(t_w)=A_2^{(j)}(t_w) = A_3^{(j)}(t_w) = 0$, $\forall j\geq 1$. Let us focus first on the lowest order. We have that $\theta^{(0)}(t) = 1$, i.e. there is no Kovacs effect if $\gamma a_0=0$. This is logical, since this condition is fulfilled if either $\gamma=0$, i.e. linear drag, or $a_0=0$, i.e. the system is at equilibrium for $t=t_w$. Neither of these situations allows for the emergence of the Kovacs effect. Second, $A_2^{(0)}$ and $A_3^{(0)}$ are obtained by solving
\begin{equation}
    \frac{d}{dt}\bm{A}^{(0)} = \bm{M} \cdot \bm{A}^{(0)},
\end{equation}
where the vector $\bm{A}^{(0)}$ and the matrix $\bm{M}$ are defined as
\begin{equation}
        \bm{A}^{(0)} \equiv \begin{pmatrix}
        A_2^{(0)} \\
        A_3^{(0)}
        \end{pmatrix}, \quad \bm{M} \equiv \begin{pmatrix}
        M_{11} & M_{12} \\
        M_{21} & M_{22}
        \end{pmatrix},
\end{equation}
where $M_{ij}$ has been given in Eq.~\eqref{eq:Mij} of the main text. The eigenvalues $\lambda_{\pm}$ of the matrix  $\bm{M}$ have been defined in Eq.~\eqref{eq:lambda-pm}, and their corresponding eigenvectors are 
\begin{equation}
    \bm{u}_{\pm} = \begin{pmatrix} M_{12} \\ \lambda_{\pm}- M_{11} \end{pmatrix}.
\end{equation}
The solution is thus given by
\begin{equation}
\label{eq:vector-cumulant}
    \bm{A}^{(0)}(t) = c_+ \bm{u_+} e^{\lambda_+ (t-t_w)} + c_- \bm{u_-} e^{\lambda_- (t-t_w)},
\end{equation}
where $c_+$ and $c_-$ are determined by imposing the initial conditions, which results in
\begin{equation}
    c_+ = \frac{M_{11}+M_{12}-\lambda_-}{(\lambda_+-\lambda_-)M_{12}}, \quad c_- = \frac{\lambda_+-M_{11}-M_{12}}{(\lambda_+-\lambda_-)M_{12}}.
\end{equation}

Once the lowest order is completed, we make use of Eq.~\eqref{eq:theta^1-a3} to compute $\theta^{(1)}(t)$, which gives the simplest theoretical prediction for the Kovacs hump. Recall that $\theta^{(0)}(t)=1$ to the lowest order, so Eq.~\eqref{eq:theta^1-a3} simplifies to
\begin{equation}
    \label{eq:theta^1-a3-v2}
        \dot{\theta}^{(1)}=-2\theta^{(1)}\left[1+\gamma (d+2)\right]-2(d+2) \frac{a_2^r}{a_0} A_2^{(0)}.
      \end{equation}
      Taking into account the initial condition $\theta^{(1)}(t_w)=0$, one gets
      \begin{equation}
        \label{eq:theta1-a3theory}
        \theta^{(1)}(t) = -2(d+2)\frac{a_2^r}{a_0} e^{-\alpha t}\int_{t_w}^t e^{\alpha u} A_2^{(0)}(u) du,
      \end{equation}
      in which $\alpha \equiv 2[1+\gamma(d+2)]$,
      and
      \begin{align} 
        A_2^{(0)}(t) = \frac{1}{\lambda_+-\lambda_-}\Big[&(M_{11}+M_{12}-\lambda_-)e^{\lambda_+(t-t_w)} \nonumber \\&+(\lambda_+-M_{11}-M_{12})e^{\lambda_-(t-t_w)}\Big].
        \label{eq:exponentials-a20}
  \end{align}
As a consequence, the Kovacs hump is given by Eq.~\eqref{eq:kovacs-hump-first-order} of the main text in  this perturbative scheme.

\bibliography{Mi-biblioteca-28-oct-2021}

\begin{thebibliography}{87}%
\makeatletter
\providecommand \@ifxundefined [1]{%
 \@ifx{#1\undefined}
}%
\providecommand \@ifnum [1]{%
 \ifnum #1\expandafter \@firstoftwo
 \else \expandafter \@secondoftwo
 \fi
}%
\providecommand \@ifx [1]{%
 \ifx #1\expandafter \@firstoftwo
 \else \expandafter \@secondoftwo
 \fi
}%
\providecommand \natexlab [1]{#1}%
\providecommand \enquote  [1]{``#1''}%
\providecommand \bibnamefont  [1]{#1}%
\providecommand \bibfnamefont [1]{#1}%
\providecommand \citenamefont [1]{#1}%
\providecommand \href@noop [0]{\@secondoftwo}%
\providecommand \href [0]{\begingroup \@sanitize@url \@href}%
\providecommand \@href[1]{\@@startlink{#1}\@@href}%
\providecommand \@@href[1]{\endgroup#1\@@endlink}%
\providecommand \@sanitize@url [0]{\catcode `\\12\catcode `\$12\catcode
  `\&12\catcode `\#12\catcode `\^12\catcode `\_12\catcode `\%12\relax}%
\providecommand \@@startlink[1]{}%
\providecommand \@@endlink[0]{}%
\providecommand \url  [0]{\begingroup\@sanitize@url \@url }%
\providecommand \@url [1]{\endgroup\@href {#1}{\urlprefix }}%
\providecommand \urlprefix  [0]{URL }%
\providecommand \Eprint [0]{\href }%
\providecommand \doibase [0]{https://doi.org/}%
\providecommand \selectlanguage [0]{\@gobble}%
\providecommand \bibinfo  [0]{\@secondoftwo}%
\providecommand \bibfield  [0]{\@secondoftwo}%
\providecommand \translation [1]{[#1]}%
\providecommand \BibitemOpen [0]{}%
\providecommand \bibitemStop [0]{}%
\providecommand \bibitemNoStop [0]{.\EOS\space}%
\providecommand \EOS [0]{\spacefactor3000\relax}%
\providecommand \BibitemShut  [1]{\csname bibitem#1\endcsname}%
\let\auto@bib@innerbib\@empty
\bibitem [{\citenamefont {Stillinger}\ and\ \citenamefont
  {Debenedetti}(2013)}]{stillinger_glass_2013}%
  \BibitemOpen
  \bibfield  {author} {\bibinfo {author} {\bibfnamefont {F.~H.}\ \bibnamefont
  {Stillinger}}\ and\ \bibinfo {author} {\bibfnamefont {P.~G.}\ \bibnamefont
  {Debenedetti}},\ }\bibfield  {title} {\bibinfo {title} {Glass {Transition}
  {Thermodynamics} and {Kinetics}},\ }\href
  {https://doi.org/10.1146/annurev-conmatphys-030212-184329} {\bibfield
  {journal} {\bibinfo  {journal} {Annual Review of Condensed Matter Physics}\
  }\textbf {\bibinfo {volume} {4}},\ \bibinfo {pages} {263} (\bibinfo {year}
  {2013})}\BibitemShut {NoStop}%
\bibitem [{\citenamefont {Lubchenko}(2015)}]{lubchenko_theory_2015}%
  \BibitemOpen
  \bibfield  {author} {\bibinfo {author} {\bibfnamefont {V.}~\bibnamefont
  {Lubchenko}},\ }\bibfield  {title} {\bibinfo {title} {Theory of the
  structural glass transition: a pedagogical review},\ }\href
  {https://doi.org/10.1080/00018732.2015.1057979} {\bibfield  {journal}
  {\bibinfo  {journal} {Advances in Physics}\ }\textbf {\bibinfo {volume}
  {64}},\ \bibinfo {pages} {283} (\bibinfo {year} {2015})}\BibitemShut
  {NoStop}%
\bibitem [{\citenamefont {Nagel}(2017)}]{nagel_experimental_2017}%
  \BibitemOpen
  \bibfield  {author} {\bibinfo {author} {\bibfnamefont {S.~R.}\ \bibnamefont
  {Nagel}},\ }\bibfield  {title} {\bibinfo {title} {Experimental soft-matter
  science},\ }\href {https://doi.org/10.1103/RevModPhys.89.025002} {\bibfield
  {journal} {\bibinfo  {journal} {Reviews of Modern Physics}\ }\textbf
  {\bibinfo {volume} {89}},\ \bibinfo {pages} {025002} (\bibinfo {year}
  {2017})}\BibitemShut {NoStop}%
\bibitem [{\citenamefont {Williams}\ and\ \citenamefont
  {Watts}(1970)}]{williams_non-symmetrical_1970}%
  \BibitemOpen
  \bibfield  {author} {\bibinfo {author} {\bibfnamefont {G.}~\bibnamefont
  {Williams}}\ and\ \bibinfo {author} {\bibfnamefont {D.~C.}\ \bibnamefont
  {Watts}},\ }\bibfield  {title} {\bibinfo {title} {Non-symmetrical dielectric
  relaxation behaviour arising from a simple empirical decay function},\
  }\href@noop {} {\bibfield  {journal} {\bibinfo  {journal} {Trans. Faraday
  Soc.}\ }\textbf {\bibinfo {volume} {66}},\ \bibinfo {pages} {80} (\bibinfo
  {year} {1970})}\BibitemShut {NoStop}%
\bibitem [{\citenamefont {Palmer}\ \emph {et~al.}(1984)\citenamefont {Palmer},
  \citenamefont {Stein}, \citenamefont {Abrahams},\ and\ \citenamefont
  {Anderson}}]{palmer_models_1984}%
  \BibitemOpen
  \bibfield  {author} {\bibinfo {author} {\bibfnamefont {R.~G.}\ \bibnamefont
  {Palmer}}, \bibinfo {author} {\bibfnamefont {D.~L.}\ \bibnamefont {Stein}},
  \bibinfo {author} {\bibfnamefont {E.}~\bibnamefont {Abrahams}},\ and\
  \bibinfo {author} {\bibfnamefont {P.~W.}\ \bibnamefont {Anderson}},\
  }\bibfield  {title} {\bibinfo {title} {Models of hierarchically constrained
  dynamics for glassy relaxation},\ }\href@noop {} {\bibfield  {journal}
  {\bibinfo  {journal} {Physical Review Letters}\ }\textbf {\bibinfo {volume}
  {53}},\ \bibinfo {pages} {958} (\bibinfo {year} {1984})}\BibitemShut
  {NoStop}%
\bibitem [{\citenamefont {Kob}\ and\ \citenamefont
  {Schilling}(1990)}]{kob_dynamics_1990}%
  \BibitemOpen
  \bibfield  {author} {\bibinfo {author} {\bibfnamefont {W.}~\bibnamefont
  {Kob}}\ and\ \bibinfo {author} {\bibfnamefont {R.}~\bibnamefont
  {Schilling}},\ }\bibfield  {title} {\bibinfo {title} {Dynamics of a
  one-dimensional ‘‘glass’’model: {Ergodicity} and nonexponential
  relaxation},\ }\href@noop {} {\bibfield  {journal} {\bibinfo  {journal}
  {Physical Review A}\ }\textbf {\bibinfo {volume} {42}},\ \bibinfo {pages}
  {2191} (\bibinfo {year} {1990})}\BibitemShut {NoStop}%
\bibitem [{\citenamefont {Brey}\ and\ \citenamefont
  {Prados}(1993)}]{brey_stretched_1993}%
  \BibitemOpen
  \bibfield  {author} {\bibinfo {author} {\bibfnamefont {J.~J.}\ \bibnamefont
  {Brey}}\ and\ \bibinfo {author} {\bibfnamefont {A.}~\bibnamefont {Prados}},\
  }\bibfield  {title} {\bibinfo {title} {Stretched exponential decay at
  intermediate times in the one-dimentional {Ising} model at low
  temperatures},\ }\href@noop {} {\bibfield  {journal} {\bibinfo  {journal}
  {Physica A}\ }\textbf {\bibinfo {volume} {197}},\ \bibinfo {pages} {569}
  (\bibinfo {year} {1993})}\BibitemShut {NoStop}%
\bibitem [{\citenamefont {Brey}\ and\ \citenamefont
  {Prados}(1996)}]{brey_low-temperature_1996}%
  \BibitemOpen
  \bibfield  {author} {\bibinfo {author} {\bibfnamefont {J.~J.}\ \bibnamefont
  {Brey}}\ and\ \bibinfo {author} {\bibfnamefont {A.}~\bibnamefont {Prados}},\
  }\bibfield  {title} {\bibinfo {title} {Low-temperature relaxation in the
  one-dimensional {Ising} model},\ }\href@noop {} {\bibfield  {journal}
  {\bibinfo  {journal} {Physical Review E}\ }\textbf {\bibinfo {volume} {53}},\
  \bibinfo {pages} {458} (\bibinfo {year} {1996})}\BibitemShut {NoStop}%
\bibitem [{\citenamefont {Angell}\ \emph {et~al.}(2000)\citenamefont {Angell},
  \citenamefont {Ngai}, \citenamefont {McKenna}, \citenamefont {McMillan},\
  and\ \citenamefont {Martin}}]{angell_relaxation_2000}%
  \BibitemOpen
  \bibfield  {author} {\bibinfo {author} {\bibfnamefont {C.~A.}\ \bibnamefont
  {Angell}}, \bibinfo {author} {\bibfnamefont {K.~L.}\ \bibnamefont {Ngai}},
  \bibinfo {author} {\bibfnamefont {G.~B.}\ \bibnamefont {McKenna}}, \bibinfo
  {author} {\bibfnamefont {P.~F.}\ \bibnamefont {McMillan}},\ and\ \bibinfo
  {author} {\bibfnamefont {S.~W.}\ \bibnamefont {Martin}},\ }\bibfield  {title}
  {\bibinfo {title} {Relaxation in glassforming liquids and amorphous solids},\
  }\href@noop {} {\bibfield  {journal} {\bibinfo  {journal} {Journal of Applied
  Physics}\ }\textbf {\bibinfo {volume} {88}},\ \bibinfo {pages} {3113}
  (\bibinfo {year} {2000})}\BibitemShut {NoStop}%
\bibitem [{\citenamefont {Brey}\ and\ \citenamefont
  {Prados}(2001)}]{brey_slow_2001}%
  \BibitemOpen
  \bibfield  {author} {\bibinfo {author} {\bibfnamefont {J.~J.}\ \bibnamefont
  {Brey}}\ and\ \bibinfo {author} {\bibfnamefont {A.}~\bibnamefont {Prados}},\
  }\bibfield  {title} {\bibinfo {title} {Slow logarithmic relaxation in models
  with hierarchically constrained dynamics},\ }\href
  {https://doi.org/10.1103/PhysRevE.63.021108} {\bibfield  {journal} {\bibinfo
  {journal} {Physical Review E}\ }\textbf {\bibinfo {volume} {63}},\ \bibinfo
  {pages} {021108} (\bibinfo {year} {2001})}\BibitemShut {NoStop}%
\bibitem [{\citenamefont {Richert}(2010)}]{richert_physical_2010}%
  \BibitemOpen
  \bibfield  {author} {\bibinfo {author} {\bibfnamefont {R.}~\bibnamefont
  {Richert}},\ }\bibfield  {title} {\bibinfo {title} {Physical {Aging} and
  {Heterogeneous} {Dynamics}},\ }\href
  {https://doi.org/10.1103/PhysRevLett.104.085702} {\bibfield  {journal}
  {\bibinfo  {journal} {Physical Review Letters}\ }\textbf {\bibinfo {volume}
  {104}},\ \bibinfo {pages} {085702} (\bibinfo {year} {2010})}\BibitemShut
  {NoStop}%
\bibitem [{\citenamefont {Paeng}\ \emph {et~al.}(2015)\citenamefont {Paeng},
  \citenamefont {Park}, \citenamefont {Hoang},\ and\ \citenamefont
  {Kaufman}}]{paeng_ideal_2015}%
  \BibitemOpen
  \bibfield  {author} {\bibinfo {author} {\bibfnamefont {K.}~\bibnamefont
  {Paeng}}, \bibinfo {author} {\bibfnamefont {H.}~\bibnamefont {Park}},
  \bibinfo {author} {\bibfnamefont {D.~T.}\ \bibnamefont {Hoang}},\ and\
  \bibinfo {author} {\bibfnamefont {L.~J.}\ \bibnamefont {Kaufman}},\
  }\bibfield  {title} {\bibinfo {title} {Ideal probe single-molecule
  experiments reveal the intrinsic dynamic heterogeneity of a supercooled
  liquid},\ }\href {https://doi.org/10.1073/pnas.1424636112} {\bibfield
  {journal} {\bibinfo  {journal} {Proceedings of the National Academy of
  Sciences}\ }\textbf {\bibinfo {volume} {112}},\ \bibinfo {pages} {4952}
  (\bibinfo {year} {2015})}\BibitemShut {NoStop}%
\bibitem [{\citenamefont {Lahini}\ \emph {et~al.}(2017)\citenamefont {Lahini},
  \citenamefont {Gottesman}, \citenamefont {Amir},\ and\ \citenamefont
  {Rubinstein}}]{lahini_nonmonotonic_2017}%
  \BibitemOpen
  \bibfield  {author} {\bibinfo {author} {\bibfnamefont {Y.}~\bibnamefont
  {Lahini}}, \bibinfo {author} {\bibfnamefont {O.}~\bibnamefont {Gottesman}},
  \bibinfo {author} {\bibfnamefont {A.}~\bibnamefont {Amir}},\ and\ \bibinfo
  {author} {\bibfnamefont {S.~M.}\ \bibnamefont {Rubinstein}},\ }\bibfield
  {title} {\bibinfo {title} {Nonmonotonic {Aging} and {Memory} {Retention} in
  {Disordered} {Mechanical} {Systems}},\ }\href
  {https://doi.org/10.1103/PhysRevLett.118.085501} {\bibfield  {journal}
  {\bibinfo  {journal} {Physical Review Letters}\ }\textbf {\bibinfo {volume}
  {118}},\ \bibinfo {pages} {085501} (\bibinfo {year} {2017})}\BibitemShut
  {NoStop}%
\bibitem [{\citenamefont {Kringle}\ \emph {et~al.}(2021)\citenamefont
  {Kringle}, \citenamefont {Thornley}, \citenamefont {Kay},\ and\ \citenamefont
  {Kimmel}}]{kringle_structural_2021}%
  \BibitemOpen
  \bibfield  {author} {\bibinfo {author} {\bibfnamefont {L.}~\bibnamefont
  {Kringle}}, \bibinfo {author} {\bibfnamefont {W.~A.}\ \bibnamefont
  {Thornley}}, \bibinfo {author} {\bibfnamefont {B.~D.}\ \bibnamefont {Kay}},\
  and\ \bibinfo {author} {\bibfnamefont {G.~A.}\ \bibnamefont {Kimmel}},\
  }\bibfield  {title} {\bibinfo {title} {Structural relaxation and
  crystallization in supercooled water from 170 to 260 {K}},\ }\href
  {https://doi.org/10.1073/pnas.2022884118} {\bibfield  {journal} {\bibinfo
  {journal} {Proceedings of the National Academy of Sciences}\ }\textbf
  {\bibinfo {volume} {118}},\ \bibinfo {pages} {e2022884118} (\bibinfo {year}
  {2021})}\BibitemShut {NoStop}%
\bibitem [{\citenamefont {Nishikawa}\ \emph {et~al.}()\citenamefont
  {Nishikawa}, \citenamefont {Ozawa}, \citenamefont {Ikeda}, \citenamefont
  {Chaudhuri},\ and\ \citenamefont {Berthier}}]{nishikawa_relaxation_2021}%
  \BibitemOpen
  \bibfield  {author} {\bibinfo {author} {\bibfnamefont {Y.}~\bibnamefont
  {Nishikawa}}, \bibinfo {author} {\bibfnamefont {M.}~\bibnamefont {Ozawa}},
  \bibinfo {author} {\bibfnamefont {A.}~\bibnamefont {Ikeda}}, \bibinfo
  {author} {\bibfnamefont {P.}~\bibnamefont {Chaudhuri}},\ and\ \bibinfo
  {author} {\bibfnamefont {L.}~\bibnamefont {Berthier}},\ }\bibfield  {title}
  {\bibinfo {title} {Relaxation dynamics in the energy landscape of
  glass-forming liquids},\ }\href {http://arxiv.org/abs/2106.01755} {\bibinfo
  {journal} {arXiv:2106.01755 [cond-mat]}\ }\BibitemShut {NoStop}%
\bibitem [{\citenamefont {Kovacs}(1963)}]{kovacs_transition_1963}%
  \BibitemOpen
\bibfield  {journal} {  }\bibfield  {author} {\bibinfo {author} {\bibfnamefont
  {A.~J.}\ \bibnamefont {Kovacs}},\ }\bibfield  {title} {\bibinfo {title}
  {Transition vitreuse dans les polymères amorphes. {Etude}
  phénoménologique},\ }\href
  {https://doi.org/https://doi.org/10.1007/BFb0050366} {\bibfield  {journal}
  {\bibinfo  {journal} {Fortschritte Der Hochpolymeren-Forschung}\ }\textbf
  {\bibinfo {volume} {3}},\ \bibinfo {pages} {394} (\bibinfo {year}
  {1963})}\BibitemShut {NoStop}%
\bibitem [{\citenamefont {Kovacs}\ \emph {et~al.}(1979)\citenamefont {Kovacs},
  \citenamefont {Aklonis}, \citenamefont {Hutchinson},\ and\ \citenamefont
  {Ramos}}]{kovacs_isobaric_1979}%
  \BibitemOpen
  \bibfield  {author} {\bibinfo {author} {\bibfnamefont {A.~J.}\ \bibnamefont
  {Kovacs}}, \bibinfo {author} {\bibfnamefont {J.~J.}\ \bibnamefont {Aklonis}},
  \bibinfo {author} {\bibfnamefont {J.~M.}\ \bibnamefont {Hutchinson}},\ and\
  \bibinfo {author} {\bibfnamefont {A.~R.}\ \bibnamefont {Ramos}},\ }\bibfield
  {title} {\bibinfo {title} {Isobaric volume and enthalpy recovery of glasses.
  {II}. {A} transparent multiparameter theory},\ }\href@noop {} {\bibfield
  {journal} {\bibinfo  {journal} {Journal of Polymer Science: Polymer Physics
  Edition}\ }\textbf {\bibinfo {volume} {17}},\ \bibinfo {pages} {1097}
  (\bibinfo {year} {1979})}\BibitemShut {NoStop}%
\bibitem [{\citenamefont {Buhot}(2003)}]{buhot_kovacs_2003}%
  \BibitemOpen
  \bibfield  {author} {\bibinfo {author} {\bibfnamefont {A.}~\bibnamefont
  {Buhot}},\ }\bibfield  {title} {\bibinfo {title} {Kovacs effect and
  fluctuation–dissipation relations in 1d kinetically constrained models},\
  }\href@noop {} {\bibfield  {journal} {\bibinfo  {journal} {Journal of Physics
  A: Mathematical and General}\ }\textbf {\bibinfo {volume} {36}},\ \bibinfo
  {pages} {12367} (\bibinfo {year} {2003})}\BibitemShut {NoStop}%
\bibitem [{\citenamefont {Bertin}\ \emph {et~al.}(2003)\citenamefont {Bertin},
  \citenamefont {Bouchaud}, \citenamefont {Drouffe},\ and\ \citenamefont
  {Godrèche}}]{bertin_kovacs_2003}%
  \BibitemOpen
  \bibfield  {author} {\bibinfo {author} {\bibfnamefont {E.~M.}\ \bibnamefont
  {Bertin}}, \bibinfo {author} {\bibfnamefont {J.~P.}\ \bibnamefont
  {Bouchaud}}, \bibinfo {author} {\bibfnamefont {J.~M.}\ \bibnamefont
  {Drouffe}},\ and\ \bibinfo {author} {\bibfnamefont {C.}~\bibnamefont
  {Godrèche}},\ }\bibfield  {title} {\bibinfo {title} {The {Kovacs} effect in
  model glasses},\ }\href@noop {} {\bibfield  {journal} {\bibinfo  {journal}
  {Journal of Physics A: Mathematical and General}\ }\textbf {\bibinfo {volume}
  {36}},\ \bibinfo {pages} {10701} (\bibinfo {year} {2003})}\BibitemShut
  {NoStop}%
\bibitem [{\citenamefont {Arenzon}\ and\ \citenamefont
  {Sellitto}(2004)}]{arenzon_kovacs_2004}%
  \BibitemOpen
  \bibfield  {author} {\bibinfo {author} {\bibfnamefont {J.~J.}\ \bibnamefont
  {Arenzon}}\ and\ \bibinfo {author} {\bibfnamefont {M.}~\bibnamefont
  {Sellitto}},\ }\bibfield  {title} {\bibinfo {title} {Kovacs effect in
  facilitated spin models of strong and fragile glasses},\ }\href@noop {}
  {\bibfield  {journal} {\bibinfo  {journal} {The European Physical Journal
  B-Condensed Matter and Complex Systems}\ }\textbf {\bibinfo {volume} {42}},\
  \bibinfo {pages} {543} (\bibinfo {year} {2004})}\BibitemShut {NoStop}%
\bibitem [{\citenamefont {Mossa}\ and\ \citenamefont
  {Sciortino}(2004)}]{mossa_crossover_2004}%
  \BibitemOpen
  \bibfield  {author} {\bibinfo {author} {\bibfnamefont {S.}~\bibnamefont
  {Mossa}}\ and\ \bibinfo {author} {\bibfnamefont {F.}~\bibnamefont
  {Sciortino}},\ }\bibfield  {title} {\bibinfo {title} {Crossover (or {Kovacs})
  effect in an aging molecular liquid},\ }\href@noop {} {\bibfield  {journal}
  {\bibinfo  {journal} {Physical Review Letters}\ }\textbf {\bibinfo {volume}
  {92}},\ \bibinfo {pages} {045504} (\bibinfo {year} {2004})}\BibitemShut
  {NoStop}%
\bibitem [{\citenamefont {Aquino}\ \emph {et~al.}(2006)\citenamefont {Aquino},
  \citenamefont {Leuzzi},\ and\ \citenamefont
  {Nieuwenhuizen}}]{aquino_kovacs_2006}%
  \BibitemOpen
  \bibfield  {author} {\bibinfo {author} {\bibfnamefont {G.}~\bibnamefont
  {Aquino}}, \bibinfo {author} {\bibfnamefont {L.}~\bibnamefont {Leuzzi}},\
  and\ \bibinfo {author} {\bibfnamefont {T.~M.}\ \bibnamefont
  {Nieuwenhuizen}},\ }\bibfield  {title} {\bibinfo {title} {Kovacs effect in a
  model for a fragile glass},\ }\href@noop {} {\bibfield  {journal} {\bibinfo
  {journal} {Physical Review B}\ }\textbf {\bibinfo {volume} {73}},\ \bibinfo
  {pages} {094205} (\bibinfo {year} {2006})}\BibitemShut {NoStop}%
\bibitem [{\citenamefont {Bouchbinder}\ and\ \citenamefont
  {Langer}(2010)}]{bouchbinder_nonequilibrium_2010}%
  \BibitemOpen
  \bibfield  {author} {\bibinfo {author} {\bibfnamefont {E.}~\bibnamefont
  {Bouchbinder}}\ and\ \bibinfo {author} {\bibfnamefont {J.~S.}\ \bibnamefont
  {Langer}},\ }\bibfield  {title} {\bibinfo {title} {Nonequilibrium
  thermodynamics of the {Kovacs} effect},\ }\href
  {https://doi.org/10.1039/c001388a} {\bibfield  {journal} {\bibinfo  {journal}
  {Soft Matter}\ }\textbf {\bibinfo {volume} {6}},\ \bibinfo {pages} {3065}
  (\bibinfo {year} {2010})}\BibitemShut {NoStop}%
\bibitem [{\citenamefont {Prados}\ and\ \citenamefont
  {Brey}(2010)}]{prados_kovacs_2010}%
  \BibitemOpen
  \bibfield  {author} {\bibinfo {author} {\bibfnamefont {A.}~\bibnamefont
  {Prados}}\ and\ \bibinfo {author} {\bibfnamefont {J.~J.}\ \bibnamefont
  {Brey}},\ }\bibfield  {title} {\bibinfo {title} {The {Kovacs} effect: a
  master equation analysis},\ }\href@noop {} {\bibfield  {journal} {\bibinfo
  {journal} {Journal of Statistical Mechanics: Theory and Experiment}\ ,\
  \bibinfo {pages} {P02009}} (\bibinfo {year} {2010})}\BibitemShut {NoStop}%
\bibitem [{\citenamefont {Diezemann}\ and\ \citenamefont
  {Heuer}(2011)}]{diezemann_memory_2011}%
  \BibitemOpen
  \bibfield  {author} {\bibinfo {author} {\bibfnamefont {G.}~\bibnamefont
  {Diezemann}}\ and\ \bibinfo {author} {\bibfnamefont {A.}~\bibnamefont
  {Heuer}},\ }\bibfield  {title} {\bibinfo {title} {Memory effects in the
  relaxation of the {Gaussian} trap model},\ }\href
  {https://doi.org/10.1103/PhysRevE.83.031505} {\bibfield  {journal} {\bibinfo
  {journal} {Physical Review E}\ }\textbf {\bibinfo {volume} {83}},\ \bibinfo
  {pages} {031505} (\bibinfo {year} {2011})}\BibitemShut {NoStop}%
\bibitem [{\citenamefont {Ruiz-García}\ and\ \citenamefont
  {Prados}(2014)}]{ruiz-garcia_kovacs_2014}%
  \BibitemOpen
  \bibfield  {author} {\bibinfo {author} {\bibfnamefont {M.}~\bibnamefont
  {Ruiz-García}}\ and\ \bibinfo {author} {\bibfnamefont {A.}~\bibnamefont
  {Prados}},\ }\bibfield  {title} {\bibinfo {title} {Kovacs effect in the
  one-dimensional {Ising} model: {A} linear response analysis},\ }\href
  {https://doi.org/10.1103/PhysRevE.89.012140} {\bibfield  {journal} {\bibinfo
  {journal} {Physical Review E}\ }\textbf {\bibinfo {volume} {89}},\ \bibinfo
  {pages} {012140} (\bibinfo {year} {2014})}\BibitemShut {NoStop}%
\bibitem [{\citenamefont {Lulli}\ \emph {et~al.}(2020)\citenamefont {Lulli},
  \citenamefont {Lee}, \citenamefont {Deng}, \citenamefont {Yip},\ and\
  \citenamefont {Lam}}]{lulli_spatial_2020}%
  \BibitemOpen
  \bibfield  {author} {\bibinfo {author} {\bibfnamefont {M.}~\bibnamefont
  {Lulli}}, \bibinfo {author} {\bibfnamefont {C.-S.}\ \bibnamefont {Lee}},
  \bibinfo {author} {\bibfnamefont {H.-Y.}\ \bibnamefont {Deng}}, \bibinfo
  {author} {\bibfnamefont {C.-T.}\ \bibnamefont {Yip}},\ and\ \bibinfo {author}
  {\bibfnamefont {C.-H.}\ \bibnamefont {Lam}},\ }\bibfield  {title} {\bibinfo
  {title} {Spatial {Heterogeneities} in {Structural} {Temperature} {Cause}
  {Kovacs}’ {Expansion} {Gap} {Paradox} in {Aging} of {Glasses}},\ }\href
  {https://doi.org/10.1103/PhysRevLett.124.095501} {\bibfield  {journal}
  {\bibinfo  {journal} {Physical Review Letters}\ }\textbf {\bibinfo {volume}
  {124}},\ \bibinfo {pages} {095501} (\bibinfo {year} {2020})}\BibitemShut
  {NoStop}%
\bibitem [{\citenamefont {Morgan}\ \emph {et~al.}(2020)\citenamefont {Morgan},
  \citenamefont {Avinery}, \citenamefont {Rahamim}, \citenamefont {Beck},\ and\
  \citenamefont {Saleh}}]{morgan_glassy_2020}%
  \BibitemOpen
  \bibfield  {author} {\bibinfo {author} {\bibfnamefont {I.~L.}\ \bibnamefont
  {Morgan}}, \bibinfo {author} {\bibfnamefont {R.}~\bibnamefont {Avinery}},
  \bibinfo {author} {\bibfnamefont {G.}~\bibnamefont {Rahamim}}, \bibinfo
  {author} {\bibfnamefont {R.}~\bibnamefont {Beck}},\ and\ \bibinfo {author}
  {\bibfnamefont {O.~A.}\ \bibnamefont {Saleh}},\ }\bibfield  {title} {\bibinfo
  {title} {Glassy {Dynamics} and {Memory} {Effects} in an {Intrinsically}
  {Disordered} {Protein} {Construct}},\ }\href
  {https://doi.org/10.1103/PhysRevLett.125.058001} {\bibfield  {journal}
  {\bibinfo  {journal} {Physical Review Letters}\ }\textbf {\bibinfo {volume}
  {125}},\ \bibinfo {pages} {058001} (\bibinfo {year} {2020})}\BibitemShut
  {NoStop}%
\bibitem [{\citenamefont {Song}\ \emph {et~al.}(2020)\citenamefont {Song},
  \citenamefont {Xu}, \citenamefont {Huo}, \citenamefont {Li}, \citenamefont
  {Wang}, \citenamefont {Ediger},\ and\ \citenamefont
  {Wang}}]{song_activation_2020}%
  \BibitemOpen
  \bibfield  {author} {\bibinfo {author} {\bibfnamefont {L.}~\bibnamefont
  {Song}}, \bibinfo {author} {\bibfnamefont {W.}~\bibnamefont {Xu}}, \bibinfo
  {author} {\bibfnamefont {J.}~\bibnamefont {Huo}}, \bibinfo {author}
  {\bibfnamefont {F.}~\bibnamefont {Li}}, \bibinfo {author} {\bibfnamefont
  {L.-M.}\ \bibnamefont {Wang}}, \bibinfo {author} {\bibfnamefont
  {M.}~\bibnamefont {Ediger}},\ and\ \bibinfo {author} {\bibfnamefont {J.-Q.}\
  \bibnamefont {Wang}},\ }\bibfield  {title} {\bibinfo {title} {Activation
  {Entropy} as a {Key} {Factor} {Controlling} the {Memory} {Effect} in
  {Glasses}},\ }\href {https://doi.org/10.1103/PhysRevLett.125.135501}
  {\bibfield  {journal} {\bibinfo  {journal} {Physical Review Letters}\
  }\textbf {\bibinfo {volume} {125}},\ \bibinfo {pages} {135501} (\bibinfo
  {year} {2020})}\BibitemShut {NoStop}%
\bibitem [{\citenamefont {Peyrard}\ and\ \citenamefont
  {Garden}(2020)}]{peyrard_memory_2020}%
  \BibitemOpen
  \bibfield  {author} {\bibinfo {author} {\bibfnamefont {M.}~\bibnamefont
  {Peyrard}}\ and\ \bibinfo {author} {\bibfnamefont {J.-L.}\ \bibnamefont
  {Garden}},\ }\bibfield  {title} {\bibinfo {title} {Memory effects in glasses:
  {Insights} into the thermodynamics of out-of-equilibrium systems revealed by
  a simple model of the {Kovacs} effect},\ }\href
  {https://doi.org/10.1103/PhysRevE.102.052122} {\bibfield  {journal} {\bibinfo
   {journal} {Physical Review E}\ }\textbf {\bibinfo {volume} {102}},\ \bibinfo
  {pages} {052122} (\bibinfo {year} {2020})}\BibitemShut {NoStop}%
\bibitem [{\citenamefont {Mandal}\ \emph {et~al.}()\citenamefont {Mandal},
  \citenamefont {Tapias},\ and\ \citenamefont {Sollich}}]{mandal_memory_2021}%
  \BibitemOpen
  \bibfield  {author} {\bibinfo {author} {\bibfnamefont {R.}~\bibnamefont
  {Mandal}}, \bibinfo {author} {\bibfnamefont {D.}~\bibnamefont {Tapias}},\
  and\ \bibinfo {author} {\bibfnamefont {P.}~\bibnamefont {Sollich}},\
  }\bibfield  {title} {\bibinfo {title} {Memory in {Non}-{Monotonic} {Stress}
  {Response} of an {Athermal} {Disordered} {Solid}},\ }\href
  {http://arxiv.org/abs/2103.14766} {\bibinfo  {journal} {arXiv:2103.14766
  [cond-mat]}\ }\BibitemShut {NoStop}%
\bibitem [{Note1()}]{Note1}%
  \BibitemOpen
\bibfield  {journal} {  }\bibinfo {note} {Figure 1 of Ref.~\cite
  {prados_kovacs-like_2014} gives a qualitative picture of the Kovacs
  protocol.}\BibitemShut {Stop}%
\bibitem [{\citenamefont {Bouchaud}(1992)}]{bouchaud_weak_1992}%
  \BibitemOpen
  \bibfield  {author} {\bibinfo {author} {\bibfnamefont {J.-P.}\ \bibnamefont
  {Bouchaud}},\ }\bibfield  {title} {\bibinfo {title} {Weak ergodicity breaking
  and aging in disordered systems},\ }\href@noop {} {\bibfield  {journal}
  {\bibinfo  {journal} {Journal de Physique I}\ }\textbf {\bibinfo {volume}
  {2}},\ \bibinfo {pages} {1705} (\bibinfo {year} {1992})}\BibitemShut
  {NoStop}%
\bibitem [{\citenamefont {Cugliandolo}\ and\ \citenamefont
  {Kurchan}(1993)}]{cugliandolo_analytical_1993}%
  \BibitemOpen
  \bibfield  {author} {\bibinfo {author} {\bibfnamefont {L.~F.}\ \bibnamefont
  {Cugliandolo}}\ and\ \bibinfo {author} {\bibfnamefont {J.}~\bibnamefont
  {Kurchan}},\ }\bibfield  {title} {\bibinfo {title} {Analytical solution of
  the off-equilibrium dynamics of a long-range spin-glass model},\ }\href
  {https://doi.org/10.1103/PhysRevLett.71.173} {\bibfield  {journal} {\bibinfo
  {journal} {Physical Review Letters}\ }\textbf {\bibinfo {volume} {71}},\
  \bibinfo {pages} {173} (\bibinfo {year} {1993})}\BibitemShut {NoStop}%
\bibitem [{\citenamefont {Prados}\ \emph {et~al.}(1997)\citenamefont {Prados},
  \citenamefont {Brey},\ and\ \citenamefont
  {Sánchez-Rey}}]{prados_aging_1997}%
  \BibitemOpen
  \bibfield  {author} {\bibinfo {author} {\bibfnamefont {A.}~\bibnamefont
  {Prados}}, \bibinfo {author} {\bibfnamefont {J.~J.}\ \bibnamefont {Brey}},\
  and\ \bibinfo {author} {\bibfnamefont {B.}~\bibnamefont {Sánchez-Rey}},\
  }\bibfield  {title} {\bibinfo {title} {Aging in the one-dimensional {Ising}
  model with {Glauber} dynamics},\ }\href@noop {} {\bibfield  {journal}
  {\bibinfo  {journal} {EPL}\ }\textbf {\bibinfo {volume} {40}},\ \bibinfo
  {pages} {13} (\bibinfo {year} {1997})}\BibitemShut {NoStop}%
\bibitem [{\citenamefont {Nicodemi}\ and\ \citenamefont
  {Coniglio}(1999)}]{nicodemi_aging_1999}%
  \BibitemOpen
  \bibfield  {author} {\bibinfo {author} {\bibfnamefont {M.}~\bibnamefont
  {Nicodemi}}\ and\ \bibinfo {author} {\bibfnamefont {A.}~\bibnamefont
  {Coniglio}},\ }\bibfield  {title} {\bibinfo {title} {Aging in
  out-of-equilibrium dynamics of models for granular media},\ }\href@noop {}
  {\bibfield  {journal} {\bibinfo  {journal} {Physical Review Letters}\
  }\textbf {\bibinfo {volume} {82}},\ \bibinfo {pages} {916} (\bibinfo {year}
  {1999})}\BibitemShut {NoStop}%
\bibitem [{\citenamefont {Ahmad}\ and\ \citenamefont
  {Puri}(2007)}]{ahmad_velocity_2007}%
  \BibitemOpen
  \bibfield  {author} {\bibinfo {author} {\bibfnamefont {S.~R.}\ \bibnamefont
  {Ahmad}}\ and\ \bibinfo {author} {\bibfnamefont {S.}~\bibnamefont {Puri}},\
  }\bibfield  {title} {\bibinfo {title} {Velocity distributions and aging in a
  cooling granular gas},\ }\href@noop {} {\bibfield  {journal} {\bibinfo
  {journal} {Physical Review E}\ }\textbf {\bibinfo {volume} {75}},\ \bibinfo
  {pages} {031302} (\bibinfo {year} {2007})}\BibitemShut {NoStop}%
\bibitem [{\citenamefont {Brey}\ \emph {et~al.}(2007)\citenamefont {Brey},
  \citenamefont {Prados}, \citenamefont {García~de Soria},\ and\ \citenamefont
  {Maynar}}]{brey_scaling_2007}%
  \BibitemOpen
  \bibfield  {author} {\bibinfo {author} {\bibfnamefont {J.~J.}\ \bibnamefont
  {Brey}}, \bibinfo {author} {\bibfnamefont {A.}~\bibnamefont {Prados}},
  \bibinfo {author} {\bibfnamefont {M.~I.}\ \bibnamefont {García~de Soria}},\
  and\ \bibinfo {author} {\bibfnamefont {P.}~\bibnamefont {Maynar}},\
  }\bibfield  {title} {\bibinfo {title} {Scaling and aging in the homogeneous
  cooling state of a granular fluid of hard particles},\ }\href@noop {}
  {\bibfield  {journal} {\bibinfo  {journal} {Journal of Physics A:
  Mathematical and Theoretical}\ }\textbf {\bibinfo {volume} {40}},\ \bibinfo
  {pages} {14331} (\bibinfo {year} {2007})}\BibitemShut {NoStop}%
\bibitem [{\citenamefont {Parravicini}\ \emph {et~al.}(2012)\citenamefont
  {Parravicini}, \citenamefont {Conti}, \citenamefont {Agranat},\ and\
  \citenamefont {DelRe}}]{parravicini_rejuvenation_2012}%
  \BibitemOpen
  \bibfield  {author} {\bibinfo {author} {\bibfnamefont {J.}~\bibnamefont
  {Parravicini}}, \bibinfo {author} {\bibfnamefont {C.}~\bibnamefont {Conti}},
  \bibinfo {author} {\bibfnamefont {A.~J.}\ \bibnamefont {Agranat}},\ and\
  \bibinfo {author} {\bibfnamefont {E.}~\bibnamefont {DelRe}},\ }\bibfield
  {title} {\bibinfo {title} {Rejuvenation in scale-free optics and enhanced
  diffraction cancellation life-time},\ }\href
  {https://doi.org/10.1364/OE.20.027382} {\bibfield  {journal} {\bibinfo
  {journal} {Optics Express}\ }\textbf {\bibinfo {volume} {20}},\ \bibinfo
  {pages} {27382} (\bibinfo {year} {2012})}\BibitemShut {NoStop}%
\bibitem [{\citenamefont {Dillavou}\ and\ \citenamefont
  {Rubinstein}(2018)}]{dillavou_nonmonotonic_2018}%
  \BibitemOpen
  \bibfield  {author} {\bibinfo {author} {\bibfnamefont {S.}~\bibnamefont
  {Dillavou}}\ and\ \bibinfo {author} {\bibfnamefont {S.~M.}\ \bibnamefont
  {Rubinstein}},\ }\bibfield  {title} {\bibinfo {title} {Nonmonotonic {Aging}
  and {Memory} in a {Frictional} {Interface}},\ }\href
  {https://doi.org/10.1103/PhysRevLett.120.224101} {\bibfield  {journal}
  {\bibinfo  {journal} {Physical Review Letters}\ }\textbf {\bibinfo {volume}
  {120}},\ \bibinfo {pages} {224101} (\bibinfo {year} {2018})}\BibitemShut
  {NoStop}%
\bibitem [{\citenamefont {Mpemba}\ and\ \citenamefont
  {Osborne}(1969)}]{mpemba_cool_1969}%
  \BibitemOpen
  \bibfield  {author} {\bibinfo {author} {\bibfnamefont {E.~B.}\ \bibnamefont
  {Mpemba}}\ and\ \bibinfo {author} {\bibfnamefont {D.~G.}\ \bibnamefont
  {Osborne}},\ }\bibfield  {title} {\bibinfo {title} {Cool?},\ }\href@noop {}
  {\bibfield  {journal} {\bibinfo  {journal} {Physics Education}\ }\textbf
  {\bibinfo {volume} {4}},\ \bibinfo {pages} {172} (\bibinfo {year}
  {1969})}\BibitemShut {NoStop}%
\bibitem [{\citenamefont {Baity-Jesi}\ \emph {et~al.}(2019)\citenamefont
  {Baity-Jesi}, \citenamefont {Calore}, \citenamefont {Cruz}, \citenamefont
  {Fernandez}, \citenamefont {Gil-Narvión}, \citenamefont {Gordillo-Guerrero},
  \citenamefont {Iñiguez}, \citenamefont {Lasanta}, \citenamefont {Maiorano},
  \citenamefont {Marinari}, \citenamefont {Martin-Mayor}, \citenamefont
  {Moreno-Gordo}, \citenamefont {Muñoz~Sudupe}, \citenamefont {Navarro},
  \citenamefont {Parisi}, \citenamefont {Perez-Gaviro}, \citenamefont
  {Ricci-Tersenghi}, \citenamefont {Ruiz-Lorenzo}, \citenamefont {Schifano},
  \citenamefont {Seoane}, \citenamefont {Tarancón}, \citenamefont
  {Tripiccione},\ and\ \citenamefont {Yllanes}}]{baity-jesi_mpemba_2019}%
  \BibitemOpen
  \bibfield  {author} {\bibinfo {author} {\bibfnamefont {M.}~\bibnamefont
  {Baity-Jesi}}, \bibinfo {author} {\bibfnamefont {E.}~\bibnamefont {Calore}},
  \bibinfo {author} {\bibfnamefont {A.}~\bibnamefont {Cruz}}, \bibinfo {author}
  {\bibfnamefont {L.~A.}\ \bibnamefont {Fernandez}}, \bibinfo {author}
  {\bibfnamefont {J.~M.}\ \bibnamefont {Gil-Narvión}}, \bibinfo {author}
  {\bibfnamefont {A.}~\bibnamefont {Gordillo-Guerrero}}, \bibinfo {author}
  {\bibfnamefont {D.}~\bibnamefont {Iñiguez}}, \bibinfo {author}
  {\bibfnamefont {A.}~\bibnamefont {Lasanta}}, \bibinfo {author} {\bibfnamefont
  {A.}~\bibnamefont {Maiorano}}, \bibinfo {author} {\bibfnamefont
  {E.}~\bibnamefont {Marinari}}, \bibinfo {author} {\bibfnamefont
  {V.}~\bibnamefont {Martin-Mayor}}, \bibinfo {author} {\bibfnamefont
  {J.}~\bibnamefont {Moreno-Gordo}}, \bibinfo {author} {\bibfnamefont
  {A.}~\bibnamefont {Muñoz~Sudupe}}, \bibinfo {author} {\bibfnamefont
  {D.}~\bibnamefont {Navarro}}, \bibinfo {author} {\bibfnamefont
  {G.}~\bibnamefont {Parisi}}, \bibinfo {author} {\bibfnamefont
  {S.}~\bibnamefont {Perez-Gaviro}}, \bibinfo {author} {\bibfnamefont
  {F.}~\bibnamefont {Ricci-Tersenghi}}, \bibinfo {author} {\bibfnamefont
  {J.~J.}\ \bibnamefont {Ruiz-Lorenzo}}, \bibinfo {author} {\bibfnamefont
  {S.~F.}\ \bibnamefont {Schifano}}, \bibinfo {author} {\bibfnamefont
  {B.}~\bibnamefont {Seoane}}, \bibinfo {author} {\bibfnamefont
  {A.}~\bibnamefont {Tarancón}}, \bibinfo {author} {\bibfnamefont
  {R.}~\bibnamefont {Tripiccione}},\ and\ \bibinfo {author} {\bibfnamefont
  {D.}~\bibnamefont {Yllanes}},\ }\bibfield  {title} {\bibinfo {title} {The
  {Mpemba} effect in spin glasses is a persistent memory effect},\ }\href
  {https://doi.org/10.1073/pnas.1819803116} {\bibfield  {journal} {\bibinfo
  {journal} {Proceedings of the National Academy of Sciences}\ }\textbf
  {\bibinfo {volume} {116}},\ \bibinfo {pages} {15350} (\bibinfo {year}
  {2019})}\BibitemShut {NoStop}%
\bibitem [{\citenamefont {Lu}\ and\ \citenamefont
  {Raz}(2017)}]{lu_nonequilibrium_2017}%
  \BibitemOpen
  \bibfield  {author} {\bibinfo {author} {\bibfnamefont {Z.}~\bibnamefont
  {Lu}}\ and\ \bibinfo {author} {\bibfnamefont {O.}~\bibnamefont {Raz}},\
  }\bibfield  {title} {\bibinfo {title} {Nonequilibrium thermodynamics of the
  {Markovian} {Mpemba} effect and its inverse},\ }\href
  {https://doi.org/10.1073/pnas.1701264114} {\bibfield  {journal} {\bibinfo
  {journal} {Proceedings of the National Academy of Sciences}\ }\textbf
  {\bibinfo {volume} {114}},\ \bibinfo {pages} {5083} (\bibinfo {year}
  {2017})}\BibitemShut {NoStop}%
\bibitem [{\citenamefont {Klich}\ \emph {et~al.}(2019)\citenamefont {Klich},
  \citenamefont {Raz}, \citenamefont {Hirschberg},\ and\ \citenamefont
  {Vucelja}}]{klich_mpemba_2019}%
  \BibitemOpen
  \bibfield  {author} {\bibinfo {author} {\bibfnamefont {I.}~\bibnamefont
  {Klich}}, \bibinfo {author} {\bibfnamefont {O.}~\bibnamefont {Raz}}, \bibinfo
  {author} {\bibfnamefont {O.}~\bibnamefont {Hirschberg}},\ and\ \bibinfo
  {author} {\bibfnamefont {M.}~\bibnamefont {Vucelja}},\ }\bibfield  {title}
  {\bibinfo {title} {Mpemba {Index} and {Anomalous} {Relaxation}},\ }\href
  {https://doi.org/10.1103/PhysRevX.9.021060} {\bibfield  {journal} {\bibinfo
  {journal} {Physical Review X}\ }\textbf {\bibinfo {volume} {9}},\ \bibinfo
  {pages} {021060} (\bibinfo {year} {2019})}\BibitemShut {NoStop}%
\bibitem [{\citenamefont {Gal}\ and\ \citenamefont
  {Raz}(2020)}]{gal_precooling_2020}%
  \BibitemOpen
  \bibfield  {author} {\bibinfo {author} {\bibfnamefont {A.}~\bibnamefont
  {Gal}}\ and\ \bibinfo {author} {\bibfnamefont {O.}~\bibnamefont {Raz}},\
  }\bibfield  {title} {\bibinfo {title} {Precooling {Strategy} {Allows}
  {Exponentially} {Faster} {Heating}},\ }\href
  {https://doi.org/10.1103/PhysRevLett.124.060602} {\bibfield  {journal}
  {\bibinfo  {journal} {Physical Review Letters}\ }\textbf {\bibinfo {volume}
  {124}},\ \bibinfo {pages} {060602} (\bibinfo {year} {2020})}\BibitemShut
  {NoStop}%
\bibitem [{\citenamefont {Kumar}\ and\ \citenamefont
  {Bechhoefer}(2020)}]{kumar_exponentially_2020}%
  \BibitemOpen
  \bibfield  {author} {\bibinfo {author} {\bibfnamefont {A.}~\bibnamefont
  {Kumar}}\ and\ \bibinfo {author} {\bibfnamefont {J.}~\bibnamefont
  {Bechhoefer}},\ }\bibfield  {title} {\bibinfo {title} {Exponentially faster
  cooling in a colloidal system},\ }\href
  {http://www.nature.com/articles/s41586-020-2560-x} {\bibfield  {journal}
  {\bibinfo  {journal} {Nature}\ }\textbf {\bibinfo {volume} {584}},\ \bibinfo
  {pages} {64} (\bibinfo {year} {2020})}\BibitemShut {NoStop}%
\bibitem [{\citenamefont {Lasanta}\ \emph {et~al.}(2017)\citenamefont
  {Lasanta}, \citenamefont {Vega~Reyes}, \citenamefont {Prados},\ and\
  \citenamefont {Santos}}]{lasanta_when_2017}%
  \BibitemOpen
  \bibfield  {author} {\bibinfo {author} {\bibfnamefont {A.}~\bibnamefont
  {Lasanta}}, \bibinfo {author} {\bibfnamefont {F.}~\bibnamefont {Vega~Reyes}},
  \bibinfo {author} {\bibfnamefont {A.}~\bibnamefont {Prados}},\ and\ \bibinfo
  {author} {\bibfnamefont {A.}~\bibnamefont {Santos}},\ }\bibfield  {title}
  {\bibinfo {title} {When the {Hotter} {Cools} {More} {Quickly}: {Mpemba}
  {Effect} in {Granular} {Fluids}},\ }\href
  {https://doi.org/10.1103/PhysRevLett.119.148001} {\bibfield  {journal}
  {\bibinfo  {journal} {Physical Review Letters}\ }\textbf {\bibinfo {volume}
  {119}},\ \bibinfo {pages} {148001} (\bibinfo {year} {2017})}\BibitemShut
  {NoStop}%
\bibitem [{\citenamefont {Torrente}\ \emph {et~al.}(2019)\citenamefont
  {Torrente}, \citenamefont {López-Castaño}, \citenamefont {Lasanta},
  \citenamefont {Reyes}, \citenamefont {Prados},\ and\ \citenamefont
  {Santos}}]{torrente_large_2019}%
  \BibitemOpen
  \bibfield  {author} {\bibinfo {author} {\bibfnamefont {A.}~\bibnamefont
  {Torrente}}, \bibinfo {author} {\bibfnamefont {M.~A.}\ \bibnamefont
  {López-Castaño}}, \bibinfo {author} {\bibfnamefont {A.}~\bibnamefont
  {Lasanta}}, \bibinfo {author} {\bibfnamefont {F.~V.}\ \bibnamefont {Reyes}},
  \bibinfo {author} {\bibfnamefont {A.}~\bibnamefont {Prados}},\ and\ \bibinfo
  {author} {\bibfnamefont {A.}~\bibnamefont {Santos}},\ }\bibfield  {title}
  {\bibinfo {title} {Large {Mpemba}-like effect in a gas of inelastic rough
  hard spheres},\ }\href {https://doi.org/10.1103/PhysRevE.99.060901}
  {\bibfield  {journal} {\bibinfo  {journal} {Physical Review E}\ }\textbf
  {\bibinfo {volume} {99}},\ \bibinfo {pages} {060901} (\bibinfo {year}
  {2019})}\BibitemShut {NoStop}%
\bibitem [{\citenamefont {Santos}\ and\ \citenamefont
  {Prados}(2020)}]{santos_mpemba_2020}%
  \BibitemOpen
  \bibfield  {author} {\bibinfo {author} {\bibfnamefont {A.}~\bibnamefont
  {Santos}}\ and\ \bibinfo {author} {\bibfnamefont {A.}~\bibnamefont
  {Prados}},\ }\bibfield  {title} {\bibinfo {title} {Mpemba effect in molecular
  gases under nonlinear drag},\ }\href {https://doi.org/10.1063/5.0016243}
  {\bibfield  {journal} {\bibinfo  {journal} {Physics of Fluids}\ }\textbf
  {\bibinfo {volume} {32}},\ \bibinfo {pages} {072010} (\bibinfo {year}
  {2020})}\BibitemShut {NoStop}%
\bibitem [{\citenamefont {Biswas}\ \emph {et~al.}(2020)\citenamefont {Biswas},
  \citenamefont {Prasad}, \citenamefont {Raz},\ and\ \citenamefont
  {Rajesh}}]{biswas_mpemba_2020}%
  \BibitemOpen
  \bibfield  {author} {\bibinfo {author} {\bibfnamefont {A.}~\bibnamefont
  {Biswas}}, \bibinfo {author} {\bibfnamefont {V.~V.}\ \bibnamefont {Prasad}},
  \bibinfo {author} {\bibfnamefont {O.}~\bibnamefont {Raz}},\ and\ \bibinfo
  {author} {\bibfnamefont {R.}~\bibnamefont {Rajesh}},\ }\bibfield  {title}
  {\bibinfo {title} {Mpemba effect in driven granular {Maxwell} gases},\ }\href
  {https://doi.org/10.1103/PhysRevE.102.012906} {\bibfield  {journal} {\bibinfo
   {journal} {Physical Review E}\ }\textbf {\bibinfo {volume} {102}},\ \bibinfo
  {pages} {012906} (\bibinfo {year} {2020})}\BibitemShut {NoStop}%
\bibitem [{\citenamefont {Biswas}\ \emph {et~al.}()\citenamefont {Biswas},
  \citenamefont {Prasad},\ and\ \citenamefont {Rajesh}}]{biswas_mpemba_2021}%
  \BibitemOpen
  \bibfield  {author} {\bibinfo {author} {\bibfnamefont {A.}~\bibnamefont
  {Biswas}}, \bibinfo {author} {\bibfnamefont {V.~V.}\ \bibnamefont {Prasad}},\
  and\ \bibinfo {author} {\bibfnamefont {R.}~\bibnamefont {Rajesh}},\
  }\bibfield  {title} {\bibinfo {title} {Mpemba effect in an anisotropically
  driven granular gas},\ }\href {http://arxiv.org/abs/2104.08730} {\bibinfo
  {journal} {arXiv:2104.08730 [cond-mat]}\ }\BibitemShut {NoStop}%
\bibitem [{\citenamefont {Gómez~González}\ \emph {et~al.}(2021)\citenamefont
  {Gómez~González}, \citenamefont {Khalil},\ and\ \citenamefont
  {Garzó}}]{gomez_gonzalez_mpemba-like_2021}%
  \BibitemOpen
\bibfield  {journal} {  }\bibfield  {author} {\bibinfo {author} {\bibfnamefont
  {R.}~\bibnamefont {Gómez~González}}, \bibinfo {author} {\bibfnamefont
  {N.}~\bibnamefont {Khalil}},\ and\ \bibinfo {author} {\bibfnamefont
  {V.}~\bibnamefont {Garzó}},\ }\bibfield  {title} {\bibinfo {title}
  {Mpemba-like effect in driven binary mixtures},\ }\href
  {https://doi.org/10.1063/5.0050530} {\bibfield  {journal} {\bibinfo
  {journal} {Physics of Fluids}\ }\textbf {\bibinfo {volume} {33}},\ \bibinfo
  {pages} {053301} (\bibinfo {year} {2021})}\BibitemShut {NoStop}%
\bibitem [{\citenamefont {Takada}\ \emph {et~al.}(2021)\citenamefont {Takada},
  \citenamefont {Hayakawa},\ and\ \citenamefont {Santos}}]{takada_mpemba_2021}%
  \BibitemOpen
  \bibfield  {author} {\bibinfo {author} {\bibfnamefont {S.}~\bibnamefont
  {Takada}}, \bibinfo {author} {\bibfnamefont {H.}~\bibnamefont {Hayakawa}},\
  and\ \bibinfo {author} {\bibfnamefont {A.}~\bibnamefont {Santos}},\
  }\bibfield  {title} {\bibinfo {title} {Mpemba effect in inertial
  suspensions},\ }\href {https://doi.org/10.1103/PhysRevE.103.032901}
  {\bibfield  {journal} {\bibinfo  {journal} {Physical Review E}\ }\textbf
  {\bibinfo {volume} {103}},\ \bibinfo {pages} {032901} (\bibinfo {year}
  {2021})}\BibitemShut {NoStop}%
\bibitem [{\citenamefont {Klimontovich}(1994)}]{klimontovich_nonlinear_1994}%
  \BibitemOpen
  \bibfield  {author} {\bibinfo {author} {\bibfnamefont {Y.~L.}\ \bibnamefont
  {Klimontovich}},\ }\bibfield  {title} {\bibinfo {title} {Nonlinear {Brownian}
  motion},\ }\href {https://doi.org/10.1070/PU1994v037n08ABEH000038} {\bibfield
   {journal} {\bibinfo  {journal} {Physics-Uspekhi}\ }\textbf {\bibinfo
  {volume} {37}},\ \bibinfo {pages} {737} (\bibinfo {year} {1994})}\BibitemShut
  {NoStop}%
\bibitem [{\citenamefont {Klimontovich}(1995)}]{klimontovich_statistical_1995}%
  \BibitemOpen
  \bibfield  {author} {\bibinfo {author} {\bibfnamefont {Y.~L.}\ \bibnamefont
  {Klimontovich}},\ }\href {https://doi.org/10.1007/978-94-011-0175-2} {\emph
  {\bibinfo {title} {Statistical {Theory} of {Open} {Systems}}}}\ (\bibinfo
  {publisher} {Springer Netherlands},\ \bibinfo {address} {Dordrecht},\
  \bibinfo {year} {1995})\BibitemShut {NoStop}%
\bibitem [{\citenamefont {Lindner}(2007)}]{lindner_diffusion_2007}%
  \BibitemOpen
  \bibfield  {author} {\bibinfo {author} {\bibfnamefont {B.}~\bibnamefont
  {Lindner}},\ }\bibfield  {title} {\bibinfo {title} {The diffusion coefficient
  of nonlinear {Brownian} motion},\ }\href
  {https://doi.org/10.1088/1367-2630/9/5/136} {\bibfield  {journal} {\bibinfo
  {journal} {New Journal of Physics}\ }\textbf {\bibinfo {volume} {9}},\
  \bibinfo {pages} {136} (\bibinfo {year} {2007})}\BibitemShut {NoStop}%
\bibitem [{\citenamefont {Goychuk}\ and\ \citenamefont
  {Pöschel}(2021)}]{goychuk_nonequilibrium_2021}%
  \BibitemOpen
  \bibfield  {author} {\bibinfo {author} {\bibfnamefont {I.}~\bibnamefont
  {Goychuk}}\ and\ \bibinfo {author} {\bibfnamefont {T.}~\bibnamefont
  {Pöschel}},\ }\bibfield  {title} {\bibinfo {title} {Nonequilibrium {Phase}
  {Transition} to {Anomalous} {Diffusion} and {Transport} in a {Basic} {Model}
  of {Nonlinear} {Brownian} {Motion}},\ }\href
  {https://doi.org/10.1103/PhysRevLett.127.110601} {\bibfield  {journal}
  {\bibinfo  {journal} {Physical Review Letters}\ }\textbf {\bibinfo {volume}
  {127}},\ \bibinfo {pages} {110601} (\bibinfo {year} {2021})}\BibitemShut
  {NoStop}%
\bibitem [{\citenamefont {Ferrari}(2007)}]{ferrari_particles_2007}%
  \BibitemOpen
  \bibfield  {author} {\bibinfo {author} {\bibfnamefont {L.}~\bibnamefont
  {Ferrari}},\ }\bibfield  {title} {\bibinfo {title} {Particles dispersed in a
  dilute gas: {Limits} of validity of the {Langevin} equation},\ }\href
  {https://doi.org/10.1016/j.chemphys.2007.05.001} {\bibfield  {journal}
  {\bibinfo  {journal} {Chemical Physics}\ }\textbf {\bibinfo {volume} {336}},\
  \bibinfo {pages} {27} (\bibinfo {year} {2007})}\BibitemShut {NoStop}%
\bibitem [{\citenamefont {Ferrari}(2014)}]{ferrari_particles_2014}%
  \BibitemOpen
  \bibfield  {author} {\bibinfo {author} {\bibfnamefont {L.}~\bibnamefont
  {Ferrari}},\ }\bibfield  {title} {\bibinfo {title} {Particles dispersed in a
  dilute gas. {II}. {From} the {Langevin} equation to a more general kinetic
  approach},\ }\href {https://doi.org/10.1016/j.chemphys.2013.10.024}
  {\bibfield  {journal} {\bibinfo  {journal} {Chemical Physics}\ }\textbf
  {\bibinfo {volume} {428}},\ \bibinfo {pages} {144} (\bibinfo {year}
  {2014})}\BibitemShut {NoStop}%
\bibitem [{\citenamefont {Hohmann}\ \emph {et~al.}(2017)\citenamefont
  {Hohmann}, \citenamefont {Kindermann}, \citenamefont {Lausch}, \citenamefont
  {Mayer}, \citenamefont {Schmidt}, \citenamefont {Lutz},\ and\ \citenamefont
  {Widera}}]{hohmann_individual_2017}%
  \BibitemOpen
  \bibfield  {author} {\bibinfo {author} {\bibfnamefont {M.}~\bibnamefont
  {Hohmann}}, \bibinfo {author} {\bibfnamefont {F.}~\bibnamefont {Kindermann}},
  \bibinfo {author} {\bibfnamefont {T.}~\bibnamefont {Lausch}}, \bibinfo
  {author} {\bibfnamefont {D.}~\bibnamefont {Mayer}}, \bibinfo {author}
  {\bibfnamefont {F.}~\bibnamefont {Schmidt}}, \bibinfo {author} {\bibfnamefont
  {E.}~\bibnamefont {Lutz}},\ and\ \bibinfo {author} {\bibfnamefont
  {A.}~\bibnamefont {Widera}},\ }\bibfield  {title} {\bibinfo {title}
  {Individual {Tracer} {Atoms} in an {Ultracold} {Dilute} {Gas}},\ }\href
  {https://doi.org/10.1103/PhysRevLett.118.263401} {\bibfield  {journal}
  {\bibinfo  {journal} {Physical Review Letters}\ }\textbf {\bibinfo {volume}
  {118}},\ \bibinfo {pages} {263401} (\bibinfo {year} {2017})}\BibitemShut
  {NoStop}%
\bibitem [{\citenamefont {Van~Noije}\ and\ \citenamefont
  {Ernst}(1998)}]{van_noije_velocity_1998}%
  \BibitemOpen
  \bibfield  {author} {\bibinfo {author} {\bibfnamefont {T.~P.~C.}\
  \bibnamefont {Van~Noije}}\ and\ \bibinfo {author} {\bibfnamefont {M.~H.}\
  \bibnamefont {Ernst}},\ }\bibfield  {title} {\bibinfo {title} {Velocity
  distributions in homogeneous granular fluids: the free and the heated case},\
  }\href@noop {} {\bibfield  {journal} {\bibinfo  {journal} {Granul. Matter}\
  }\textbf {\bibinfo {volume} {1}},\ \bibinfo {pages} {57} (\bibinfo {year}
  {1998})}\BibitemShut {NoStop}%
\bibitem [{\citenamefont {Montanero}\ and\ \citenamefont
  {Santos}(2000)}]{montanero_computer_2000}%
  \BibitemOpen
  \bibfield  {author} {\bibinfo {author} {\bibfnamefont {J.~M.}\ \bibnamefont
  {Montanero}}\ and\ \bibinfo {author} {\bibfnamefont {A.}~\bibnamefont
  {Santos}},\ }\bibfield  {title} {\bibinfo {title} {Computer simulation of
  uniformly heated granular fluids},\ }\href
  {https://doi.org/10.1007/s100350050035} {\bibfield  {journal} {\bibinfo
  {journal} {Granular Matter}\ }\textbf {\bibinfo {volume} {2}},\ \bibinfo
  {pages} {53} (\bibinfo {year} {2000})}\BibitemShut {NoStop}%
\bibitem [{\citenamefont {Pöschel}\ and\ \citenamefont
  {Luding}(2001)}]{poschel_granular_2001}%
  \BibitemOpen
  \bibinfo {editor} {\bibfnamefont {T.}~\bibnamefont {Pöschel}}\ and\ \bibinfo
  {editor} {\bibfnamefont {S.}~\bibnamefont {Luding}},\ eds.,\ \href@noop {}
  {\emph {\bibinfo {title} {Granular Gases}}},\ Lecture Notes in Physics 564\
  (\bibinfo  {publisher} {Springer-Verlag Berlin Heidelberg},\ \bibinfo {year}
  {2001})\BibitemShut {NoStop}%
\bibitem [{\citenamefont {García~de Soria}\ \emph {et~al.}(2012)\citenamefont
  {García~de Soria}, \citenamefont {Maynar},\ and\ \citenamefont
  {Trizac}}]{garcia_de_soria_universal_2012}%
  \BibitemOpen
  \bibfield  {author} {\bibinfo {author} {\bibfnamefont {M.~I.}\ \bibnamefont
  {García~de Soria}}, \bibinfo {author} {\bibfnamefont {P.}~\bibnamefont
  {Maynar}},\ and\ \bibinfo {author} {\bibfnamefont {E.}~\bibnamefont
  {Trizac}},\ }\bibfield  {title} {\bibinfo {title} {Universal reference state
  in a driven homogeneous granular gas},\ }\href
  {https://doi.org/10.1103/PhysRevE.85.051301} {\bibfield  {journal} {\bibinfo
  {journal} {Physical Review E}\ }\textbf {\bibinfo {volume} {85}},\ \bibinfo
  {pages} {051301} (\bibinfo {year} {2012})}\BibitemShut {NoStop}%
\bibitem [{\citenamefont {Marconi}\ \emph {et~al.}(2013)\citenamefont
  {Marconi}, \citenamefont {Puglisi},\ and\ \citenamefont
  {Vulpiani}}]{marconi_about_2013}%
  \BibitemOpen
  \bibfield  {author} {\bibinfo {author} {\bibfnamefont {U.~M.~B.}\
  \bibnamefont {Marconi}}, \bibinfo {author} {\bibfnamefont {A.}~\bibnamefont
  {Puglisi}},\ and\ \bibinfo {author} {\bibfnamefont {A.}~\bibnamefont
  {Vulpiani}},\ }\bibfield  {title} {\bibinfo {title} {About an {H}-theorem for
  systems with non-conservative interactions},\ }\href
  {https://doi.org/10.1088/1742-5468/2013/08/P08003} {\bibfield  {journal}
  {\bibinfo  {journal} {Journal of Statistical Mechanics: Theory and
  Experiment}\ ,\ \bibinfo {pages} {P08003}} (\bibinfo {year}
  {2013})}\BibitemShut {NoStop}%
\bibitem [{\citenamefont {Prados}\ and\ \citenamefont
  {Trizac}(2014)}]{prados_kovacs-like_2014}%
  \BibitemOpen
  \bibfield  {author} {\bibinfo {author} {\bibfnamefont {A.}~\bibnamefont
  {Prados}}\ and\ \bibinfo {author} {\bibfnamefont {E.}~\bibnamefont
  {Trizac}},\ }\bibfield  {title} {\bibinfo {title} {Kovacs-{Like} {Memory}
  {Effect} in {Driven} {Granular} {Gases}},\ }\href
  {https://doi.org/10.1103/PhysRevLett.112.198001} {\bibfield  {journal}
  {\bibinfo  {journal} {Physical Review Letters}\ }\textbf {\bibinfo {volume}
  {112}},\ \bibinfo {pages} {198001} (\bibinfo {year} {2014})}\BibitemShut
  {NoStop}%
\bibitem [{\citenamefont {Abramowitz}\ \emph {et~al.}(1988)\citenamefont
  {Abramowitz}, \citenamefont {Stegun},\ and\ \citenamefont
  {Romer}}]{abramowitz_handbook_1988}%
  \BibitemOpen
  \bibfield  {author} {\bibinfo {author} {\bibfnamefont {M.}~\bibnamefont
  {Abramowitz}}, \bibinfo {author} {\bibfnamefont {I.~A.}\ \bibnamefont
  {Stegun}},\ and\ \bibinfo {author} {\bibfnamefont {R.~H.}\ \bibnamefont
  {Romer}},\ }\bibfield  {title} {\bibinfo {title} {Handbook of {Mathematical}
  {Functions} with {Formulas}, {Graphs}, and {Mathematical} {Tables}},\
  }\href@noop {} {\bibfield  {journal} {\bibinfo  {journal} {Am. J. Phys.}\
  }\textbf {\bibinfo {volume} {56}},\ \bibinfo {pages} {958} (\bibinfo {year}
  {1988})}\BibitemShut {NoStop}%
\bibitem [{Note2()}]{Note2}%
  \BibitemOpen
  \bibinfo {note} {This approximation was employed in Ref.~\cite
  {santos_mpemba_2020} to analytically investigate the Mpemba
  effect.}\BibitemShut {Stop}%
\bibitem [{\citenamefont {Resibois}\ and\ \citenamefont
  {de~Leener}(1977)}]{resibois_classical_1977}%
  \BibitemOpen
  \bibfield  {author} {\bibinfo {author} {\bibfnamefont {P.}~\bibnamefont
  {Resibois}}\ and\ \bibinfo {author} {\bibfnamefont {M.}~\bibnamefont
  {de~Leener}},\ }\href@noop {} {\emph {\bibinfo {title} {Classical {Kinetic}
  {Theory} of {Fluids}}}}\ (\bibinfo  {publisher} {John Wiley \& Sons},\
  \bibinfo {year} {1977})\BibitemShut {NoStop}%
\bibitem [{\citenamefont {Meg\'{\i}as}\ \emph {et~al.}()\citenamefont
  {Meg\'{\i}as}, \citenamefont {Santos},\ and\ \citenamefont
  {Prados}}]{megias_unpublished}%
  \BibitemOpen
  \bibfield  {author} {\bibinfo {author} {\bibfnamefont {A.}~\bibnamefont
  {Meg\'{\i}as}}, \bibinfo {author} {\bibfnamefont {A.}~\bibnamefont
  {Santos}},\ and\ \bibinfo {author} {\bibfnamefont {A.}~\bibnamefont
  {Prados}},\ }\bibinfo {note} {(in preparation)}\BibitemShut {NoStop}%
\bibitem [{Note3()}]{Note3}%
  \BibitemOpen
  \bibinfo {note} {For the ultracold gas mixture considered in Ref.~\cite
  {hohmann_individual_2017}, $\xi \simeq 674$, and the system thus corresponds
  to this limit.}\BibitemShut {Stop}%
\bibitem [{Note4()}]{Note4}%
  \BibitemOpen
  \bibinfo {note} {In the three-dimensional case, $\gamma =0.1$ for
  self-diffusion (equal masses). For the ultracold gas mixture considered in
  Ref.~\cite {hohmann_individual_2017}, $\gamma \simeq 0.067$.}\BibitemShut
  {Stop}%
\bibitem [{\citenamefont {Ritort}\ and\ \citenamefont
  {Sollich}(2003)}]{ritort_glassy_2003}%
  \BibitemOpen
  \bibfield  {author} {\bibinfo {author} {\bibfnamefont {F.}~\bibnamefont
  {Ritort}}\ and\ \bibinfo {author} {\bibfnamefont {P.}~\bibnamefont
  {Sollich}},\ }\bibfield  {title} {\bibinfo {title} {Glassy dynamics of
  kinetically constrained models},\ }\href@noop {} {\bibfield  {journal}
  {\bibinfo  {journal} {Advances in Physics}\ }\textbf {\bibinfo {volume}
  {52}},\ \bibinfo {pages} {219} (\bibinfo {year} {2003})}\BibitemShut
  {NoStop}%
\bibitem [{Note5()}]{Note5}%
  \BibitemOpen
  \bibinfo {note} {Let us note that this limit is analogous to the ``cooling''
  protocol considered in the literature to investigate the emergence of the
  Kovacs hump in a uniformly heated granular gas~\cite
  {prados_kovacs-like_2014}.}\BibitemShut {Stop}%
\bibitem [{Note6()}]{Note6}%
  \BibitemOpen
  \bibinfo {note} {These dominant terms correspond to the quadratic in $\theta
  $ ones in Eq.~\protect \textup {\hbox {\mathsurround \z@ \protect \normalfont
  (\ignorespaces \ref {evol-eq-T-a3}\unskip \@@italiccorr )}} and linear in
  $\theta $ ones in Eqs.~\protect \textup {\hbox {\mathsurround \z@ \protect
  \normalfont (\ignorespaces \ref {evol-eq-a2-a3}\unskip \@@italiccorr )}} and
  \protect \textup {\hbox {\mathsurround \z@ \protect \normalfont
  (\ignorespaces \ref {evol-eq-a3-a3}\unskip \@@italiccorr )}}.}\BibitemShut
  {Stop}%
\bibitem [{\citenamefont {Bird}(1994)}]{bird_g_a_molecular_1994}%
  \BibitemOpen
  \bibfield  {author} {\bibinfo {author} {\bibfnamefont {G.~A.}\ \bibnamefont
  {Bird}},\ }\href@noop {} {\emph {\bibinfo {title} {Molecular {Gas} {Dynamics}
  and the {Direct} {Simulation} of {Gas} {Flows}}}}\ (\bibinfo  {publisher}
  {Clarendon Press},\ \bibinfo {address} {Oxford},\ \bibinfo {year}
  {1994})\BibitemShut {NoStop}%
\bibitem [{\citenamefont {Montanero}\ and\ \citenamefont
  {Santos}(1996)}]{montanero_monte_1996}%
  \BibitemOpen
  \bibfield  {author} {\bibinfo {author} {\bibfnamefont {J.~M.}\ \bibnamefont
  {Montanero}}\ and\ \bibinfo {author} {\bibfnamefont {A.}~\bibnamefont
  {Santos}},\ }\bibfield  {title} {\bibinfo {title} {Monte {Carlo} simulation
  method for the {Enskog} equation},\ }\href@noop {} {\bibfield  {journal}
  {\bibinfo  {journal} {Physical Review E}\ }\textbf {\bibinfo {volume} {54}},\
  \bibinfo {pages} {438} (\bibinfo {year} {1996})}\BibitemShut {NoStop}%
\bibitem [{Note7()}]{Note7}%
  \BibitemOpen
  \bibinfo {note} {For a more detailed discussion of this issue, see
  Appendix~\ref {app:Sonine-expansion}.}\BibitemShut {Stop}%
\bibitem [{Note8()}]{Note8}%
  \BibitemOpen
  \bibinfo {note} {For the values of the parameters in Fig.~\ref
  {fig:universal-behaviour-temp}, this takes place for very small values of
  $\theta /\theta _{\protect \text {i}}$, namely $\theta /\theta _{\protect
  \text {i}}\lesssim 0.04$ ($1/Y\gtrsim 25$).}\BibitemShut {Stop}%
\bibitem [{Note9()}]{Note9}%
  \BibitemOpen
  \bibinfo {note} {A similar tendency of the cooling rate with the excess
  kurtosis has been found in other systems described at a kinetic level, both
  with inelastic and elastic collisions~\cite
  {lasanta_when_2017,torrente_large_2019,gomez_gonzalez_mpemba-like_2021,santos_mpemba_2020,takada_mpemba_2021}.}\BibitemShut
  {Stop}%
\bibitem [{\citenamefont {Pontryagin}(1987)}]{pontryagin_mathematical_1987}%
  \BibitemOpen
  \bibfield  {author} {\bibinfo {author} {\bibfnamefont {L.~S.}\ \bibnamefont
  {Pontryagin}},\ }\href@noop {} {\emph {\bibinfo {title} {Mathematical
  {Theory} of {Optimal} {Processes}}}}\ (\bibinfo  {publisher} {CRC Press},\
  \bibinfo {year} {1987})\BibitemShut {NoStop}%
\bibitem [{\citenamefont {Liberzon}(2012)}]{liberzon_calculus_2012}%
  \BibitemOpen
  \bibfield  {author} {\bibinfo {author} {\bibfnamefont {D.}~\bibnamefont
  {Liberzon}},\ }\href@noop {} {\emph {\bibinfo {title} {Calculus of
  {Variations} and {Optimal} {Control} {Theory}: {A} {Concise}
  {Introduction}}}}\ (\bibinfo  {publisher} {Princeton University Press},\
  \bibinfo {year} {2012})\BibitemShut {NoStop}%
\bibitem [{\citenamefont {Patr\'on}\ and\ \citenamefont
  {Prados}()}]{patron_unpublished}%
  \BibitemOpen
  \bibfield  {author} {\bibinfo {author} {\bibfnamefont {A.}~\bibnamefont
  {Patr\'on}}\ and\ \bibinfo {author} {\bibfnamefont {A.}~\bibnamefont
  {Prados}},\ }\bibinfo {note} {(in preparation)}\BibitemShut {NoStop}%
\bibitem [{Note10()}]{Note10}%
  \BibitemOpen
  \bibinfo {note} {Again, the optimal procedure would be to follow a heating
  protocol such that $a_{2\protect \text {i},A}=a_2^{\protect \qopname \relax
  m{max}}$ but, since $a_{2}^{\protect \qopname \relax m{max}}$ is very small,
  our initial preparation is nearly optimal and more practical.}\BibitemShut
  {Stop}%
\bibitem [{Note11()}]{Note11}%
  \BibitemOpen
  \bibinfo {note} {In the relaxation experiment and the Mpemba memory effect,
  the unit of temperature was formally the steady temperature
  $T_s$.}\BibitemShut {Stop}%
\bibitem [{Note12()}]{Note12}%
  \BibitemOpen
  \bibinfo {note} {Our theory underestimates $a_2^r$ by roughly 15 per cent, as
  shown by Figure~\ref {fig:LLNES-curves}.}\BibitemShut {Stop}%
\bibitem [{Note13()}]{Note13}%
  \BibitemOpen
  \bibinfo {note} {$a_2(t)<0$ when the system is cooled, as observed in
  Fig.~\ref {fig:phase-1-gamma-approach}, so that $a_2(t_w)<0$ and the effect
  remains to be normal.}\BibitemShut {Stop}%
\end{thebibliography}%

\end{document}